%% file: Yastrebov_arxiv.tex
\documentclass[11pt]{article}

\usepackage[table]{xcolor}
\usepackage{graphicx,pslatex,float,color,array,amssymb,amsmath,euscript}
\usepackage[paperwidth=21.0cm,paperheight=29.7cm,textwidth=16cm,textheight=24.5cm,top=3cm,left=2.5cm]{geometry}
\usepackage[utf8]{inputenc}
\usepackage[OT1]{fontenc}
\definecolor{navy}{rgb}{0,0,0.5}
\usepackage[%
           pdftex,
           hyperindex=true,
           colorlinks=true,
           linkcolor=navy,
           anchorcolor=magenta,
           citecolor=navy,
           urlcolor=navy,
           unicode,
           implicit=true]{hyperref}
\usepackage[%
        backend=bibtex,
        giveninits=true,
	isbn=false,
	url=false,
	style=numeric-comp,
	maxcitenames=1,
	maxbibnames=99]{biblatex}
	
\usepackage[table]{xcolor}
\usepackage{pgf}
\usepackage{fancyhdr}
\usepackage{lastpage}

\newcommand{\ddp}[2]{\frac{\partial{#1}}{\partial{#2}}} 

\title{Wave propagation through an\\elastically-asymmetric architected material}

\author{\large Vladislav A. Yastrebov \href{https://orcid.org/0000-0002-4052-3557}{\includegraphics[height=10pt]{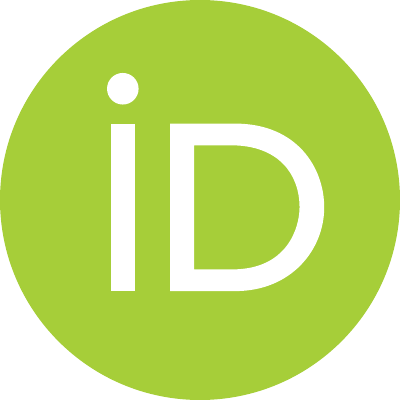}}}
\date{\footnotesize\textit{MINES ParisTech, PSL University, Centre des Mat\'eriaux, CNRS UMR 7633, BP 87, 91003, Evry, France}}

\begin{document}

\maketitle

\begin{abstract} 
\noindent A one-dimensional wave propagation through elastically asymmetric media is investigated. A class of metamaterials possessing an arbitrary elastic asymmetry is proposed. This asymmetry results in different wave speeds of tensile and compressive components of elastic waves. The faster component can overtake the slower one resulting in their dissipative annihilation through energy cascades. Efficient absorbing assemblies are presented and analysed numerically. The length of the asymmetric part needed to damp a harmonic signal is determined analytically and validated numerically. Transmission properties for random self-affine wave-packets are studied: a universal scaling for the transmission factor variation with the length of the asymmetric part was established.

\begin{flushleft}
 \textbf{Keywords:} elastic asymmetry, bimodular material, wave damping,  wave propagation,  wave annihilation,  architected material,  internal contact
\end{flushleft}

\vspace{10pt}
\hrule
\end{abstract}

\section{Introduction}

Elastically asymmetric materials are rather common in nature and include granular materials, soils, materials with internal flaws (e.g. cracked rocks, concrete) and others~\cite{jones1977stress}. 
Natural materials are in general stiffer in compression than in tension due to respective closing and opening of internal cracks~\cite{mauge1994effective}. 
However, there is a class of structures and materials which have the opposite asymmetry, i.e. they can be hard in tension and soft in compression, like wires/ropes and fibre networks, whose elements buckle under compressive loads~\cite{lake2012mechanics,abhilash2014remodeling,dirrenberger2014towards}.
In soft matter, living cells, due to elastic asymmetry of fibrin and collagen, are able to self-adjust in response to external loads~\cite{wang2001mechanical,janmey2007negative,notbohm2015microbuckling,ronceray2016fiber,du2016new}.
However, for most \emph{solid} materials the elastic asymmetry is rather small, it could be also centred not at zero deformation, and is hard to control.
In contrast, the asymmetry could be infinitely high in granular chains or granular crystals, which represent a particular class of artificial assemblies whose asymmetry and non-linearity comes from the contact interaction~\cite{boechler2010prl,jayaprakash2011nodyn,leonard2014granmat}.
In addition to elastic or inelastic asymmetry~\cite{guillemer2011cyclic,warner2008deformation}, the asymmetry of the interfacial behaviour is relevant for many systems: for example, the asymmetry of the skin drag in marine fauna~\cite{dean2010shark} or asymmetry of friction in natural (e.g., fur, snake skin) or artificial (e.g., ski-tour skis, kirigami) systems~\cite{rafsanjani2018kirigami,rafsanjani2017buckling}.
Finally, a particular type of asymmetry, which is non-centred at zero, exists in mechanical response of rate-independent elasto-plastic materials at the yield surface, and also in materials experiencing twinning~\cite{guillemer2011cyclic}, however, these types of \emph{inelastic} asymmetry are out of scope of the present discussion.

Mechanical behaviour of elastically asymmetric materials depends on the ``directionality'' of the strain tensor, i.e. on the signs of principal strains. 
Here, we define an elastically asymmetric material as a material for which exist such orientations of the strain tensor $\bar{\boldsymbol\varepsilon}$ that the following statement holds
\begin{equation}
  \forall \delta \ll 1: \; \boldsymbol\sigma(\delta\bar{\boldsymbol\varepsilon}) \ne -\boldsymbol\sigma(-\delta\bar{\boldsymbol\varepsilon}),\label{eq:0}
\end{equation}
where $\boldsymbol\sigma$ is the Cauchy stress tensor, $\delta$ is the amplitude factor, and $\delta\bar{\boldsymbol\varepsilon}$ is the infinitesimal strain tensor.
Materials with such asymmetry are called heteromodular, bimodulus or bimodular.
Therefore, the Young's, shear and bulk moduli of such materials depend on the loading direction and on its sign. 
Obviously, because of this asymmetry only ``one-sided'' derivatives or directional derivatives for the elastic tensor can be defined: 
\[
\boldsymbol C(\bar{\boldsymbol\varepsilon})= \left.\frac{\partial \boldsymbol\sigma}{\partial\boldsymbol\varepsilon}\right|_{\mathrm{dir}=\bar{\boldsymbol\varepsilon}}
\]
These elastically asymmetric materials, while retaining their strong asymmetry, can remain linear in the sense of argument multiplication (amplitude independent) : 
\begin{equation}\boldsymbol\sigma(\alpha\boldsymbol\varepsilon) = \alpha\boldsymbol\sigma(\boldsymbol\varepsilon), \;\forall \alpha \ge 0,\label{eq:1}
 \end{equation}
but not in the sense of superposition 
$$\boldsymbol\sigma(\boldsymbol\varepsilon_1+\boldsymbol\varepsilon_2) \ne \boldsymbol\sigma(\boldsymbol\varepsilon_1)+\boldsymbol\sigma(\boldsymbol\varepsilon_2).$$

In addition to this asymmetry, elastic anisotropy often comes into play~\cite{desmorat2007nonlocal,desmorat2008modeling}. 
For elastically asymmetric and isotropic materials, in particular case, the relation between the stresses and strains can be formulated in terms of principal values~\cite{ambartsumyan1966basic,jones1977stress}, with the anisotropic and asymmetric materials, the situation is more complicated: for example, a  formulation of constitutive relations for orthotropic materials can be found in~\cite{tabaddor1969two,tabaddor1972constitutive}.
However, for the best of author's knowledge, a general theory of elastic anisotropic asymmetry in which the anisotropy comes from a particular asymmetry directions, which could be further combined with the independent anisotropy of the material itself,  does not exist. 
A more general class of materials, for which, at given configuration $\{\boldsymbol\varepsilon_0,\boldsymbol\sigma\}$, the following statement holds:
$$
 -\boldsymbol\sigma(\boldsymbol\varepsilon_0 - \delta\bar{\boldsymbol\varepsilon}) \ne \boldsymbol\sigma(\boldsymbol\varepsilon_0 + \delta\bar{\boldsymbol\varepsilon}), \mbox{ for }\|\delta\bar{\boldsymbol\varepsilon}\|\to0,
$$
cannot be called elastically asymmetric.
Since such materials for $\forall\boldsymbol\varepsilon_0\ne0$ do not demonstrate asymmetric behaviour centred at zero strain,  the linearity in the sense of~\eqref{eq:1} is lost. 
Since this property is essential feature for the following analysis, this class of materials is out of scope of the current study.

Theoretical studies of quasi-static mechanical behaviour of such materials 
can be found in~\cite{ambartsumyan1966basic,jones1977stress,curnier1994conewise,sun2010review,nemat2013micromechanics}. 
However, the most interesting and intriguing effects these materials demonstrate in dynamics. Because of the strong and amplitude independent non-linearity, these materials obey a strongly unconventional wave propagation and vibrational patterns. Nevertheless, such behaviour has not yet been fully investigated:
the propagation of elastic waves in asymmetric media was studied in~\cite{maslov1985general,Lepri2011prl,gavrilov2012wave,radostin2013propagation,nazarov2017stationary}.
Since the propagation of compressive and tensile parts follow a simple linear non-dispersive equation, the particularity of such studies reduces to the understanding of propagation of signotons, the points at which the deformation changes its sign, and their motion, emergence and annihilation follows very non-trivial behaviour~\cite{maslov1985general}.
Vibrational analyses of materials with internal impacts, which render them elastically asymmetric, were conducted in~\cite{thompson1982chaos,natsiavas1993dynamics,goldstein2015study}, more results are summarized in the monograph~\cite{awrejcewicz2003bifurcation}. Here, we focus on a particular property of the overlap of compressive and tensile components, which could result in a very efficient annihilation of the waves if some damping is present in the system. This phenomenon was already studied theoretically in~\cite{radostin2013propagation}, here we carry out a numerical analysis for simple and, more complex, realistic incident wave patterns and obtain a more general conclusions for the energy damping and signal deformation.

The structure of the paper is the following, we present a new class of materials with controllable and arbitrary asymmetry centred at zero, this asymmetry comes from internally architected contacts (Section~\ref{sec:2}).
After presenting the governing equations and the solution methods in Sections~\ref{sec:3},\ref{sec:4}, to start, we show simple simulations of a mono-harmonic elastic wave propagating in bimodular one-dimensional medium (Section~\ref{sec:5}).
The damping properties of simple assemblies of bimodular materials are studied in Sections~\ref{sec:6},\ref{sec:7}. A more general study of the damping properties for random nominally self-affine and compact incident waves packets is conducted in Section~\ref{sec:8}, in Section~\ref{sec:9} we develop a simple geometrical model to explain the observed behaviour. We conclude in Section~\ref{sec:10}.

\begin{figure}[ht!]
 \includegraphics[width=1\textwidth]{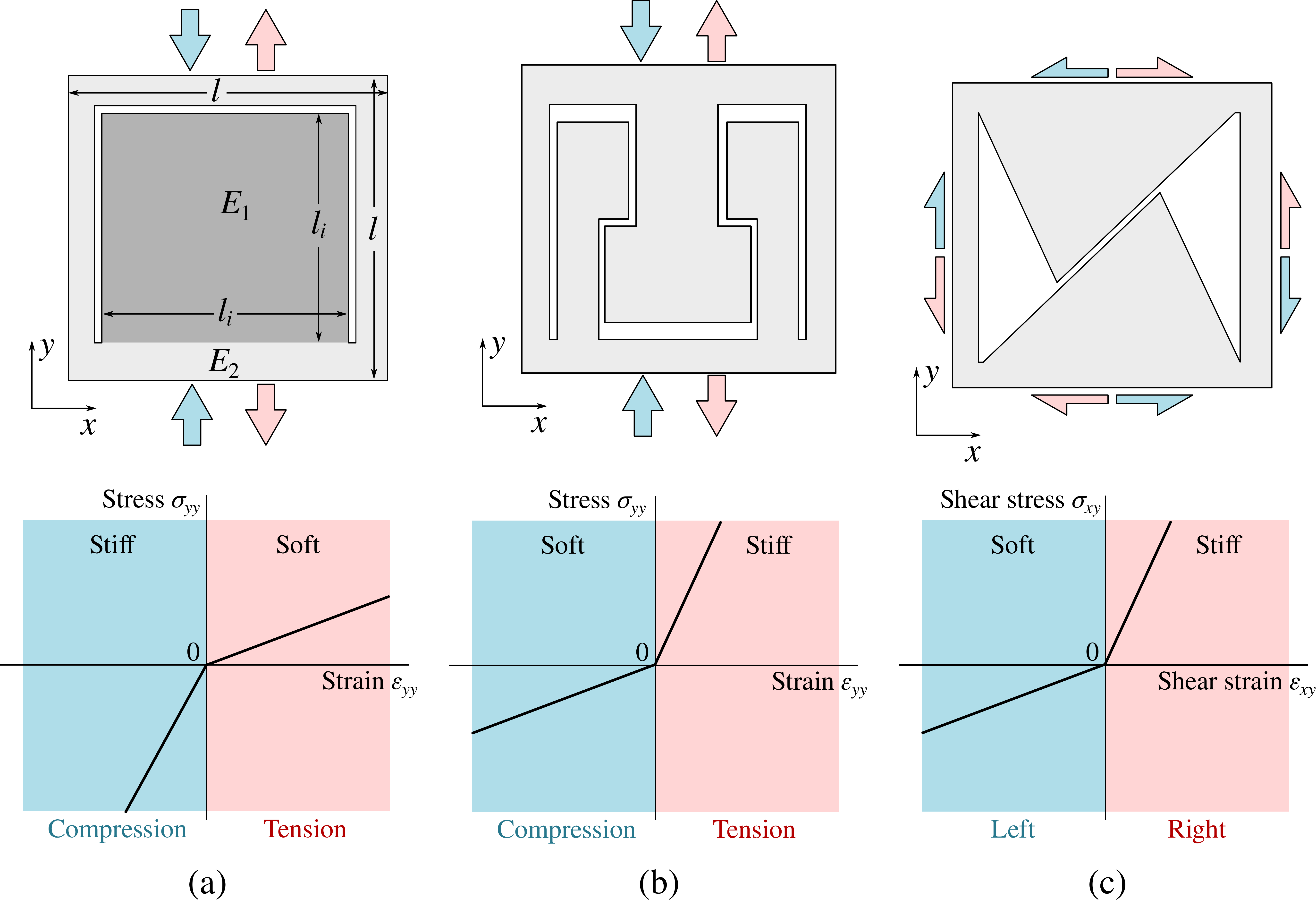}
 \caption{\label{fig:0}Examples of design of elemental architected cells with asymmetric elastic properties: the effective stiffness depends on the sign of the axial strain component (compression/tension): architecture (a) is soft-in-tension and stiff-in-compression; architecture (b) is stiff-in-tension and soft-in-compression (within a certain limit); architecture (c) is shear-asymmetric.
 The associated deformation curves are also shown.
 }
\end{figure}

\begin{figure}[t]
 \centering\includegraphics[width=1\textwidth]{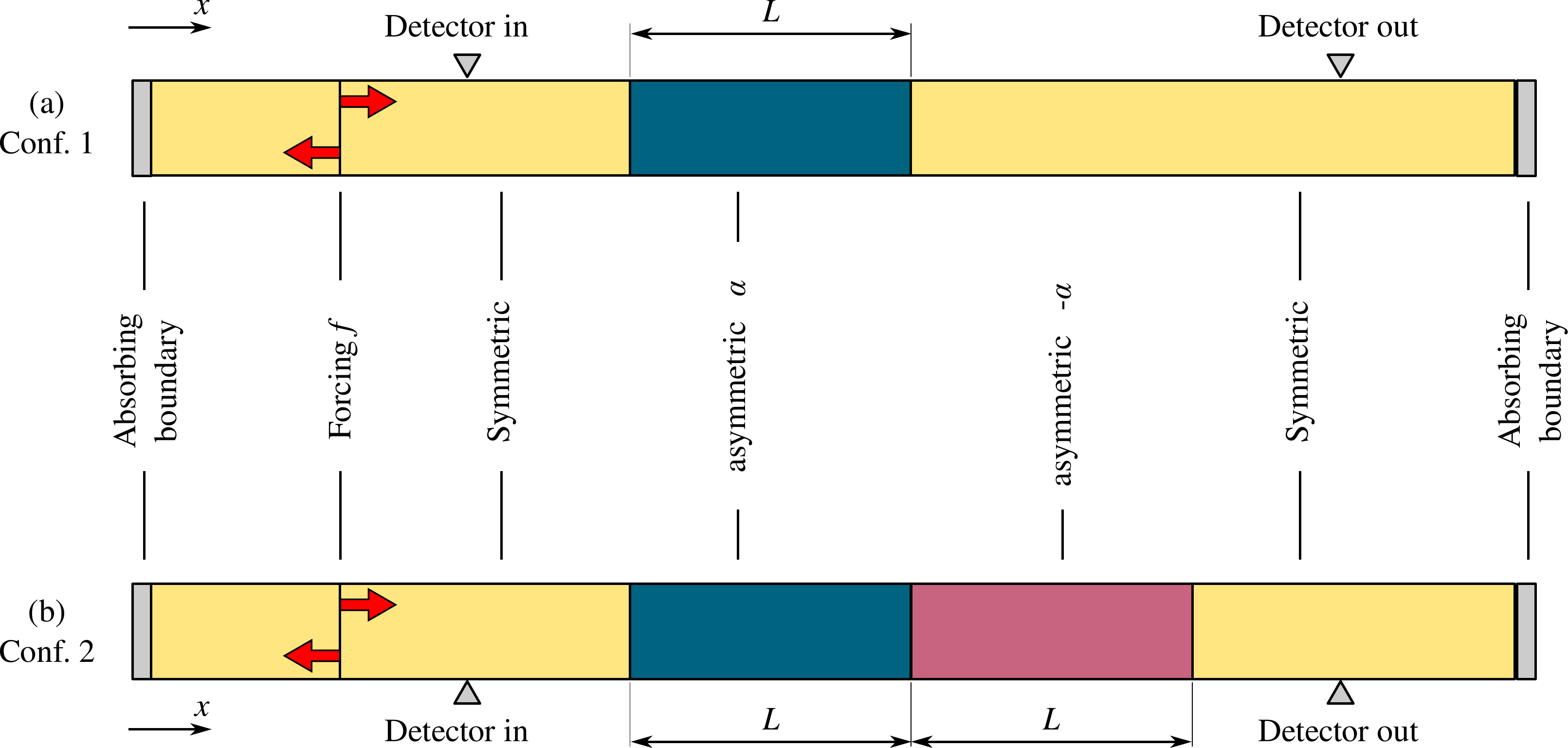}
 \caption{\label{fig:0a}
 Configurations 1 and 2, which are used to study 1D wave propagation in asymmetric media, are depicted in (a) and (b), respectively.
 }
\end{figure}

\section{Architected materials with internal contacts\label{sec:2}}

First, we propose a class of architected materials with an elastic asymmetry, which emerges from internal contacts between parts of the elemental cell (see Fig.~\ref{fig:0}). 
This asymmetry is controllable and reversible, non-destructive and arbitrary-strong, contrary to a marginal asymmetry occurring in natural solid materials.
The idea behind this novel architecture is that the non-adhesive contact can bear only compressive loads and opens in case of tensile ones. 
This non-smooth and amplitude-independent non-linearity renders the internal contact a good candidate for widening or enhancing novel and meta properties of architectured materials (see e.g.~\cite{kochmann2009dynamic,wang2014harnessing,florijn2014programmable}).
For example, if a zero-thickness cut is introduced in the material, as shown in Fig.~\ref{fig:0}(a), the resulting elastic modulus along, for example, $OY$ direction depends on the sign of the applied deformation: in case of tension, only thin ligaments bear the load and thus macroscopically the material behaves as a soft one; on the contrary, in compression, the contact in the cut is closed and can fully bear the load, thus resulting in a stiffer elastic behaviour. The elastic asymmetry can be controlled by the dimensions of the cut and by materials used in the central and peripheral zones. For the simple heterogeneous design presented in Fig.~\ref{fig:0}(a), a rough estimation of the ratio of the Young's moduli in tension $E^+$ and compression $E^-$ can be given from simple geometrical consideration as
$$\frac{E^+}{E^-} \approx \left(1-\frac{l_i}{l}\right)\frac{E_2}{E_1},$$ 
where $l$ is the square-cell size, $l_i$ is the side-length of $\Pi$-shape cut, and $E_{1,2}$ are Young's moduli of the inner and outer materials:
to amplify the asymmetry, materials can be chosen such that $E_1 > E_2$. 
Thus, the resulting asymmetry can greatly overpass the asymmetry of existing materials, which are stiffer in compression than in tension, such as rocks and concrete.
The opposite asymmetry also occurs in Nature in fibrous materials~\cite{lourie1998buckling,sears2004macroscopic,meza2014strong,dirrenberger2014towards} and living cells~\cite{notbohm2015microbuckling}: this asymmetry is based on local buckling of fibres under compression. 
In contrast to this mechanism, our architecture uses contact interaction to achieve comparable asymmetry.
A novel stiff-in-tension and soft-in-compression architecture is presented in Fig.~\ref{fig:0}(b): the contact is activated in tension and renders the material stiffer than in compression; in the latter case only thin ligaments bear the load as far as the gap remains open. 
The asymmetry of this material can be also enhanced by combination of stiff and soft materials in the architecture. 
The shear-enhanced asymmetry can be obtained, for example, through the design presented in Fig.~\ref{fig:0}(c).

Such materials demonstrate unusual properties in dynamics both in vibration and wave propagation~\cite{natsiavas1993dynamics,tournat2009nonlinear,radostin2013propagation}. 
The latter presents the main topic of this paper. 
Namely, we study propagation of elastic waves through a one-dimensional assembly of symmetric and asymmetric materials, the focus is put on the damping properties and further on the signal form change. 
We show that the elastic asymmetry modifies the energy dissipation mechanisms, ensuring rapid damping even for low-frequency signals.

\section{Wave equation\label{sec:3}}

A one-dimensional wave equation for longitudinal waves propagating through a bimodular elastic material~\cite{maslov1985general} with viscous dissipation of Kelvin-Voigt type (an elastic spring connected in parallel with a damper)~\cite{eringen1980mechanics} is of the form:
\begin{equation}
\displaystyle u_{,tt} = \frac{E}{\rho}(u_{,x}+\alpha|u_{,x}|)_{,x} + \frac{\mu}{\rho} u_{,xxt}, \quad -1 < \alpha < 1,
\end{equation}
where $u$ is the axial displacement, lower indices after comma denote partial derivation $\bullet_{,x}$ and $\bullet_{,t}$ with respect to coordinate and time, respectively; $\rho$ is the mass density (kg/m$^3$), $E$ is the elastic modulus (Pa), and $\mu$ is the viscosity (Pa$\cdot$s). The dimensionless factor $\alpha$ determines the material asymmetry: the elastic moduli are equal to 
$$E^+=E(1+\alpha), \quad E^-=E(1-\alpha)$$ 
in tension ($u_{,x}>0$) and compression ($u_{,x}<0$), respectively. 
Thus, the tensile and compressive components of elastic waves propagate at different speeds given by
$$c^\pm = \sqrt{E^\pm/\rho}.$$
Note that for a signal which has either purely compressive or tensile deformation, the wave propagation is governed by a linear Kelvin-Voigt model, however the superposition principle for deformation of different signs does not hold. 
The behaviour is independent of the signal amplitude since the non-linearity is localized in a single point on deformation curve (change of sign), which is centred at zero deformation. 
For one-dimensional systems, \emph{elastic asymmetry} reduces to \emph{bimodular} material model, thus hereinafter these two terms will be used interchangeably.

\begin{figure}[t]
 \centering\includegraphics[width=1\textwidth]{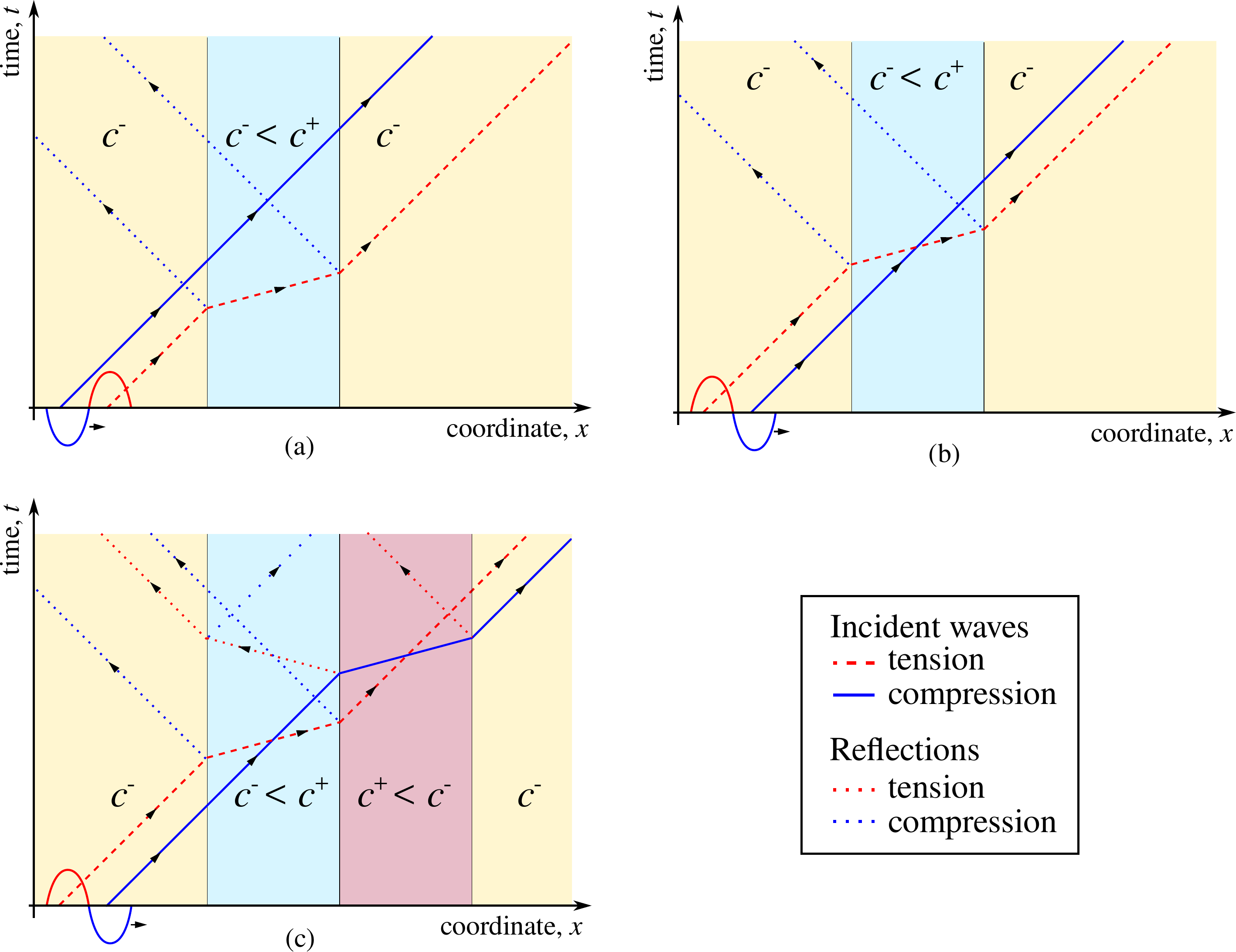}
 \caption{\label{fig:0c}
 Spatio-temporal wave tracing is presented in (a) for a tensile component (dashed line) followed by a compressive one (solid line), (b) inverse incident signal, (c) inverse incident signal passing through a configuration with sections of opposite asymmetries; (a,b) and (c) correspond to Configuration 1 and 2, respectively [see Fig.~\ref{fig:0a}(a,b)].
 }
\end{figure}
 
 \section{Methods\label{sec:4}}
 
We consider propagation of elastic waves in structures made from materials shown in Fig~\ref{fig:0}(a,b); the structure includes one or two segments with bimodular materials.
Oscillations are induced at an excitation point in an elastically symmetric segment of a bar by applying a harmonic force $f = f_0\sin(\omega_0 t)$ or a more complicated force as will be discussed in Section~\ref{sec:8}.
Forcing frequency $\omega_0$ is chosen in the interval in which viscous effects are almost negligible, i.e. $\mu \omega_0 / \rho c^{{\pm}^2} \ll 1$. 
The induced elastic waves propagate to the left and to the right: on the left they are absorbed by an absorbing layer. On the right they pass through   a single segment of length $L$ of bimodular material [\emph{Configuration 1}, see Fig.~\ref{fig:0}(c)], which without loss of generality can be considered to be stiff-in-tension and soft-in-compression, 
i.e. $\alpha_1 > 0$ and $E_1^+ > E_1^-$.
In addition, we consider \emph{Configuration 2} [see Fig.~\ref{fig:0a}(b)], which has an extra segment of a bimodular material with the opposite asymmetry, i.e. it is soft-in-tension and stiff-in-compression ($\alpha_2 = -\alpha_1 < 0$, $E_2^+ < E_2^-$). 
The bimodular segment(s), are followed by a symmetric elastic bar with an absorbing layer on its extremity. 
The equations of motion are solved numerically using finite differences and St\"ormer-Verlet integrator.
Since all the absorbing boundaries bound symmetric segments, we enforce there $u_{,t} = \pm c_0  u_{,x}$ with the minus and plus sign at the right and left boundaries, respectively, see implementation in~\cite{Supplementary}.
The spatial discretization unit $l$ reflects the size of the architectured elemental cell, thus the boundaries between symmetric and asymmetric segments pass between two springs separated by the concentrated mass. If $\lambda_0 \gg l$ (where $\lambda_0$ is a signal wavelength), such a homogenized model represents an effective medium for asymmetric architectured materials with internal contacts. The model is less reliable for short wavelengths $\lambda/l \sim 1$, but can be viewed as a first order approximation.

\begin{figure}[t!]
 \centering\includegraphics[width=0.5\textwidth]{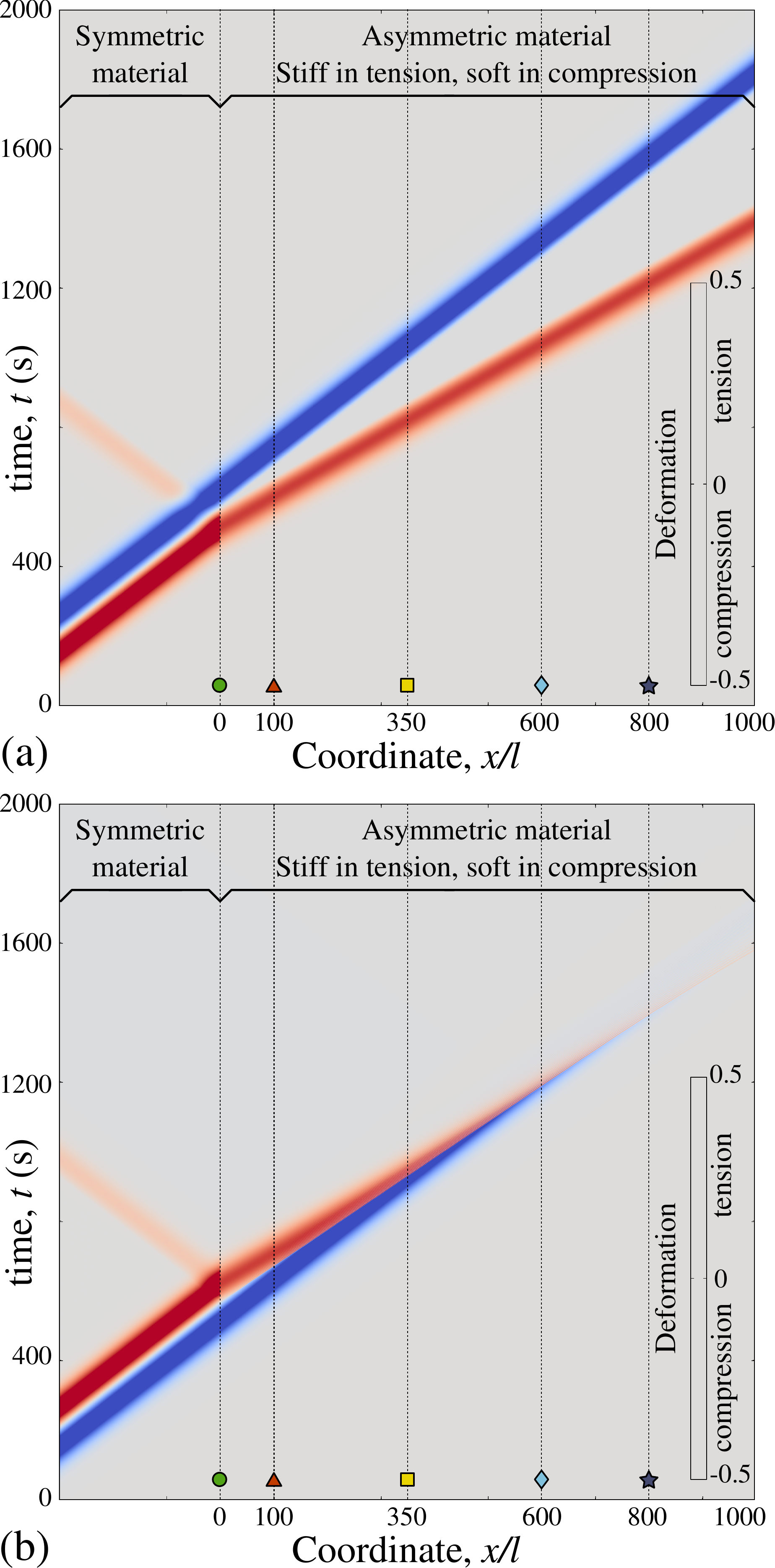}
 \caption{\label{fig:1} Simulation results for $\alpha=0.3$, $\mu/\rho$ = 0.01 m$^2$/s, $E/\rho$ = 1 m$^2$/s$^2$:
 (a) and (b) represent spatio-temporal deformation map ($u_{,x}$) and correspond to the diagram (a) and (b) in Fig.~\ref{fig:0c}, respectively; $x/l=0$ separates symmetric and asymmetric segments.
 }
\end{figure}

 \section{Simple examples\label{sec:5}}
 
Consider Configuration 1 with a single wave 
$$u_{,t}(x,t) \approx 
\begin{cases}
 -v_0
\sin\left(\omega_0\left(\frac{x}{c_0}-t\right)\right),&\mbox{ if } 0 \le t - \frac{x}{c_0} \le \frac{2\pi}{\omega_0}\\
0,&\mbox{ elsewhere.}
\end{cases}
$$
propagating from the forcing point to the right ($\omega_0>0$) at speed $c_0$ in the elastically symmetric segment. 
To avoid high frequency oscillations at extremities of the signal, we used a Gaussian smoothing to initiate this localized wave, i.e. the following forcing was applied 
$$
  f = \begin{cases}
	\mathrm{sign}(\omega_0)\left[e(t,3) - e(t,5)
\right],&\mbox{  if }0 \le t \le 4\pi/|\omega_0|\\
	0,&\mbox{ otherwise,}
      \end{cases}
$$ 
where $e(t,t'_0) = \exp\left(-(2|\omega_0| t/\pi-t'_0)^2\right)$.
In vicinity of $t=2\pi/\omega_0$ such forcing is very similar to $f = \sin(\omega_0 t-\pi)$, but it produces a smoother signal.
In this wave propagating to the right, the tensile component is followed by a compressive one.
We assume, that the elastic modulus of the symmetric segment is equal to the compressive modulus of the bimodular material $E_0 = E^-$ and $c_0 = c^-$.
When this single wave enters the bimodular segment, the tensile component accelerates abruptly and propagates at speed $c^+ > c^-$,
whereas the compressive component continues to propagate at the same speed $c^-$ and passes smoothly from symmetric to bimodular segment, and further to the other symmetric segment [see Fig.~\ref{fig:0c}(a) and  Fig.~\ref{fig:1}(a)].
In contrast, the tensile component is partly reflected back towards the emitter because of the elastic contrast: it occurs in entering and escaping the bimodular segment. 
Within the bimodular segment, because of the difference in wave speed of the compressive and tensile components, at re-entering into elastically symmetric segment, these two components are separated by $\Delta t = L/(c^+-c^-)$ in time and by $\Delta x = L/(1-c^-/c^+)$ in space.
These properties can be used to construct a wave filter, which (1) due to reflection can partly attenuate the passage of either tensile of compressive wave components, and (2) due to contrast in speeds can separate the tensile and compressive components in space/time. 

If the order of wave components is reversed [Fig.~\ref{fig:0c}(b)], i.e. $\omega_0 < 0$, then the compressive component precedes the tensile one, and the system dynamics becomes more complex. 
The slow leading component is overtaken by 
the faster tensile one and they start to interfere. Note that contrary to the purely symmetric case considered in~\cite{radostin2013propagation}, 
the ratio of amplitudes of the tensile and compressive components propagating in the bimodular segment is given by $$A^+/A^- = 2/(1+c^+/c^-)$$ (due to reflection), and their wavelengths relate as $$\lambda^-/\lambda^+ = c^-/c^+.$$
The overlap process creates a discontinuity in deformation and thus leads to emergence of high frequency oscillations and accompanying viscous dissipation, which results in partial or almost complete annihilation of tensile and compressive wave components [see Fig.~\ref{fig:1}(b)]. 
However, these oscillations do not necessarily imply high-frequency alteration between tension and compression. 
In the first stage, high frequency oscillations superpose with the compressive wave~\footnote{Note that these oscillations do not emerge if tensile and compressive waves going in opposite directions pass through each other.}. 
Since in Kelvin-Voigt model the amplitude of a wave with a real wavenumber $k$ decays in time $t$ as $\exp(-\mu k^2t/2\rho)$;
in the limit of high wavenumber $k > 2c^\pm/\mu$, the waves are overdamped and the harmonic part fully disappears.
At later stages, the oscillations produced by the overlap of tensile and compressive components lead to frequent alteration of the deformation sign.
As known from~\cite{thompson1982chaos,natsiavas1993dynamics}, oscillators with piecewise smooth characteristics posses sub-harmonic resonances at higher frequencies, which ensure relatively high amplitude and enhanced dissipation.

It is straightforward to find the propagation distance $L_o$ needed for the tensile and compressive parts of initially harmonic signal to superpose completely. 
We introduce the following notations: 
\[
E_{\max} = \max\{E^+,E^-\},\quad E_{\min} = \min\{E^+,E^-\}
\] 
and
\[
c_{\max} = \max\{c^+,c^-\},\quad c_{\min} = \min\{c^+,c^-\},
\]
the material contrast is then denoted by $$\gamma = E_{\max}/E_{\min} = (1+|\alpha|)/(1-|\alpha|) > 1$$ and $$c_{\max}/c_{\min} = \sqrt\gamma.$$ 
The overlap distance depends on wave speeds in the bimodular segment, on the amplitudes and the initial separation in time of tensile and compressive components, which is equal to a half of the period $\Delta T = \pi/\omega_0$ in the symmetric segment (where the oscillations are forced). Equating the length needed for faster and slower waves to travel to the same spatio-temporal point and requiring the full-period overlap, we obtain the overlap distance
\begin{equation}
 L_o = \frac{2\pi c_{\max}}{\omega_0\left(\sqrt{\gamma} - 1\right)}.
\label{eq:2}
\end{equation}
The full geometrical overlap is needed because, as will be shown later, the speed of the signoton separating compressive and tensile components for equal amplitude waves, travels at the average speed of the two components.
If the symmetric segment is stiff $c_0 = c_{\max}$, then the overlap distance is given by 
\[
L_o = \lambda_0/\left(\sqrt\gamma-1\right);
\]
if $c_0=c_{\min}$ then 
\[
L_o = \lambda_0+\lambda_0/\left(\sqrt\gamma-1\right).
\]
 
\begin{figure}[t!]
 \includegraphics[width=1\textwidth]{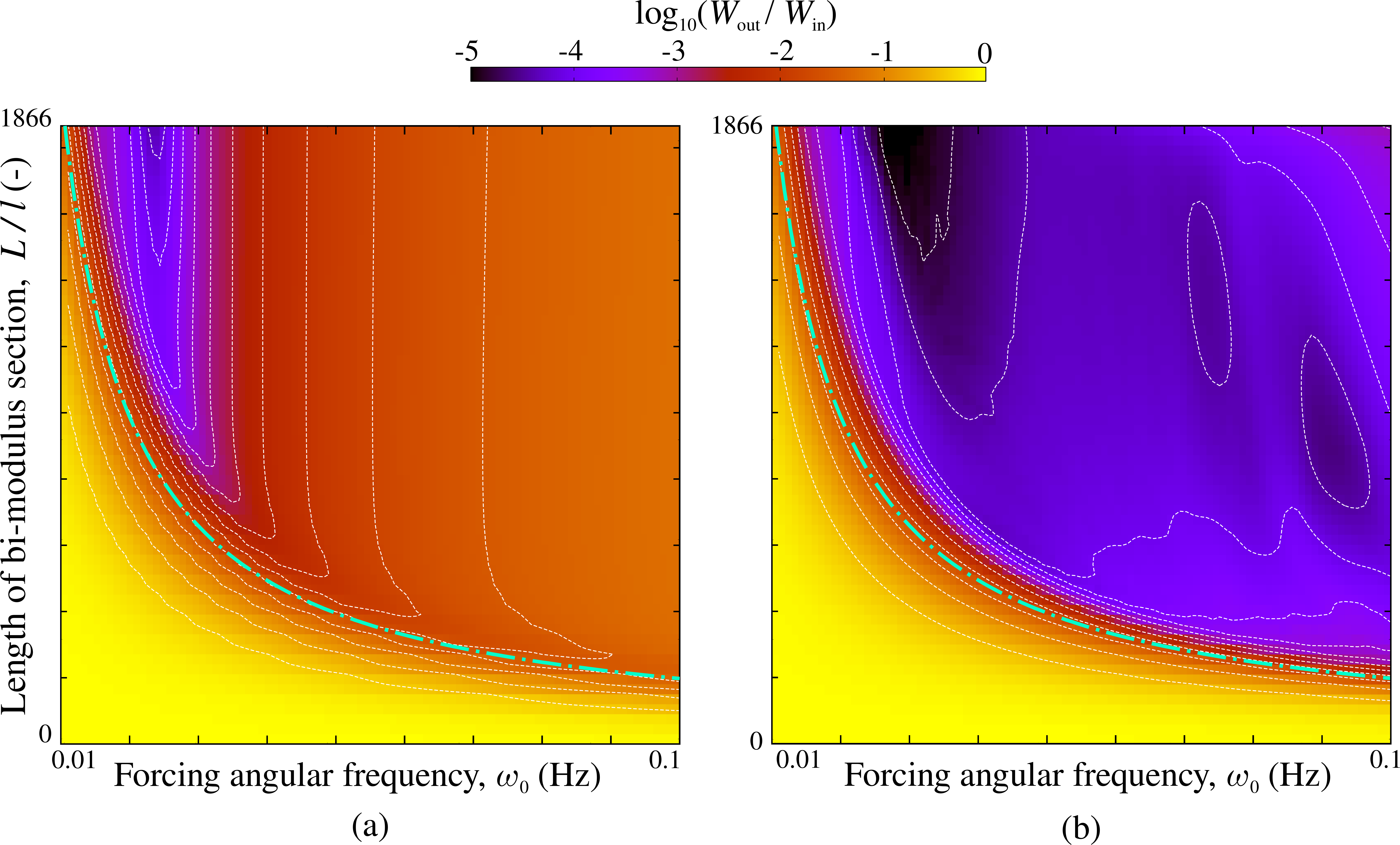}
 \caption{\label{fig:2}The decimal logarithm of the ratio of transmitted to incident energy is plotted for different forcing frequencies $\omega_0$ and different lengths of bimodular segment $L$: (a) Configuration 1 [see Figs.~\ref{fig:0a}(a),\ref{fig:0c}(b)], (b) Configuration 2 [see  Figs.~\ref{fig:0a}(b),\ref{fig:0c}(c)]. Dashed curves mark transmission iso-levels, dash-dotted curves represent Eq.~(\ref{eq:2}).
 }
\end{figure}

\section{Energy-absorbing properties\label{sec:6}}

The high-frequency cascades emerging in annihilation of tensile and compressive waves present a powerful dissipative or energy-absorbing mechanism. 
To test its properties we analyse the energy passing through the bimodular segment(s) as a function of its length and forcing frequency $\omega_0$. The injected energy is computed as the work of the forcing
\begin{equation}
W_{\mbox{\tiny in}} = \frac12\int\limits_0^T f(t) u_{,t}dt,
\label{eq:inject}
\end{equation}
where $T$ is the forcing time. 
The factor $1/2$ appears here since only a half of the energy goes towards the bimodular segment, the other part goes to the left towards the absorbing layer.
The transmitted energy $W_{\mbox{\tiny out}}$ is computed right after the bimodular segment(s) (see detectors in Fig.~\ref{fig:0}); it is computed as the energy  passing through a point $x_d$ and computed as:
\begin{equation}
W_{\mbox{\tiny out}}=A\int\limits_{0}^{T_{\text{sim}}} \left.\left[ \rho \left(\ddp{u}{t}\right)^2 + E \left(\ddp{u}{x} \right)^2 \right]\right|_{x=x_d} c_0 dt,
\label{eq:out}
\end{equation}
where $A$ is the area of the section, $c_0$ is the wave velocity in the symmetric segment and $x_d$ is the location of the detector.
The time interval $[0,T_{\text{sim}}]$ spans the entire duration of the simulation to grasp all waves and most of reflections excited by incident waves.
Because of the absorbing layer on the far-right end of the structure, 
the energy integrated over time is equal to the energy passed from left to right within the given time interval.
The calculation of the injected energy $W_{\mbox{\tiny in}}$ is also verified by the same calculations for a detector located prior to the bimodular segment(s), which slightly differ from the accurate Eq.~\eqref{eq:inject}, because Eq.~\eqref{eq:out} accounts for the reflected waves from the interfaces.

The logarithm of the ratio of transmitted to incident signal energy $\log_{10}(W_{\mbox{\tiny out}}/W_{\mbox{\tiny in}})$ is plotted in Fig.~\ref{fig:2}(a,b) as a function of the forcing frequency
$\omega_0$ and the length of the bimodular segment $L$, for Configurations 1 and 2, respectively. The following parameters were used $E/\rho = 1$ (m$^2$/s$^2$), elemental cell length is
$l=1$ (m), section area $A=1$ (m$^2$), $\alpha=0.3$, $\mu/\rho = 0.01$ (m$^2$/s).  For Configuration 1, the transmitted energy reduces significantly if the length of bimodular segment
is greater than the overlap length $L_o$ given in Eq.~(\ref{eq:2}). For Configuration 2 [see the scheme in Fig.~\ref{fig:0c}(c)], the same qualitative effect is observed, however, it is greatly amplified~\footnote{In average, the double layer enables to absorb energy in such a way that the remaining energy is two orders of magnitude less than in the case of a single layer.} by the bimodular segment of the same length $L$ but with the opposite asymmetry, which is introduced right after the first bimodular segment. 
After the first (partial) annihilation of tensile and compressive waves, the remnants of the tensile component precedes the compressive ones. 
The second segment with the inverse asymmetry serves to collide them again and dissipate their energy. 
Multiple combinations of antisymmetric segments can be used to ensure even more efficient damping as long as within every segment tensile and compressive components overtake each other, i.e. the length of the segment is greater than the overlap length $L\ge L_o$ (see Eq.~(\ref{eq:2})).
From Eq.~(\ref{eq:2}) and the simulation data it follows that  the absorption is very efficient for high and low frequencies as long as  the following condition is met:
\begin{equation}
  \omega_0 \ge \frac{2\pi c_{\max}}{L\left(\sqrt\gamma-1\right)}.
\end{equation}
A comparable damping mechanism occur for a random wave packet, which contains a roughly equal proportion of tensile and compressive components which ``annihilate'' via the same mechanism. It will be investigated in detail in Section~\ref{sec:8}.

\begin{figure}[t!]
 \includegraphics[width=1\textwidth]{Spectral_analysis_new2}
 \caption{\label{fig:3} Spectral density $\Phi(\omega)$ at locations marked in Fig.~\ref{fig:1}(b). (a) Configuration 1, scenario of Fig.~\ref{fig:0c}(b).
 Initial peak at the incident frequency $\omega_0$ is visible, as well as an emerging high frequency peak at sub-resonance frequency $2\omega_r$.
 Energy cascade with an exponent in the interval $[0.9,1]$ is highlighted.
 (b) the spectral density remains almost unchanged for Configuration 1 and scenario of non-overlapping tensile and compressive wave components, see Fig.~\ref{fig:0c}(a), Fig.~\ref{fig:1}(a).
 }
\end{figure}

\section{Spectral analysis\label{sec:7}}

Presence of high-frequency energy cascades can be shown through spectral analysis. In Fig.~\ref{fig:3} the evolution of the spectral density of deformation $u_{,x}$ is shown for the cases depicted for the simulations presented in Fig.~\ref{fig:1}. 
The power spectral density is computed at several locations $x$ along the bimodular segment (marked with dashed lines equipped with a marker in Fig.~\ref{fig:1}) for the whole time history as $\Phi(\omega,x) = \hat u_{,x}\hat u_{,x}^*$, where 
\[
\hat u_{,x}(\omega,x) = \int\limits_{0}^{T_{\text{sim}}} e^{-i \omega t}u_{,x}(x,t) dt.
\]
is the temporal Fourier transform, where $\hat \bullet^*$ denotes the conjugate value, and $\omega$ is the angular frequency. We assume that at $t<0$ and $t > T_{\text{sim}}$ the studied system is at rest. 
The system is forced mainly at frequency $\omega_0=0.03$ Hz, resulting in a smooth peak in $\Phi$ for the signal entering the bimodular segment at $x=0$.
Note that the resonance frequencies of the bimodular element is given by 
\[ 
 \omega_r = \frac{2j\sqrt{1-\alpha^2}\sqrt{E_0/\rho}}{l\left(\sqrt{1-\alpha}+\sqrt{1+\alpha}\right)}
\]
(see,~\cite{goldstein2015study}), 
where the main resonance occurs for $j=1$ and high-frequency sub-harmonic resonances occur at $j > 1$, $j\in \mathbb N$.
In the signal spectrum [Fig.~\ref{fig:3}(a)] for the case of wave overlap [Fig.~\ref{fig:1}(b)], a second peak emerges at $j=2$ sub-harmonic frequency of elemental cells.
This peak, which grows with the propagation distance, presents a sink for the energy transmitted from low frequencies. 
The energy spectrum thus contains two peaks connected via a power law decay segment with the exponent between $-1.0$ and $-0.9$ as shown in Fig.~\ref{fig:3}(c). 
Note that the energy decay of the main frequency $\omega_0$, as well as the rise and subsequent decay of sub-frequencies $n\omega_0$, for $n\in\mathbb N, n >1$ is consistent with the findings presented in~\cite{radostin2013propagation}, where the authors grasped the main features of this dissipation mechanism.
In contrast, when tensile and compressive components separate [Fig.~\ref{fig:0c}(a) and Fig.~\ref{fig:1}(a)], the spectrum depicted in Fig.~\ref{fig:3}(b) does not present any particularities and the energy spectrum remains stable.

\section{Random incident wave-packet\label{sec:8}}

\begin{figure}[htb!]
\begin{minipage}[t]{0.49\textwidth}
\centering \includegraphics[width=1\textwidth]{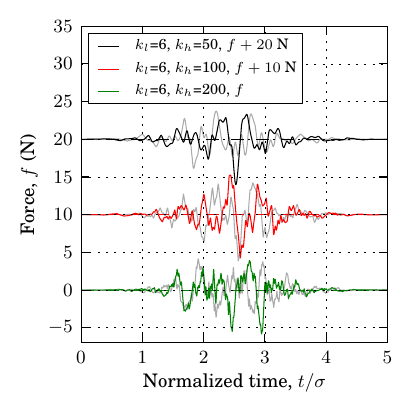}\\
 (a)
\end{minipage}
\hfill
\begin{minipage}[t]{0.49\textwidth}
 \centering \includegraphics[width=1\textwidth]{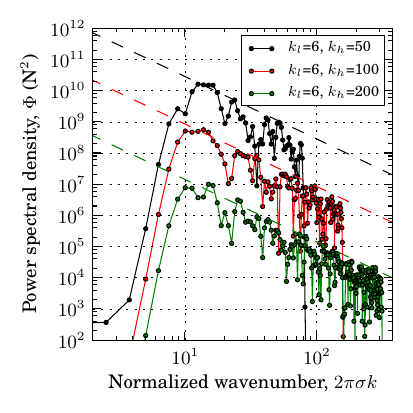}\\
 (b)
\end{minipage}
 \caption{\label{fig:wavepacket}Examples of random self-affine force used to generate incident wave packets according to Eq.~\eqref{eq:random_force}: (a) generated forces for $k_l=6$ and $k_h=\{50,100,200\}$ and $H=0.5$ are presented, two random realization are shown for every combination of parameters; (b) the corresponding power spectral densities of the coloured signals, to guide the eye, dashed curves show a power law $\sim k^{-(1+2H)}$. For the representation purpose, the plots are shifted in $Y$ axis.}
\end{figure}

\begin{figure}[htb!]
 \centering\includegraphics[width=1\textwidth]{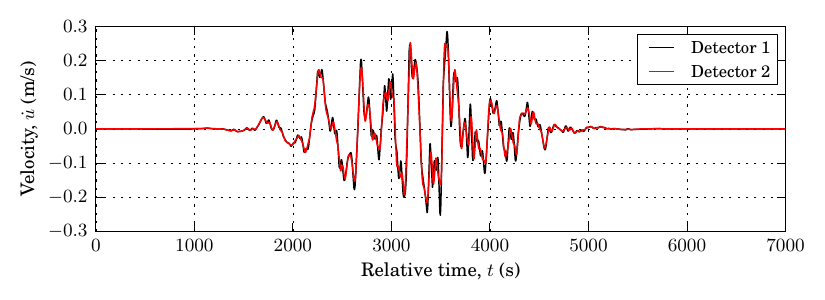}
 \caption{\label{fig:compare}The velocity at detector 2 is compared with the velocity at detector 1 to visualize how the signal changes after passing through the simulation domain due to viscous dissipation but in absence of the elastic contrast,  i.e. $\alpha=0$. The signal at detector 2 is shifted by $19053$ s to superpose with the signal at detector 1. The corresponding ratio of signal energies is $\approx 93$ \%.}
\end{figure}

Here, we consider a random incident wave-packet containing many harmonics but compact in space and time. The wave-packet is induced by a force which follows a  nominal self-affine evolution localized in time/space:
\begin{equation}
 f(t) = \exp\left(-\frac{(t-t_0)^2}{\sigma_i^2}\right) \sum\limits_{k=k_l}^{k_h} A_k \left(\frac{k}{k_l}\right)^{-(0.5+H)} \sin\left(\frac{2\pi k t}{T_i} + \varphi_k\right),
 \label{eq:random_force}
\end{equation}
where $A_k$ is a random amplitude given by $A_k = 1 + 0.2r_k$ (N), where $r_k \in [-1,1]$ is a random variable with a uniform probability density; $\varphi_k \in [0,2\pi]$ is a random phase which is also selected from a uniform distribution, 
$k_l$ is the lower summation number and represents a normalized lower cut-off wavenumber, $k_h>k_l$ is the upper summation limit and represents the upper cut-off wavenumber, so $k_h-k_l$
is the number of modes present in the signal, $\sigma_i$ (s) determines the duration of the signal and $t_0$ is the centred signal time, $T_i$ is the basic period and is selected to be $T_i = 4\sigma_i$ to avoid repetition of patterns in the incident signal; $0 < H < 1$ is the Hurst exponent controlling the self-affinity, so the spectrum of the signal follows a power law decay $\langle \Phi_k \rangle \sim k^{-(1+2H)} = k^{-(5-2D_f)}$, where the average $\langle\bullet\rangle$ represents the ensemble average and $D_f = 2-H$ is the fractal dimension, $D_f \in [1,2]$. Examples of random wave-packets and their spectra are shown in Fig.~\ref{fig:wavepacket}.
Because of the viscous dissipation in the system, to study the effect of bimodular segments, one has to select such parameters for the incident signal so that in absence of such a segment, the signal passes through the entire system without significant filtering of high frequency modes. The selected parameters are summarized in Tab.~\ref{table:1}, for such parameters the ratio of transmitted to injected energy measured at the two detectors is $W_{\mbox{\tiny out}}/W_{\mbox{\tiny in}} \approx 0.93$; the signal at detector 2 is compared with the signal at detector 1 in Fig.~\ref{fig:compare} for $\alpha=0$, other parameters remain the same as in Tab.~\ref{table:1}.
The simulation domain and its duration are kept rather big because of the rich content of the signal.

\begin{table}[ht]
\begin{center}
 \begin{tabular}{lccc}
  \textbf{Parameter} & \textbf{notation} & \textbf{value} & \textbf{units}\\
  \hline
  Signal lower cut-off & $k_l$ & 6 & -\\
  Signal higher cut-off & $k_h$ & 100 & -\\
  Standard deviation & $\sigma_i$ & 1000 & s\\
  Basic period & $T_i$ & 4000 & s\\
  Central time & $t_0$ & 3000 & s\\
  Hurst exponent & $H$ & 0.5 & - \\
  Elastic contrast & $\alpha$ & [-0.3,0.3] & -\\
  Kinematic viscosity & $\mu/\rho$ & 0.003 & m$^2$/s\\
  Reduced stiffness of symmetric layer & $E_0/\rho$ & 1 & m$^2$/s$^2$\\
  Reduced tension stiffness in first meta-segment & $E^+/\rho$ & 1 & m$^2$/s$^2$\\
  Simulated normalized length & $L/l$ & 20\,000 & -\\
  Simulated time & $T$ & 30\,000 & s\\
  Time step & $dt$ & 0.2 & s\\[5pt]
\end{tabular}
\end{center}
\caption{\label{table:1}Parameters used in simulation of a transmission of a random incident wave-packet, see Eq.~\eqref{eq:random_force} for notations.}
\end{table}

\subsection{Transmission of energy}

In Fig.~\ref{fig:energies}(a) we plot the energy ratio $W_{\mbox{\tiny out}}/W_{\mbox{\tiny in}}$ (transmission factor) as a function of bimodular length $L$ for Configuration 1 (without second segment with the opposite asymmetry) and different elastic contrasts $\alpha$. 
As seen from the figure, all the data could be described by the same function with an exponential decay (to a certain positive limit) with respect to the length of the bimodular segment raised to the power of $3/2$:
\begin{equation}
 \mathcal T(L) = \frac{ W_{\mbox{\tiny out}}}{ W_{\mbox{\tiny in}} } = \mathcal T_{\min} + (\mathcal T_{\max} - \mathcal T_{\min}) \exp\left(-(L/L_*)^{3/2}\right),
 \label{eq:transmission_fit0}
\end{equation}
where $\mathcal T_{\min},\mathcal T_{\max}$ are the minimal and the maximal values obtained for saturated transmitted energy and transmitted energy in absence of bimodular segment, respectively;
$L_*$ is the characteristic length which characterizes the reduction in transmitted energy.
However, it would be reasonable to suggest a renormalization of the bimodular-segment length in order to account for the difference is elastic moduli used for positive and negative values of $\alpha$:  the characteristic length of intersection would scale as $L \sim c^{\max} - c^{\min}$, therefore it would be reasonable to plot the transmission factor with respect to the renormalized bimodular length $L'$
\[
L' = \begin{cases}
     \displaystyle \frac{L}{l}\frac{2}{1 - \sqrt{(1-\alpha)/(1+\alpha)}}, & \text{ if } \alpha > 0,\\
     \displaystyle \frac{L}{l}\frac{2}{\sqrt{(1+\alpha)/(1-\alpha)}-1}, & \text{ if } \alpha < 0,\\
     \end{cases}
\]
The mean value of the transmission factor and the error bars (root-mean-square deviation) are obtained for 10 simulations carried out for different realizations of the incident signal (see Supplementary material~\cite{Supplementary}).
Contrary to a mono-harmonic signal, for a random wave-packet the transmission factor cannot be reduced below $\approx 10$ \% because of eventual full elimination of either tensile or compressive wave components,
so that the remaining component, accordingly compressive or tensile, persists. In order to obtain a more precise expression, it would be reasonable to take into account the reflected part of the signal, i.e. the part of the signal that does not travel through the bimodular segment, so the ultimate form of the transmission factor with the normalization of the bimodular-segment length takes the form:
\begin{equation}
 \mathcal T'(L') = \frac{ W_{\mbox{\tiny out}} + W_{\mbox{\tiny refl}}}{ W_{\mbox{\tiny in}} } = \mathcal T'_{\min} + (\mathcal T'_{\max} - \mathcal T'_{\min}) \exp\left(-(L'/L'_*)^{3/2}\right) + \frac{W_{\mbox{\tiny refl}}}{ W_{\mbox{\tiny in}} }.
 \label{eq:transmission_fit}
\end{equation}
The last term can be estimates as $W_{\mbox{\tiny refl}} / W_{\mbox{\tiny in}} \approx \left(1 - 2/(1+\sqrt{(1-\alpha)/(1+\alpha)})\right)^2$ for $\alpha < 0$, for $\alpha>0$ the reflected waves have a much smaller energy and can be neglected. Even for the negative $\alpha$, the value of the reflected energy is relatively small as could be seen in Fig.~\ref{fig:energies}(a) as the shift between curves in the asymptotic limit of $L\to 0$.
Note that this simplification~\eqref{eq:transmission_fit} is possible because the reflected energy is independent of the length of bimodular segment. The transmission factor taking into account the reflected energy $\mathcal T'$ is shown in Fig.~\ref{fig:energies}(b) and the relevant normalization of the bimodular segment length $L'$ allows to collapse all the data points on a single master curve with the following parameters: 
fixed $\mathcal T'_{\max} = 0.93$ and those obtained by the mean least square fit $\mathcal T'_{\min} = 0.1128$, $L'_* = 438.37$.

\begin{figure}[ht!]
\begin{center}
\input{RandomTransmission_DifAlpha_NoNormalizationShift_NoReverso_.pgf}\\(a)\\
\input{RandomTransmission_DifAlpha_NormalizationShift_NoReverso_.pgf}\\(b)
\end{center}
\caption{\label{fig:energies}(a) Transmission factor $\mathcal{T}$ (ratio of transmitted to injected energy) is plotted with respect to the length of the bimodular segment $L$ for different elastic contrasts $\alpha$; the curves represent Eq.~\eqref{eq:transmission_fit0} with parameters identified by the mean least square fit, dashed curves correspond to $\alpha<0$, solid curves to $\alpha>0$.
(b) Transmission factor taking into account reflected energy $\mathcal{T}'$ and plotted with respect to the normalized length of the bimodular segment $L'$, the master curve represent Eq.~\eqref{eq:transmission_fit} with parameters identified by the mean least square fit; the same data points as in (a) are plotted; the points represent the average data computed over 10 realizations and the error bars are equal to the standard deviation.}
\end{figure}

\subsection{Signal change}

Apart from the reduction in the transmitted energy, it is important to understand how the shape of the velocity signal changes after passing through the bimodular segment. In Fig.~\ref{fig:in_out_signals} we show examples of the input and output signal for the length of bimodular segment $L/l = 5\,000$ and $\alpha = \pm 1/4$. The most evident observation is the change in standard deviation $\sigma$ and average ``sign'' of the signal.
For $\alpha > 0$ the output signal contains rather negative velocities, and vice versa, for $\alpha < 0$ the signal contains rather positive components.
This change is important for understanding of the energy absorption in the bimodular segment, because such trend in converting the signal into purely mono-sign signal does not permit further overlap and annihilation, moreover, this ``sign-polarization'' results in a net motion to the right or to the left of the entire system.

\begin{figure}
\begin{center}
 \input{SignalsInOut_k_1.66667_seed_128.pgf}\\(a)\\
 \input{SignalsInOut_k_1.66667_seed_129.pgf}\\(b)
 \end{center}
 \caption{\label{fig:in_out_signals}Illustration of signal's change for two random realizations. The incident signal (upper panels in (a,b)) deforms passing through the bimodular segment of length $L/l=5\,000$ and takes the form of sign-polarized signal with mainly positive components for $\alpha > 0$ and negative components for $\alpha < 0$, see middle and lower panels in (a,b).}
\end{figure}

\begin{figure}[htb!]
 \begin{minipage}[t]{0.45\textwidth}
  \centering
  \includegraphics[width=1\textwidth]{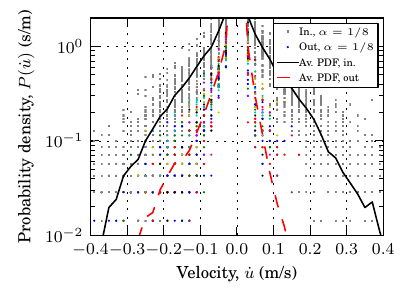}\\(a)
  \includegraphics[width=1\textwidth]{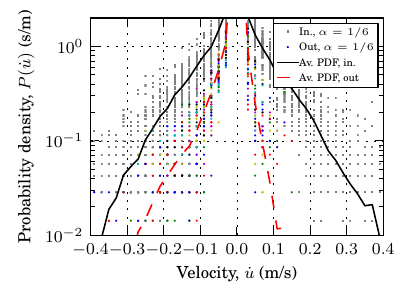}\\(c)\\
   \includegraphics[width=1\textwidth]{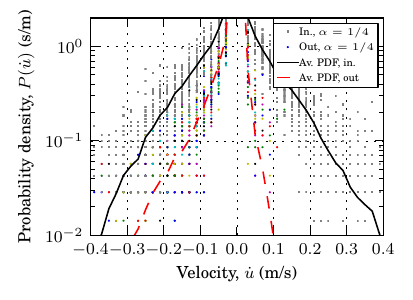}\\(e)\\
 \end{minipage}
 \hfill
 \begin{minipage}[t]{0.45\textwidth}
 \centering
 \includegraphics[width=1\textwidth]{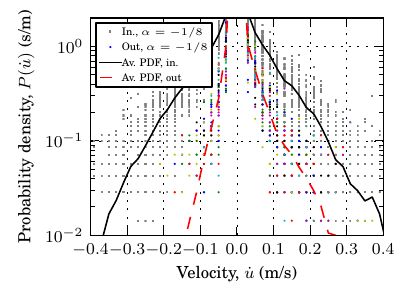}\\(b)
 \includegraphics[width=1\textwidth]{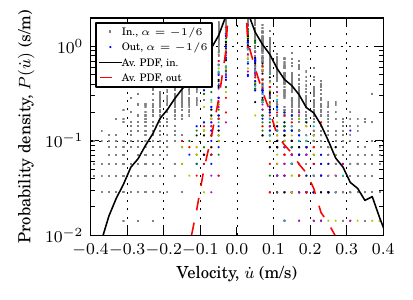}\\(d)\\
  \includegraphics[width=1\textwidth]{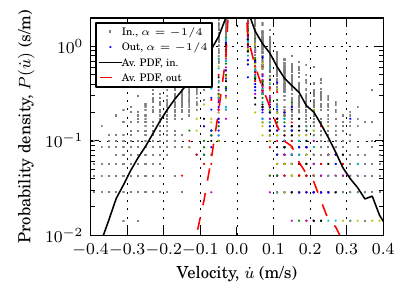}\\(f)\\
 \end{minipage}
 \caption{\label{fig:in_out_pdfs}Probability density (PDF) of velocities of the incident signal and of the output signal which passed through the bimodular segment of length $L/l = 5\,000$, individual points represent PDF values of a particular test and the curves are the average data computed over 50 realizations, both distributions are computed over 50 bins: (a,b) $\alpha=\pm 1/8$, (c,d) $\alpha=\pm 1/6$, (e,f) $\alpha=\pm 1/4$. All the data are available in Supplementary material~\cite{Supplementary}.}
\end{figure}

The ``sign'' of the signal can be characterized by skewness  of the distribution $\tilde\mu_3 = \mu_3/\sigma^3$, where $\mu_3$ is the third moment of the probability distribution. For our data, the distributions are computed over an input signal in the window of $t\in[0,7\sigma_i]$  and for the output signal in $t\in[t_{\max}-3.5\sigma,t_{\max}+3.5\sigma]$, where $t_{\max}$ corresponds to the $\max{|\dot u|}$ of the output signal. The standard deviation of the signal reduces by a factor of three, and the skewness becomes considerably positive for negative $\alpha$ and vice versa, i.e. the skewness is negative for positive $\alpha$, see Table~\ref{tab:skew}(a). For $L/l = 5\,000$ the dependence on the level of elastic contrast $\alpha$ is quite marginal, whereas for shorter lengths, for example, $L/l=1000$ this dependence becomes pronounced since such a length is not sufficient to ensure overlap of positive and negative parts, see Table~\ref{tab:skew}(b).  Note that in the Configuration 2, in presence of the segment of the inverse asymmetry, the skewness remains almost the same and the standard deviation is slightly reduced by $\approx 12.5$ \%.
In Fig.~\ref{fig:in_out_pdfs} we show velocity probability densities of the incident signal and of the output signal which passed through the bimodular of length $L/l = 5\,000$, the data are shown for different values of the elastic contrast $\alpha$.
Note that incident signals are the same for different $\alpha$, the small difference in PDFs is explained by different reflections from the bimodular interface.

\begin{table}[ht]
\begin{center}
 \begin{tabular}{ccccc}
  & \multicolumn{2}{c}{Input signal} & \multicolumn{2}{c}{Output signal}\\
  $\alpha$ & std $\sigma$ & skewness $\tilde\mu_3$  & std $\sigma$ & skewness $\tilde\mu_3$\\[3pt]
  \hline 
 -1/4 & 0.0634 & 0.1882 & 0.0214 & 5.2150 $\vphantom{A^{A^A}}$\\ 
 -1/5 & 0.0631 & 0.1579 & 0.0210 & 4.9662\\ 
 -1/6 & 0.0630 & 0.1399 & 0.0207 & 4.7409\\ 
 -1/7 & 0.0623 & -0.1475 & 0.0210 & 4.4319\\
 -1/8 & 0.0628 & 0.1184 & 0.0206 & 4.3772\\  
  1/8 & 0.0622 & -0.0146 & 0.0205 & -5.0246\\ 
  1/7 & 0.0620 & -0.2617 & 0.0193 & -4.6584\\
  1/6 & 0.0621 & -0.0355 & 0.0208 & -5.2974\\ 
  1/5 & 0.0621 & -0.0525 & 0.0212 & -5.4514\\ 
  1/4 & 0.0620 & -0.0803 & 0.0216 & -5.8283\\[5pt] 
 \end{tabular}\\
\centering (a) $L/l = 5\,000$\\[5pt]
 \begin{tabular}{ccccc}
 & \multicolumn{2}{c}{Input signal} & \multicolumn{2}{c}{Output signal}\\
 $\alpha$ & std $\sigma$ & skewness $\tilde\mu_3$  & std $\sigma$ & skewness $\tilde\mu_3$\\[3pt]
 \hline 
-1/4 & 0.0627 & -0.0225 & 0.0397 & 0.8776 $\vphantom{A^{A^A}}$\\
-1/5 & 0.0626 & -0.0600 & 0.0443 & 0.4105\\
-1/6 & 0.0625 & -0.0836 & 0.0470 & 0.1804\\
-1/7 & 0.0624 & -0.1011 & 0.0489 & 0.0718\\
-1/8 & 0.0623 & -0.1139 & 0.0503 & 0.0239\\
1/8 & 0.0620 & -0.2894 & 0.0486 & -0.7302\\
1/7 & 0.0620 & -0.2979 & 0.0462 & -0.7522\\
1/6 & 0.0620 & -0.3085 & 0.0432 & -0.9362\\
1/5 & 0.0621 & -0.3222 & 0.0383 & -1.3727\\
1/4 & 0.0621 & -0.3406 & 0.0320 & -2.0355\\[5pt]
 \end{tabular}\\
\centering (b) $L/l = 1\,000$\\[5pt]
\end{center}
\caption{\label{tab:skew}Averaged standard deviation and skewness of the input and output  signals for different vales of $\alpha$ and for (a) $L/l=5\,000$ averaged over 50 realizations, (b) $L/l=1\,000$ averaged over 10 realizations.}
\end{table}

\begin{figure}[htb!]
\begin{minipage}[t]{0.49\textwidth}
\centering \includegraphics[width=0.75\textwidth]{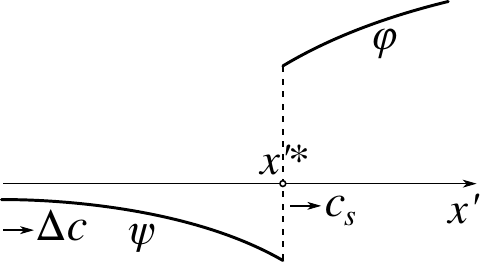}\\
 (a)
\end{minipage}
\hfill
\begin{minipage}[t]{0.49\textwidth}
\centering \includegraphics[width=0.75\textwidth]{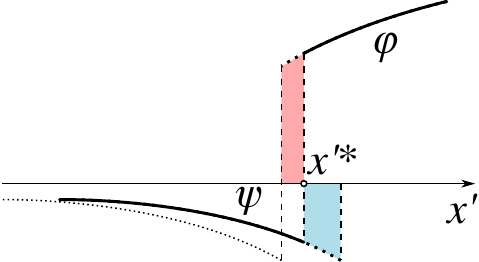}\\
 (b)
\end{minipage}
 \caption{\label{fig:explication}Motion of a signoton $x'^*$ at speed $c_s$ in coordinate system at which the positive part $\varphi$ remains at rest: configuration (a) at $t_0$ and (b) at $t_0+\Delta t$, dotted lines represent the signal at time $t_0$. Dashed areas under the curves for slowly varying $\varphi$ and $\psi$ and for $\Delta t \to 0$ are given by $\phi(x'^*)c_s\Delta t$ for the tensile part (reddish one) and $|\psi(x'^*)|(\Delta c-c_s)\Delta t$, these areas are equal and they correspond to the deformation amplitude loss.}
\end{figure}

\section{Geometrical model\label{sec:9}}

A simple geometrical explication could be given to the observed decrease in the transmitted energy. Let us consider a wave-packet in a symmetric non-dispersive medium travelling to the right at wave speed $c_0$ as a superposition of the positive and negative parts $u_{,x}(x-c_0t) = \langle u_{,x}(x-c_0t)\rangle - \langle -u_{,x}(x-c_0t)\rangle$, where $\langle x \rangle = \max\{0,x\}$, within the bimodular segment the speed of propagation for these parts would be different, moreover, the positive and negative parts will change due to their gradual superposition accompanied by energy cascades:
\[
 u_{,x}(x,t) = \langle u_{,x}(x-c_0t,t)\rangle - \langle -u_{,x}(x-c_0\sqrt{(1-\alpha)/(1+\alpha)} t,t)\rangle,
\]
by changing the variable $x' = x - c_0t$ and introducing new notations $\varphi(x',t) = \langle u_{,x'}(x',t)\rangle$ and $\psi(x'-\Delta c t, t) = -\langle -u_{,x'}(x',t)\rangle$, where $\Delta c = c_0(\sqrt{(1-\alpha)/(1+\alpha)} - 1)$ ,we obtain the following equation to represent the signal:
\[
 u_{,x'}(x',t) = \varphi(x',t) + \psi(x'-\Delta c t, t),
\]
of course, far from signotons, i.e. the points $x'^*(t)$ at which the function changes sign $u_{,x'}(x'^*+,t)u_{,x'}(x'^*-,t)<0$, functions $\varphi$ and $\psi$ preserves their form, i.e. their time derivatives take the form $\varphi_{,t} = 0$ and 
$\psi_{,t} = -\Delta c \psi_{,x'}$. However, at signotons the signal looses some amplitude following (see Fig.~\ref{fig:explication}):
\[
 \int\limits_{-\infty}^{\infty} u_{,x't} dx = 2\Delta c \sum_{x'^*} \frac{\varphi(x'^*)|\psi(x'^*)|}{\varphi(x'^*)+|\psi(x'^*)|},
\]
where the sum is over all signotons and it always holds that $\varphi\ge0$ and $\psi\le 0$. This result could be obtained by simple geometrical interpretation sketched in Fig.~\ref{fig:explication} for a signoton moving at speed $c_s$:
if we assume small change of deformation at distance $c_s \Delta t$, i.e. $|\varphi(x) - \varphi(x+c_s\Delta t)| \ll 1$ and $|\psi(x) - \psi(x-c_s\Delta t)| \ll 1$ then in the course of its propagation from instance $t_0$ to instance $t_0+\Delta t$, the signoton will release the integral of deformation of the positive part given by $c_s \varphi(x'^*)\Delta t$ which appears to be equal to the lost deformation of the negative part $|\psi(x'^*)|(\Delta c - c_s)\Delta t$. 
By equating these two terms the speed of the signoton $c_s$ can be found, it depends on the asymmetry of the jump and in the moving coordinate system $x'$ it is given by:
\[
 c_s(x'^*) = \Delta c\left. \frac{|\psi|}{\varphi+|\psi|}\right|_{x'=x'^*}.
\]
To obtain its absolute speed, one needs to simply add $c^+$ speed.
As follows from this equation, the speed of the signoton can be non-constant even though the positive and negative parts propagate at constant speeds. 
For equal values of $\varphi$ and $\psi$, the speed of the signoton is simply $c_s = \Delta c/2$, which was used in Section~\ref{sec:5} to deduce the full overlap distance.
The squared values of all deformation losses with a Young's modulus factor and integrating them in time and in section $A$ would give us the total loss in potential energy predicted by this simplified geometric model:
\begin{equation}
  W_{\text{loss}}(t) =\frac{A(E^+ + E^-)}{2} \Delta c \int\limits_0^t \sum\limits_{x'^*(t')}  \left( \frac{\varphi(x'^*)\psi(x'^*)}{\varphi(x'^*)+|\psi(x'^*)|} \right)^2 dt'.
  \label{eq:geom_loss}
\end{equation}
Even though this equation is not easy to use with a random signal, it could serve to obtain an asymptotic solution for small overlaps, it could be also helpful for a numerical treatment or for the analysis of simple signals.

\begin{figure}[htb!]
\begin{minipage}[t]{0.49\textwidth}
\centering 
\input{AnnihilationSignal_1.pgf}\\(a)\\
\input{AnnihilationSignal_2.pgf}\\(c)
\end{minipage}
\hfill
\begin{minipage}[t]{0.49\textwidth}
\centering 
\input{AnnihilationSignal_1_model.pgf}\\(b)\\
\input{AnnihilationSignal_2_model.pgf}\\(d)
\end{minipage}
 \caption{\label{fig:simple_annihilation}Evolution of the deformation signal $u_{,x'}(x',t)$ in time in the reference coordinate system $x'$ moving at speed $c^+$, this evolution is simulated with a geometrical overlap model for (a) $a^+ = a^-$ and (c) for $a^-/a^+ = 1.5$; the remaining parameters are $\rho=2$ kg/m$^2$, $E^+ = 1$ Pa, $E^-=2$ Pa, $c^+ = 1/\sqrt{2}$ m/s, $c^- = 1$ m/s, $\Delta c = 1-1/\sqrt{2}$ m/s, $\lambda=1$ m, $k = 4\pi/\lambda$, $dt=\lambda/(N\Delta c)$, where $N=4\,000$ is the spatial discretization. Every snapshot taken every 100 time steps $dt$ is shifted vertically by the value of $500dt$.
 In (b) and (d) the evolution of the remaining potential energy in the signal compared to the initial energy is computed using the geometrical overlap model and compared with analytical solution Eq.~\eqref{eq:geom_model_sym} in (b); in (d) solid, dashed and dash-dotted lines represent the remaining energy computed by Eq.~\eqref{eq:remaining_energy}.}
\end{figure}

Let us consider, for example, a signal $\varphi = \langle a^+ \sin(k(x-c^+t))\rangle$ and $\psi = -\langle - a^- \sin(k(x-c^-t))\rangle$, values of $\varphi$ and $\psi$ at signoton in the moving coordinate system $x' = x - c^+t$ will be:
\[
 \varphi(x'^*,t) = a^+ \sin\left(k x'^*\right), \quad \psi(x'^*,t) = -a^- \sin\left(k \left(\Delta c t- x'^*\right)\right), \quad x'^* = \int\limits_0^t c_s(t) dt
\]
then the following integral equation can be formulated for the signoton speed:
\[
 c_s(t) = \Delta c \frac{a^- \sin\left(k \left(\Delta c t- x'^*\right)\right)}{a^-\sin\left(k \left(\Delta c t- x'^*\right)\right) + a^+ \sin\left(k x'^*\right)}, \quad x'^* = \int\limits_0^t c_s(t) dt
\]
which greatly simplifies for $a^+ = a^- = a$ because in this case $c_s = \Delta c/2$ and 
\[
 \varphi(x'^*,t) = a \sin\left(k \Delta c t /2\right) = - \psi(x'^*,t).
\]
For this simple case (see Fig.~\ref{fig:simple_annihilation}(a)) the energy loss per one signoton would be given by:
\[
   W_{\text{loss}}(x'^*,t) = \Delta c  \frac{A(E^+ + E^-)a^2}{4} \int\limits_0^t  \sin^2(k  \Delta c_s t)  dt' = \frac{A(E^+ + E^-)a^2}{8} \left( L - \frac{\sin(kL)}{k}\right), 
\]
which results in the following evolution of the energy loss reformulated in length of the bimodular layer following $L = \Delta c t = 2 c_s t$ and expressed as the remaining energy normalized by the initial energy $(W_0-W_{\text{loss}}(L))/W_0$:
\begin{equation}
 1 - \frac{W_{\text{loss}}(L)}{W_0} = \begin{cases} \displaystyle
                                       1- \frac{kL - \sin(kL)}{2\pi}, & \text{ if } L < 2\pi/k\\
                                       0, & \text{ otherwise}.
                                      \end{cases}
                                      \label{eq:geom_model_sym}
\end{equation}
In Fig.~\ref{fig:simple_annihilation}(a) the evolution of the signal within this simplified geometrical model is shown for the case $a^+ = a^-$ (see Supplementary material~\cite{Supplementary} for the code).
In this case the signal fully disappears. The evolution of the transmitted energy is shown in Fig.~\ref{fig:simple_annihilation}(b) and compared with Eq.~\eqref{eq:geom_model_sym}. 
In case when $a^+ \ne a^-$ (Fig.~\ref{fig:simple_annihilation}(c)), the evolution of the energy loss is not easy to compute, but the remaining energy can be readily computed.
The integral deformation for the considered case is given as $2a^{\pm}/k$ for positive and negative parts of a single wave, so the absolute value of their difference is $2|a^- - a^+|/k$ which
is nothing but the remaining integral deformation of the part with the greater amplitude, say $a^- > a^+$, i.e.
\[
 \int\limits_0^y a^- \sin(kx')dx' = 2(a^- - a^+)/k,
\]
where a non-zero signal will remain only within the intervals $[0+2\pi,y+2\pi]$. From this equation, we can find the value of $y$ as:
\begin{equation}
 y = \frac{1}{k}\arccos(2a^+/a^- - 1).
 \label{eq:yc}
\end{equation}
Therefore, the remaining energy per wavelength will be given by
\begin{equation}
 1 - \frac{W_{\text{loss}}}{W_0} = \frac{2k E^{-} a^{-2}}{\pi(E^+ a^{+2} + E^- a^{-2})} \int\limits_0^y \sin^2(kx) dx =  \frac{k E^{-} a^{-2}}{\pi(E^+ a^{+2} + E^- a^{-2})} \left(y - \frac{\sin(2ky)}{2k}\right),
  \label{eq:remaining_energy}
\end{equation}
where $y$ should be substituted from Eq.~\eqref{eq:yc}. 
In Fig.~\ref{fig:simple_annihilation}(c) we show the evolution of the signal shape for $a^-/a^+ = 1.5$, the change in signoton speed as well as the remnant non zero signal of the compressive component can be easily observed.
In Fig.~\ref{fig:simple_annihilation}(d) the evolution of the energy obtained using the geometrical model is plotted for different amplitudes $a^-/a^+ = \{1.2, 1.5, 2\}$. Naturally, the bigger the amplitude difference, the more energy remains in the signal. In the same figure, the remaining energy is estimated using Eq.~\eqref{eq:remaining_energy} for the same amplitude ratios and it is shown with horizontal lines, the match between the model and the equation is perfect. Note that qualitatively the shape of transmission factor obtained for this simple signal is very similar to those observed for a random self-affine signal shown in Fig.~\ref{fig:energies}. Of course, the geometrical model cannot capture the full complexity of energy cascades and viscous damping, but it permits to understand better and predict rather accurately the change in the signal after it passes through a bimodular section.

 \section{Conclusions\label{sec:10}}

A new concept for architected materials was developed, in which the elastic asymmetry can be finely adjusted by combining internal contacts and components of different stiffness. 
Propagation of one-dimensional elastic waves in the resulting elastically asymmetric media was studied.
We found that the overlap of tensile and compressive wave components propagating at different speeds results in emergence of energy cascades leading to a partial or, in particular cases, almost complete annihilation of tensile and compressive wave components. This annihilation mechanism presents a novel and powerful signal damping mechanism. 
Such a damping is drastically enhanced if a segment with one asymmetry is followed by another segment with the opposite asymmetry. 
The efficient wave damping is ensured if the bimodular segment is chosen longer than the wave-overlap distance, Eq.~\eqref{eq:2}.
The ratio of the overlap length to the incident wavelength scales as $L_o/\lambda_0 \sim 1/(\sqrt{\gamma}-1)$. 
The key advantage of the proposed architected materials consists in fact that the elastic asymmetry $\gamma$ can be adjusted to be very high, which would enable to keep the damping device relatively small compared to the incident wavelength.

In more realistic situation of an incident wave containing many modes only partial annihilation can occur, however, the signal after passing through a relatively long bimodular segment appears ``polarized'' either to positive or negative deformation. This observation was made in the study of self-affine incident wave packets (Gaussian envelope) passing through a bimodular section. This analysis permitted to obtain a rather universal form of the ratio of transmitted to injected energy (transmission factor) with respect to the length of bimodular segment and also to obtain relevant normalizations. In addition, a simple model of geometrical overlap was developed and some simple analytical results for the transmission factor were obtained.

The demonstrated efficient damping and sign-polarization mechanisms can be used in shock absorbing and wave filtering systems, and, hypothetically, in seismic protection from surface waves. 
For further investigation, the one-dimensional model should be extended to two and three dimensional cases, where the compressive/tensile elastic asymmetry should be enhanced with shear asymmetry and complemented by elastic anisotropy. 

The computational code for simulation of the one-dimensional wave propagation in asymmetric medium with absorbing boundaries is available in supplementary material~\cite{Supplementary} as well as the data of hundreds of simulations and the scripts for their post-processing and figure plotting.

\section*{Conflicts of interest}

The author declares no competing financial interest.

\section*{Acknowledgements}

The author is grateful to Lev Truskinovsky for valuable discussions, to Samuel Forest for his encouragement, and to Arsen Subashiev for helpful remarks to the text.

\section*{Supplementary data}

Supplementary material for this article (documented computational code, simulation data, scripts for the geometrical model and scripts for Figs.~(7,8,9,10,11,13)) is accessible at DOI:\href{https://doi.org/10.5281/zenodo.4461652}{10.5281/zenodo.4461652}.

\section*{References}
\printbibliography[heading=none]

\end{document}

%% file: AnnihilationSignal_1_model.pgf
\begingroup%
\makeatletter%
\begin{pgfpicture}%
\pgfpathrectangle{\pgfpointorigin}{\pgfqpoint{2.750000in}{2.750000in}}%
\pgfusepath{use as bounding box}%
\begin{pgfscope}%
\pgfsetrectcap%
\pgfsetroundjoin%
\definecolor{currentfill}{rgb}{1.000000,1.000000,1.000000}%
\pgfsetfillcolor{currentfill}%
\pgfsetlinewidth{0.000000pt}%
\definecolor{currentstroke}{rgb}{1.000000,1.000000,1.000000}%
\pgfsetstrokecolor{currentstroke}%
\pgfsetdash{}{0pt}%
\pgfpathmoveto{\pgfqpoint{0.000000in}{0.000000in}}%
\pgfpathlineto{\pgfqpoint{2.750000in}{0.000000in}}%
\pgfpathlineto{\pgfqpoint{2.750000in}{2.750000in}}%
\pgfpathlineto{\pgfqpoint{0.000000in}{2.750000in}}%
\pgfpathclose%
\pgfusepath{fill}%
\end{pgfscope}%
\begin{pgfscope}%
\pgfsetrectcap%
\pgfsetroundjoin%
\definecolor{currentfill}{rgb}{1.000000,1.000000,1.000000}%
\pgfsetfillcolor{currentfill}%
\pgfsetlinewidth{0.000000pt}%
\definecolor{currentstroke}{rgb}{0.000000,0.000000,0.000000}%
\pgfsetstrokecolor{currentstroke}%
\pgfsetdash{}{0pt}%
\pgfpathmoveto{\pgfqpoint{0.539608in}{0.464787in}}%
\pgfpathlineto{\pgfqpoint{2.533228in}{0.464787in}}%
\pgfpathlineto{\pgfqpoint{2.533228in}{2.574873in}}%
\pgfpathlineto{\pgfqpoint{0.539608in}{2.574873in}}%
\pgfpathclose%
\pgfusepath{fill}%
\end{pgfscope}%
\begin{pgfscope}%
\pgfpathrectangle{\pgfqpoint{0.539608in}{0.464787in}}{\pgfqpoint{1.993620in}{2.110085in}} %
\pgfusepath{clip}%
\pgfsetrectcap%
\pgfsetroundjoin%
\pgfsetlinewidth{1.003750pt}%
\definecolor{currentstroke}{rgb}{0.000000,0.000000,0.000000}%
\pgfsetstrokecolor{currentstroke}%
\pgfsetdash{}{0pt}%
\pgfpathmoveto{\pgfqpoint{0.539608in}{2.574873in}}%
\pgfpathlineto{\pgfqpoint{0.625376in}{2.573771in}}%
\pgfpathlineto{\pgfqpoint{0.671253in}{2.570909in}}%
\pgfpathlineto{\pgfqpoint{0.710148in}{2.566305in}}%
\pgfpathlineto{\pgfqpoint{0.745053in}{2.559990in}}%
\pgfpathlineto{\pgfqpoint{0.777965in}{2.551799in}}%
\pgfpathlineto{\pgfqpoint{0.808881in}{2.541865in}}%
\pgfpathlineto{\pgfqpoint{0.838800in}{2.529976in}}%
\pgfpathlineto{\pgfqpoint{0.868720in}{2.515672in}}%
\pgfpathlineto{\pgfqpoint{0.897642in}{2.499403in}}%
\pgfpathlineto{\pgfqpoint{0.926564in}{2.480614in}}%
\pgfpathlineto{\pgfqpoint{0.955486in}{2.459206in}}%
\pgfpathlineto{\pgfqpoint{0.985405in}{2.434223in}}%
\pgfpathlineto{\pgfqpoint{1.015324in}{2.406296in}}%
\pgfpathlineto{\pgfqpoint{1.046241in}{2.374310in}}%
\pgfpathlineto{\pgfqpoint{1.077157in}{2.339139in}}%
\pgfpathlineto{\pgfqpoint{1.109071in}{2.299517in}}%
\pgfpathlineto{\pgfqpoint{1.142980in}{2.253793in}}%
\pgfpathlineto{\pgfqpoint{1.177885in}{2.202939in}}%
\pgfpathlineto{\pgfqpoint{1.214786in}{2.145190in}}%
\pgfpathlineto{\pgfqpoint{1.253681in}{2.080155in}}%
\pgfpathlineto{\pgfqpoint{1.295568in}{2.005743in}}%
\pgfpathlineto{\pgfqpoint{1.341444in}{1.919650in}}%
\pgfpathlineto{\pgfqpoint{1.393304in}{1.817505in}}%
\pgfpathlineto{\pgfqpoint{1.456134in}{1.688652in}}%
\pgfpathlineto{\pgfqpoint{1.560852in}{1.467856in}}%
\pgfpathlineto{\pgfqpoint{1.654599in}{1.272206in}}%
\pgfpathlineto{\pgfqpoint{1.714437in}{1.152367in}}%
\pgfpathlineto{\pgfqpoint{1.764303in}{1.057263in}}%
\pgfpathlineto{\pgfqpoint{1.809182in}{0.976301in}}%
\pgfpathlineto{\pgfqpoint{1.850071in}{0.906945in}}%
\pgfpathlineto{\pgfqpoint{1.888966in}{0.845285in}}%
\pgfpathlineto{\pgfqpoint{1.925867in}{0.790960in}}%
\pgfpathlineto{\pgfqpoint{1.960773in}{0.743501in}}%
\pgfpathlineto{\pgfqpoint{1.993684in}{0.702371in}}%
\pgfpathlineto{\pgfqpoint{2.025598in}{0.665907in}}%
\pgfpathlineto{\pgfqpoint{2.056514in}{0.633819in}}%
\pgfpathlineto{\pgfqpoint{2.087431in}{0.604910in}}%
\pgfpathlineto{\pgfqpoint{2.117350in}{0.579927in}}%
\pgfpathlineto{\pgfqpoint{2.147269in}{0.557828in}}%
\pgfpathlineto{\pgfqpoint{2.177189in}{0.538528in}}%
\pgfpathlineto{\pgfqpoint{2.207108in}{0.521917in}}%
\pgfpathlineto{\pgfqpoint{2.237027in}{0.507862in}}%
\pgfpathlineto{\pgfqpoint{2.266946in}{0.496205in}}%
\pgfpathlineto{\pgfqpoint{2.297863in}{0.486491in}}%
\pgfpathlineto{\pgfqpoint{2.330774in}{0.478510in}}%
\pgfpathlineto{\pgfqpoint{2.365680in}{0.472389in}}%
\pgfpathlineto{\pgfqpoint{2.403578in}{0.468048in}}%
\pgfpathlineto{\pgfqpoint{2.447459in}{0.465362in}}%
\pgfpathlineto{\pgfqpoint{2.504306in}{0.464302in}}%
\pgfpathlineto{\pgfqpoint{2.533228in}{0.464260in}}%
\pgfpathlineto{\pgfqpoint{2.533228in}{0.464260in}}%
\pgfusepath{stroke}%
\end{pgfscope}%
\begin{pgfscope}%
\pgfpathrectangle{\pgfqpoint{0.539608in}{0.464787in}}{\pgfqpoint{1.993620in}{2.110085in}} %
\pgfusepath{clip}%
\pgfsetbuttcap%
\pgfsetroundjoin%
\pgfsetlinewidth{0.250937pt}%
\definecolor{currentstroke}{rgb}{0.000000,0.000000,0.000000}%
\pgfsetstrokecolor{currentstroke}%
\pgfsetdash{{1.000000pt}{3.000000pt}}{0.000000pt}%
\pgfpathmoveto{\pgfqpoint{0.539608in}{0.464787in}}%
\pgfpathlineto{\pgfqpoint{0.539608in}{2.574873in}}%
\pgfusepath{stroke}%
\end{pgfscope}%
\begin{pgfscope}%
\pgfsetbuttcap%
\pgfsetroundjoin%
\definecolor{currentfill}{rgb}{0.000000,0.000000,0.000000}%
\pgfsetfillcolor{currentfill}%
\pgfsetlinewidth{0.501875pt}%
\definecolor{currentstroke}{rgb}{0.000000,0.000000,0.000000}%
\pgfsetstrokecolor{currentstroke}%
\pgfsetdash{}{0pt}%
\pgfsys@defobject{currentmarker}{\pgfqpoint{0.000000in}{0.000000in}}{\pgfqpoint{0.000000in}{0.055556in}}{%
\pgfpathmoveto{\pgfqpoint{0.000000in}{0.000000in}}%
\pgfpathlineto{\pgfqpoint{0.000000in}{0.055556in}}%
\pgfusepath{stroke,fill}%
}%
\begin{pgfscope}%
\pgfsys@transformshift{0.539608in}{0.464787in}%
\pgfsys@useobject{currentmarker}{}%
\end{pgfscope}%
\end{pgfscope}%
\begin{pgfscope}%
\pgfsetbuttcap%
\pgfsetroundjoin%
\definecolor{currentfill}{rgb}{0.000000,0.000000,0.000000}%
\pgfsetfillcolor{currentfill}%
\pgfsetlinewidth{0.501875pt}%
\definecolor{currentstroke}{rgb}{0.000000,0.000000,0.000000}%
\pgfsetstrokecolor{currentstroke}%
\pgfsetdash{}{0pt}%
\pgfsys@defobject{currentmarker}{\pgfqpoint{0.000000in}{-0.055556in}}{\pgfqpoint{0.000000in}{0.000000in}}{%
\pgfpathmoveto{\pgfqpoint{0.000000in}{0.000000in}}%
\pgfpathlineto{\pgfqpoint{0.000000in}{-0.055556in}}%
\pgfusepath{stroke,fill}%
}%
\begin{pgfscope}%
\pgfsys@transformshift{0.539608in}{2.574873in}%
\pgfsys@useobject{currentmarker}{}%
\end{pgfscope}%
\end{pgfscope}%
\begin{pgfscope}%
\pgftext[left,bottom,x=0.457529in,y=0.328676in,rotate=0.000000]{{\rmfamily\fontsize{9.000000}{10.800000}\selectfont \(\displaystyle 0.0\)}}
\end{pgfscope}%
\begin{pgfscope}%
\pgfpathrectangle{\pgfqpoint{0.539608in}{0.464787in}}{\pgfqpoint{1.993620in}{2.110085in}} %
\pgfusepath{clip}%
\pgfsetbuttcap%
\pgfsetroundjoin%
\pgfsetlinewidth{0.250937pt}%
\definecolor{currentstroke}{rgb}{0.000000,0.000000,0.000000}%
\pgfsetstrokecolor{currentstroke}%
\pgfsetdash{{1.000000pt}{3.000000pt}}{0.000000pt}%
\pgfpathmoveto{\pgfqpoint{0.938332in}{0.464787in}}%
\pgfpathlineto{\pgfqpoint{0.938332in}{2.574873in}}%
\pgfusepath{stroke}%
\end{pgfscope}%
\begin{pgfscope}%
\pgfsetbuttcap%
\pgfsetroundjoin%
\definecolor{currentfill}{rgb}{0.000000,0.000000,0.000000}%
\pgfsetfillcolor{currentfill}%
\pgfsetlinewidth{0.501875pt}%
\definecolor{currentstroke}{rgb}{0.000000,0.000000,0.000000}%
\pgfsetstrokecolor{currentstroke}%
\pgfsetdash{}{0pt}%
\pgfsys@defobject{currentmarker}{\pgfqpoint{0.000000in}{0.000000in}}{\pgfqpoint{0.000000in}{0.055556in}}{%
\pgfpathmoveto{\pgfqpoint{0.000000in}{0.000000in}}%
\pgfpathlineto{\pgfqpoint{0.000000in}{0.055556in}}%
\pgfusepath{stroke,fill}%
}%
\begin{pgfscope}%
\pgfsys@transformshift{0.938332in}{0.464787in}%
\pgfsys@useobject{currentmarker}{}%
\end{pgfscope}%
\end{pgfscope}%
\begin{pgfscope}%
\pgfsetbuttcap%
\pgfsetroundjoin%
\definecolor{currentfill}{rgb}{0.000000,0.000000,0.000000}%
\pgfsetfillcolor{currentfill}%
\pgfsetlinewidth{0.501875pt}%
\definecolor{currentstroke}{rgb}{0.000000,0.000000,0.000000}%
\pgfsetstrokecolor{currentstroke}%
\pgfsetdash{}{0pt}%
\pgfsys@defobject{currentmarker}{\pgfqpoint{0.000000in}{-0.055556in}}{\pgfqpoint{0.000000in}{0.000000in}}{%
\pgfpathmoveto{\pgfqpoint{0.000000in}{0.000000in}}%
\pgfpathlineto{\pgfqpoint{0.000000in}{-0.055556in}}%
\pgfusepath{stroke,fill}%
}%
\begin{pgfscope}%
\pgfsys@transformshift{0.938332in}{2.574873in}%
\pgfsys@useobject{currentmarker}{}%
\end{pgfscope}%
\end{pgfscope}%
\begin{pgfscope}%
\pgftext[left,bottom,x=0.856253in,y=0.328676in,rotate=0.000000]{{\rmfamily\fontsize{9.000000}{10.800000}\selectfont \(\displaystyle 0.1\)}}
\end{pgfscope}%
\begin{pgfscope}%
\pgfpathrectangle{\pgfqpoint{0.539608in}{0.464787in}}{\pgfqpoint{1.993620in}{2.110085in}} %
\pgfusepath{clip}%
\pgfsetbuttcap%
\pgfsetroundjoin%
\pgfsetlinewidth{0.250937pt}%
\definecolor{currentstroke}{rgb}{0.000000,0.000000,0.000000}%
\pgfsetstrokecolor{currentstroke}%
\pgfsetdash{{1.000000pt}{3.000000pt}}{0.000000pt}%
\pgfpathmoveto{\pgfqpoint{1.337056in}{0.464787in}}%
\pgfpathlineto{\pgfqpoint{1.337056in}{2.574873in}}%
\pgfusepath{stroke}%
\end{pgfscope}%
\begin{pgfscope}%
\pgfsetbuttcap%
\pgfsetroundjoin%
\definecolor{currentfill}{rgb}{0.000000,0.000000,0.000000}%
\pgfsetfillcolor{currentfill}%
\pgfsetlinewidth{0.501875pt}%
\definecolor{currentstroke}{rgb}{0.000000,0.000000,0.000000}%
\pgfsetstrokecolor{currentstroke}%
\pgfsetdash{}{0pt}%
\pgfsys@defobject{currentmarker}{\pgfqpoint{0.000000in}{0.000000in}}{\pgfqpoint{0.000000in}{0.055556in}}{%
\pgfpathmoveto{\pgfqpoint{0.000000in}{0.000000in}}%
\pgfpathlineto{\pgfqpoint{0.000000in}{0.055556in}}%
\pgfusepath{stroke,fill}%
}%
\begin{pgfscope}%
\pgfsys@transformshift{1.337056in}{0.464787in}%
\pgfsys@useobject{currentmarker}{}%
\end{pgfscope}%
\end{pgfscope}%
\begin{pgfscope}%
\pgfsetbuttcap%
\pgfsetroundjoin%
\definecolor{currentfill}{rgb}{0.000000,0.000000,0.000000}%
\pgfsetfillcolor{currentfill}%
\pgfsetlinewidth{0.501875pt}%
\definecolor{currentstroke}{rgb}{0.000000,0.000000,0.000000}%
\pgfsetstrokecolor{currentstroke}%
\pgfsetdash{}{0pt}%
\pgfsys@defobject{currentmarker}{\pgfqpoint{0.000000in}{-0.055556in}}{\pgfqpoint{0.000000in}{0.000000in}}{%
\pgfpathmoveto{\pgfqpoint{0.000000in}{0.000000in}}%
\pgfpathlineto{\pgfqpoint{0.000000in}{-0.055556in}}%
\pgfusepath{stroke,fill}%
}%
\begin{pgfscope}%
\pgfsys@transformshift{1.337056in}{2.574873in}%
\pgfsys@useobject{currentmarker}{}%
\end{pgfscope}%
\end{pgfscope}%
\begin{pgfscope}%
\pgftext[left,bottom,x=1.254977in,y=0.328676in,rotate=0.000000]{{\rmfamily\fontsize{9.000000}{10.800000}\selectfont \(\displaystyle 0.2\)}}
\end{pgfscope}%
\begin{pgfscope}%
\pgfpathrectangle{\pgfqpoint{0.539608in}{0.464787in}}{\pgfqpoint{1.993620in}{2.110085in}} %
\pgfusepath{clip}%
\pgfsetbuttcap%
\pgfsetroundjoin%
\pgfsetlinewidth{0.250937pt}%
\definecolor{currentstroke}{rgb}{0.000000,0.000000,0.000000}%
\pgfsetstrokecolor{currentstroke}%
\pgfsetdash{{1.000000pt}{3.000000pt}}{0.000000pt}%
\pgfpathmoveto{\pgfqpoint{1.735780in}{0.464787in}}%
\pgfpathlineto{\pgfqpoint{1.735780in}{2.574873in}}%
\pgfusepath{stroke}%
\end{pgfscope}%
\begin{pgfscope}%
\pgfsetbuttcap%
\pgfsetroundjoin%
\definecolor{currentfill}{rgb}{0.000000,0.000000,0.000000}%
\pgfsetfillcolor{currentfill}%
\pgfsetlinewidth{0.501875pt}%
\definecolor{currentstroke}{rgb}{0.000000,0.000000,0.000000}%
\pgfsetstrokecolor{currentstroke}%
\pgfsetdash{}{0pt}%
\pgfsys@defobject{currentmarker}{\pgfqpoint{0.000000in}{0.000000in}}{\pgfqpoint{0.000000in}{0.055556in}}{%
\pgfpathmoveto{\pgfqpoint{0.000000in}{0.000000in}}%
\pgfpathlineto{\pgfqpoint{0.000000in}{0.055556in}}%
\pgfusepath{stroke,fill}%
}%
\begin{pgfscope}%
\pgfsys@transformshift{1.735780in}{0.464787in}%
\pgfsys@useobject{currentmarker}{}%
\end{pgfscope}%
\end{pgfscope}%
\begin{pgfscope}%
\pgfsetbuttcap%
\pgfsetroundjoin%
\definecolor{currentfill}{rgb}{0.000000,0.000000,0.000000}%
\pgfsetfillcolor{currentfill}%
\pgfsetlinewidth{0.501875pt}%
\definecolor{currentstroke}{rgb}{0.000000,0.000000,0.000000}%
\pgfsetstrokecolor{currentstroke}%
\pgfsetdash{}{0pt}%
\pgfsys@defobject{currentmarker}{\pgfqpoint{0.000000in}{-0.055556in}}{\pgfqpoint{0.000000in}{0.000000in}}{%
\pgfpathmoveto{\pgfqpoint{0.000000in}{0.000000in}}%
\pgfpathlineto{\pgfqpoint{0.000000in}{-0.055556in}}%
\pgfusepath{stroke,fill}%
}%
\begin{pgfscope}%
\pgfsys@transformshift{1.735780in}{2.574873in}%
\pgfsys@useobject{currentmarker}{}%
\end{pgfscope}%
\end{pgfscope}%
\begin{pgfscope}%
\pgftext[left,bottom,x=1.653701in,y=0.328676in,rotate=0.000000]{{\rmfamily\fontsize{9.000000}{10.800000}\selectfont \(\displaystyle 0.3\)}}
\end{pgfscope}%
\begin{pgfscope}%
\pgfpathrectangle{\pgfqpoint{0.539608in}{0.464787in}}{\pgfqpoint{1.993620in}{2.110085in}} %
\pgfusepath{clip}%
\pgfsetbuttcap%
\pgfsetroundjoin%
\pgfsetlinewidth{0.250937pt}%
\definecolor{currentstroke}{rgb}{0.000000,0.000000,0.000000}%
\pgfsetstrokecolor{currentstroke}%
\pgfsetdash{{1.000000pt}{3.000000pt}}{0.000000pt}%
\pgfpathmoveto{\pgfqpoint{2.134504in}{0.464787in}}%
\pgfpathlineto{\pgfqpoint{2.134504in}{2.574873in}}%
\pgfusepath{stroke}%
\end{pgfscope}%
\begin{pgfscope}%
\pgfsetbuttcap%
\pgfsetroundjoin%
\definecolor{currentfill}{rgb}{0.000000,0.000000,0.000000}%
\pgfsetfillcolor{currentfill}%
\pgfsetlinewidth{0.501875pt}%
\definecolor{currentstroke}{rgb}{0.000000,0.000000,0.000000}%
\pgfsetstrokecolor{currentstroke}%
\pgfsetdash{}{0pt}%
\pgfsys@defobject{currentmarker}{\pgfqpoint{0.000000in}{0.000000in}}{\pgfqpoint{0.000000in}{0.055556in}}{%
\pgfpathmoveto{\pgfqpoint{0.000000in}{0.000000in}}%
\pgfpathlineto{\pgfqpoint{0.000000in}{0.055556in}}%
\pgfusepath{stroke,fill}%
}%
\begin{pgfscope}%
\pgfsys@transformshift{2.134504in}{0.464787in}%
\pgfsys@useobject{currentmarker}{}%
\end{pgfscope}%
\end{pgfscope}%
\begin{pgfscope}%
\pgfsetbuttcap%
\pgfsetroundjoin%
\definecolor{currentfill}{rgb}{0.000000,0.000000,0.000000}%
\pgfsetfillcolor{currentfill}%
\pgfsetlinewidth{0.501875pt}%
\definecolor{currentstroke}{rgb}{0.000000,0.000000,0.000000}%
\pgfsetstrokecolor{currentstroke}%
\pgfsetdash{}{0pt}%
\pgfsys@defobject{currentmarker}{\pgfqpoint{0.000000in}{-0.055556in}}{\pgfqpoint{0.000000in}{0.000000in}}{%
\pgfpathmoveto{\pgfqpoint{0.000000in}{0.000000in}}%
\pgfpathlineto{\pgfqpoint{0.000000in}{-0.055556in}}%
\pgfusepath{stroke,fill}%
}%
\begin{pgfscope}%
\pgfsys@transformshift{2.134504in}{2.574873in}%
\pgfsys@useobject{currentmarker}{}%
\end{pgfscope}%
\end{pgfscope}%
\begin{pgfscope}%
\pgftext[left,bottom,x=2.052425in,y=0.328676in,rotate=0.000000]{{\rmfamily\fontsize{9.000000}{10.800000}\selectfont \(\displaystyle 0.4\)}}
\end{pgfscope}%
\begin{pgfscope}%
\pgfpathrectangle{\pgfqpoint{0.539608in}{0.464787in}}{\pgfqpoint{1.993620in}{2.110085in}} %
\pgfusepath{clip}%
\pgfsetbuttcap%
\pgfsetroundjoin%
\pgfsetlinewidth{0.250937pt}%
\definecolor{currentstroke}{rgb}{0.000000,0.000000,0.000000}%
\pgfsetstrokecolor{currentstroke}%
\pgfsetdash{{1.000000pt}{3.000000pt}}{0.000000pt}%
\pgfpathmoveto{\pgfqpoint{2.533228in}{0.464787in}}%
\pgfpathlineto{\pgfqpoint{2.533228in}{2.574873in}}%
\pgfusepath{stroke}%
\end{pgfscope}%
\begin{pgfscope}%
\pgfsetbuttcap%
\pgfsetroundjoin%
\definecolor{currentfill}{rgb}{0.000000,0.000000,0.000000}%
\pgfsetfillcolor{currentfill}%
\pgfsetlinewidth{0.501875pt}%
\definecolor{currentstroke}{rgb}{0.000000,0.000000,0.000000}%
\pgfsetstrokecolor{currentstroke}%
\pgfsetdash{}{0pt}%
\pgfsys@defobject{currentmarker}{\pgfqpoint{0.000000in}{0.000000in}}{\pgfqpoint{0.000000in}{0.055556in}}{%
\pgfpathmoveto{\pgfqpoint{0.000000in}{0.000000in}}%
\pgfpathlineto{\pgfqpoint{0.000000in}{0.055556in}}%
\pgfusepath{stroke,fill}%
}%
\begin{pgfscope}%
\pgfsys@transformshift{2.533228in}{0.464787in}%
\pgfsys@useobject{currentmarker}{}%
\end{pgfscope}%
\end{pgfscope}%
\begin{pgfscope}%
\pgfsetbuttcap%
\pgfsetroundjoin%
\definecolor{currentfill}{rgb}{0.000000,0.000000,0.000000}%
\pgfsetfillcolor{currentfill}%
\pgfsetlinewidth{0.501875pt}%
\definecolor{currentstroke}{rgb}{0.000000,0.000000,0.000000}%
\pgfsetstrokecolor{currentstroke}%
\pgfsetdash{}{0pt}%
\pgfsys@defobject{currentmarker}{\pgfqpoint{0.000000in}{-0.055556in}}{\pgfqpoint{0.000000in}{0.000000in}}{%
\pgfpathmoveto{\pgfqpoint{0.000000in}{0.000000in}}%
\pgfpathlineto{\pgfqpoint{0.000000in}{-0.055556in}}%
\pgfusepath{stroke,fill}%
}%
\begin{pgfscope}%
\pgfsys@transformshift{2.533228in}{2.574873in}%
\pgfsys@useobject{currentmarker}{}%
\end{pgfscope}%
\end{pgfscope}%
\begin{pgfscope}%
\pgftext[left,bottom,x=2.451149in,y=0.328676in,rotate=0.000000]{{\rmfamily\fontsize{9.000000}{10.800000}\selectfont \(\displaystyle 0.5\)}}
\end{pgfscope}%
\begin{pgfscope}%
\pgftext[left,bottom,x=0.532518in,y=0.133011in,rotate=0.000000]{{\rmfamily\fontsize{9.000000}{10.800000}\selectfont Normalized length, \(\displaystyle L/\lambda=\Delta c t/\lambda\)}}
\end{pgfscope}%
\begin{pgfscope}%
\pgfpathrectangle{\pgfqpoint{0.539608in}{0.464787in}}{\pgfqpoint{1.993620in}{2.110085in}} %
\pgfusepath{clip}%
\pgfsetbuttcap%
\pgfsetroundjoin%
\pgfsetlinewidth{0.250937pt}%
\definecolor{currentstroke}{rgb}{0.000000,0.000000,0.000000}%
\pgfsetstrokecolor{currentstroke}%
\pgfsetdash{{1.000000pt}{3.000000pt}}{0.000000pt}%
\pgfpathmoveto{\pgfqpoint{0.539608in}{0.464787in}}%
\pgfpathlineto{\pgfqpoint{2.533228in}{0.464787in}}%
\pgfusepath{stroke}%
\end{pgfscope}%
\begin{pgfscope}%
\pgfsetbuttcap%
\pgfsetroundjoin%
\definecolor{currentfill}{rgb}{0.000000,0.000000,0.000000}%
\pgfsetfillcolor{currentfill}%
\pgfsetlinewidth{0.501875pt}%
\definecolor{currentstroke}{rgb}{0.000000,0.000000,0.000000}%
\pgfsetstrokecolor{currentstroke}%
\pgfsetdash{}{0pt}%
\pgfsys@defobject{currentmarker}{\pgfqpoint{0.000000in}{0.000000in}}{\pgfqpoint{0.055556in}{0.000000in}}{%
\pgfpathmoveto{\pgfqpoint{0.000000in}{0.000000in}}%
\pgfpathlineto{\pgfqpoint{0.055556in}{0.000000in}}%
\pgfusepath{stroke,fill}%
}%
\begin{pgfscope}%
\pgfsys@transformshift{0.539608in}{0.464787in}%
\pgfsys@useobject{currentmarker}{}%
\end{pgfscope}%
\end{pgfscope}%
\begin{pgfscope}%
\pgfsetbuttcap%
\pgfsetroundjoin%
\definecolor{currentfill}{rgb}{0.000000,0.000000,0.000000}%
\pgfsetfillcolor{currentfill}%
\pgfsetlinewidth{0.501875pt}%
\definecolor{currentstroke}{rgb}{0.000000,0.000000,0.000000}%
\pgfsetstrokecolor{currentstroke}%
\pgfsetdash{}{0pt}%
\pgfsys@defobject{currentmarker}{\pgfqpoint{-0.055556in}{0.000000in}}{\pgfqpoint{0.000000in}{0.000000in}}{%
\pgfpathmoveto{\pgfqpoint{0.000000in}{0.000000in}}%
\pgfpathlineto{\pgfqpoint{-0.055556in}{0.000000in}}%
\pgfusepath{stroke,fill}%
}%
\begin{pgfscope}%
\pgfsys@transformshift{2.533228in}{0.464787in}%
\pgfsys@useobject{currentmarker}{}%
\end{pgfscope}%
\end{pgfscope}%
\begin{pgfscope}%
\pgftext[left,bottom,x=0.319894in,y=0.424510in,rotate=0.000000]{{\rmfamily\fontsize{9.000000}{10.800000}\selectfont \(\displaystyle 0.0\)}}
\end{pgfscope}%
\begin{pgfscope}%
\pgfpathrectangle{\pgfqpoint{0.539608in}{0.464787in}}{\pgfqpoint{1.993620in}{2.110085in}} %
\pgfusepath{clip}%
\pgfsetbuttcap%
\pgfsetroundjoin%
\pgfsetlinewidth{0.250937pt}%
\definecolor{currentstroke}{rgb}{0.000000,0.000000,0.000000}%
\pgfsetstrokecolor{currentstroke}%
\pgfsetdash{{1.000000pt}{3.000000pt}}{0.000000pt}%
\pgfpathmoveto{\pgfqpoint{0.539608in}{0.886805in}}%
\pgfpathlineto{\pgfqpoint{2.533228in}{0.886805in}}%
\pgfusepath{stroke}%
\end{pgfscope}%
\begin{pgfscope}%
\pgfsetbuttcap%
\pgfsetroundjoin%
\definecolor{currentfill}{rgb}{0.000000,0.000000,0.000000}%
\pgfsetfillcolor{currentfill}%
\pgfsetlinewidth{0.501875pt}%
\definecolor{currentstroke}{rgb}{0.000000,0.000000,0.000000}%
\pgfsetstrokecolor{currentstroke}%
\pgfsetdash{}{0pt}%
\pgfsys@defobject{currentmarker}{\pgfqpoint{0.000000in}{0.000000in}}{\pgfqpoint{0.055556in}{0.000000in}}{%
\pgfpathmoveto{\pgfqpoint{0.000000in}{0.000000in}}%
\pgfpathlineto{\pgfqpoint{0.055556in}{0.000000in}}%
\pgfusepath{stroke,fill}%
}%
\begin{pgfscope}%
\pgfsys@transformshift{0.539608in}{0.886805in}%
\pgfsys@useobject{currentmarker}{}%
\end{pgfscope}%
\end{pgfscope}%
\begin{pgfscope}%
\pgfsetbuttcap%
\pgfsetroundjoin%
\definecolor{currentfill}{rgb}{0.000000,0.000000,0.000000}%
\pgfsetfillcolor{currentfill}%
\pgfsetlinewidth{0.501875pt}%
\definecolor{currentstroke}{rgb}{0.000000,0.000000,0.000000}%
\pgfsetstrokecolor{currentstroke}%
\pgfsetdash{}{0pt}%
\pgfsys@defobject{currentmarker}{\pgfqpoint{-0.055556in}{0.000000in}}{\pgfqpoint{0.000000in}{0.000000in}}{%
\pgfpathmoveto{\pgfqpoint{0.000000in}{0.000000in}}%
\pgfpathlineto{\pgfqpoint{-0.055556in}{0.000000in}}%
\pgfusepath{stroke,fill}%
}%
\begin{pgfscope}%
\pgfsys@transformshift{2.533228in}{0.886805in}%
\pgfsys@useobject{currentmarker}{}%
\end{pgfscope}%
\end{pgfscope}%
\begin{pgfscope}%
\pgftext[left,bottom,x=0.319894in,y=0.846527in,rotate=0.000000]{{\rmfamily\fontsize{9.000000}{10.800000}\selectfont \(\displaystyle 0.2\)}}
\end{pgfscope}%
\begin{pgfscope}%
\pgfpathrectangle{\pgfqpoint{0.539608in}{0.464787in}}{\pgfqpoint{1.993620in}{2.110085in}} %
\pgfusepath{clip}%
\pgfsetbuttcap%
\pgfsetroundjoin%
\pgfsetlinewidth{0.250937pt}%
\definecolor{currentstroke}{rgb}{0.000000,0.000000,0.000000}%
\pgfsetstrokecolor{currentstroke}%
\pgfsetdash{{1.000000pt}{3.000000pt}}{0.000000pt}%
\pgfpathmoveto{\pgfqpoint{0.539608in}{1.308822in}}%
\pgfpathlineto{\pgfqpoint{2.533228in}{1.308822in}}%
\pgfusepath{stroke}%
\end{pgfscope}%
\begin{pgfscope}%
\pgfsetbuttcap%
\pgfsetroundjoin%
\definecolor{currentfill}{rgb}{0.000000,0.000000,0.000000}%
\pgfsetfillcolor{currentfill}%
\pgfsetlinewidth{0.501875pt}%
\definecolor{currentstroke}{rgb}{0.000000,0.000000,0.000000}%
\pgfsetstrokecolor{currentstroke}%
\pgfsetdash{}{0pt}%
\pgfsys@defobject{currentmarker}{\pgfqpoint{0.000000in}{0.000000in}}{\pgfqpoint{0.055556in}{0.000000in}}{%
\pgfpathmoveto{\pgfqpoint{0.000000in}{0.000000in}}%
\pgfpathlineto{\pgfqpoint{0.055556in}{0.000000in}}%
\pgfusepath{stroke,fill}%
}%
\begin{pgfscope}%
\pgfsys@transformshift{0.539608in}{1.308822in}%
\pgfsys@useobject{currentmarker}{}%
\end{pgfscope}%
\end{pgfscope}%
\begin{pgfscope}%
\pgfsetbuttcap%
\pgfsetroundjoin%
\definecolor{currentfill}{rgb}{0.000000,0.000000,0.000000}%
\pgfsetfillcolor{currentfill}%
\pgfsetlinewidth{0.501875pt}%
\definecolor{currentstroke}{rgb}{0.000000,0.000000,0.000000}%
\pgfsetstrokecolor{currentstroke}%
\pgfsetdash{}{0pt}%
\pgfsys@defobject{currentmarker}{\pgfqpoint{-0.055556in}{0.000000in}}{\pgfqpoint{0.000000in}{0.000000in}}{%
\pgfpathmoveto{\pgfqpoint{0.000000in}{0.000000in}}%
\pgfpathlineto{\pgfqpoint{-0.055556in}{0.000000in}}%
\pgfusepath{stroke,fill}%
}%
\begin{pgfscope}%
\pgfsys@transformshift{2.533228in}{1.308822in}%
\pgfsys@useobject{currentmarker}{}%
\end{pgfscope}%
\end{pgfscope}%
\begin{pgfscope}%
\pgftext[left,bottom,x=0.319894in,y=1.268544in,rotate=0.000000]{{\rmfamily\fontsize{9.000000}{10.800000}\selectfont \(\displaystyle 0.4\)}}
\end{pgfscope}%
\begin{pgfscope}%
\pgfpathrectangle{\pgfqpoint{0.539608in}{0.464787in}}{\pgfqpoint{1.993620in}{2.110085in}} %
\pgfusepath{clip}%
\pgfsetbuttcap%
\pgfsetroundjoin%
\pgfsetlinewidth{0.250937pt}%
\definecolor{currentstroke}{rgb}{0.000000,0.000000,0.000000}%
\pgfsetstrokecolor{currentstroke}%
\pgfsetdash{{1.000000pt}{3.000000pt}}{0.000000pt}%
\pgfpathmoveto{\pgfqpoint{0.539608in}{1.730839in}}%
\pgfpathlineto{\pgfqpoint{2.533228in}{1.730839in}}%
\pgfusepath{stroke}%
\end{pgfscope}%
\begin{pgfscope}%
\pgfsetbuttcap%
\pgfsetroundjoin%
\definecolor{currentfill}{rgb}{0.000000,0.000000,0.000000}%
\pgfsetfillcolor{currentfill}%
\pgfsetlinewidth{0.501875pt}%
\definecolor{currentstroke}{rgb}{0.000000,0.000000,0.000000}%
\pgfsetstrokecolor{currentstroke}%
\pgfsetdash{}{0pt}%
\pgfsys@defobject{currentmarker}{\pgfqpoint{0.000000in}{0.000000in}}{\pgfqpoint{0.055556in}{0.000000in}}{%
\pgfpathmoveto{\pgfqpoint{0.000000in}{0.000000in}}%
\pgfpathlineto{\pgfqpoint{0.055556in}{0.000000in}}%
\pgfusepath{stroke,fill}%
}%
\begin{pgfscope}%
\pgfsys@transformshift{0.539608in}{1.730839in}%
\pgfsys@useobject{currentmarker}{}%
\end{pgfscope}%
\end{pgfscope}%
\begin{pgfscope}%
\pgfsetbuttcap%
\pgfsetroundjoin%
\definecolor{currentfill}{rgb}{0.000000,0.000000,0.000000}%
\pgfsetfillcolor{currentfill}%
\pgfsetlinewidth{0.501875pt}%
\definecolor{currentstroke}{rgb}{0.000000,0.000000,0.000000}%
\pgfsetstrokecolor{currentstroke}%
\pgfsetdash{}{0pt}%
\pgfsys@defobject{currentmarker}{\pgfqpoint{-0.055556in}{0.000000in}}{\pgfqpoint{0.000000in}{0.000000in}}{%
\pgfpathmoveto{\pgfqpoint{0.000000in}{0.000000in}}%
\pgfpathlineto{\pgfqpoint{-0.055556in}{0.000000in}}%
\pgfusepath{stroke,fill}%
}%
\begin{pgfscope}%
\pgfsys@transformshift{2.533228in}{1.730839in}%
\pgfsys@useobject{currentmarker}{}%
\end{pgfscope}%
\end{pgfscope}%
\begin{pgfscope}%
\pgftext[left,bottom,x=0.319894in,y=1.690561in,rotate=0.000000]{{\rmfamily\fontsize{9.000000}{10.800000}\selectfont \(\displaystyle 0.6\)}}
\end{pgfscope}%
\begin{pgfscope}%
\pgfpathrectangle{\pgfqpoint{0.539608in}{0.464787in}}{\pgfqpoint{1.993620in}{2.110085in}} %
\pgfusepath{clip}%
\pgfsetbuttcap%
\pgfsetroundjoin%
\pgfsetlinewidth{0.250937pt}%
\definecolor{currentstroke}{rgb}{0.000000,0.000000,0.000000}%
\pgfsetstrokecolor{currentstroke}%
\pgfsetdash{{1.000000pt}{3.000000pt}}{0.000000pt}%
\pgfpathmoveto{\pgfqpoint{0.539608in}{2.152856in}}%
\pgfpathlineto{\pgfqpoint{2.533228in}{2.152856in}}%
\pgfusepath{stroke}%
\end{pgfscope}%
\begin{pgfscope}%
\pgfsetbuttcap%
\pgfsetroundjoin%
\definecolor{currentfill}{rgb}{0.000000,0.000000,0.000000}%
\pgfsetfillcolor{currentfill}%
\pgfsetlinewidth{0.501875pt}%
\definecolor{currentstroke}{rgb}{0.000000,0.000000,0.000000}%
\pgfsetstrokecolor{currentstroke}%
\pgfsetdash{}{0pt}%
\pgfsys@defobject{currentmarker}{\pgfqpoint{0.000000in}{0.000000in}}{\pgfqpoint{0.055556in}{0.000000in}}{%
\pgfpathmoveto{\pgfqpoint{0.000000in}{0.000000in}}%
\pgfpathlineto{\pgfqpoint{0.055556in}{0.000000in}}%
\pgfusepath{stroke,fill}%
}%
\begin{pgfscope}%
\pgfsys@transformshift{0.539608in}{2.152856in}%
\pgfsys@useobject{currentmarker}{}%
\end{pgfscope}%
\end{pgfscope}%
\begin{pgfscope}%
\pgfsetbuttcap%
\pgfsetroundjoin%
\definecolor{currentfill}{rgb}{0.000000,0.000000,0.000000}%
\pgfsetfillcolor{currentfill}%
\pgfsetlinewidth{0.501875pt}%
\definecolor{currentstroke}{rgb}{0.000000,0.000000,0.000000}%
\pgfsetstrokecolor{currentstroke}%
\pgfsetdash{}{0pt}%
\pgfsys@defobject{currentmarker}{\pgfqpoint{-0.055556in}{0.000000in}}{\pgfqpoint{0.000000in}{0.000000in}}{%
\pgfpathmoveto{\pgfqpoint{0.000000in}{0.000000in}}%
\pgfpathlineto{\pgfqpoint{-0.055556in}{0.000000in}}%
\pgfusepath{stroke,fill}%
}%
\begin{pgfscope}%
\pgfsys@transformshift{2.533228in}{2.152856in}%
\pgfsys@useobject{currentmarker}{}%
\end{pgfscope}%
\end{pgfscope}%
\begin{pgfscope}%
\pgftext[left,bottom,x=0.319894in,y=2.112578in,rotate=0.000000]{{\rmfamily\fontsize{9.000000}{10.800000}\selectfont \(\displaystyle 0.8\)}}
\end{pgfscope}%
\begin{pgfscope}%
\pgfpathrectangle{\pgfqpoint{0.539608in}{0.464787in}}{\pgfqpoint{1.993620in}{2.110085in}} %
\pgfusepath{clip}%
\pgfsetbuttcap%
\pgfsetroundjoin%
\pgfsetlinewidth{0.250937pt}%
\definecolor{currentstroke}{rgb}{0.000000,0.000000,0.000000}%
\pgfsetstrokecolor{currentstroke}%
\pgfsetdash{{1.000000pt}{3.000000pt}}{0.000000pt}%
\pgfpathmoveto{\pgfqpoint{0.539608in}{2.574873in}}%
\pgfpathlineto{\pgfqpoint{2.533228in}{2.574873in}}%
\pgfusepath{stroke}%
\end{pgfscope}%
\begin{pgfscope}%
\pgfsetbuttcap%
\pgfsetroundjoin%
\definecolor{currentfill}{rgb}{0.000000,0.000000,0.000000}%
\pgfsetfillcolor{currentfill}%
\pgfsetlinewidth{0.501875pt}%
\definecolor{currentstroke}{rgb}{0.000000,0.000000,0.000000}%
\pgfsetstrokecolor{currentstroke}%
\pgfsetdash{}{0pt}%
\pgfsys@defobject{currentmarker}{\pgfqpoint{0.000000in}{0.000000in}}{\pgfqpoint{0.055556in}{0.000000in}}{%
\pgfpathmoveto{\pgfqpoint{0.000000in}{0.000000in}}%
\pgfpathlineto{\pgfqpoint{0.055556in}{0.000000in}}%
\pgfusepath{stroke,fill}%
}%
\begin{pgfscope}%
\pgfsys@transformshift{0.539608in}{2.574873in}%
\pgfsys@useobject{currentmarker}{}%
\end{pgfscope}%
\end{pgfscope}%
\begin{pgfscope}%
\pgfsetbuttcap%
\pgfsetroundjoin%
\definecolor{currentfill}{rgb}{0.000000,0.000000,0.000000}%
\pgfsetfillcolor{currentfill}%
\pgfsetlinewidth{0.501875pt}%
\definecolor{currentstroke}{rgb}{0.000000,0.000000,0.000000}%
\pgfsetstrokecolor{currentstroke}%
\pgfsetdash{}{0pt}%
\pgfsys@defobject{currentmarker}{\pgfqpoint{-0.055556in}{0.000000in}}{\pgfqpoint{0.000000in}{0.000000in}}{%
\pgfpathmoveto{\pgfqpoint{0.000000in}{0.000000in}}%
\pgfpathlineto{\pgfqpoint{-0.055556in}{0.000000in}}%
\pgfusepath{stroke,fill}%
}%
\begin{pgfscope}%
\pgfsys@transformshift{2.533228in}{2.574873in}%
\pgfsys@useobject{currentmarker}{}%
\end{pgfscope}%
\end{pgfscope}%
\begin{pgfscope}%
\pgftext[left,bottom,x=0.319894in,y=2.534595in,rotate=0.000000]{{\rmfamily\fontsize{9.000000}{10.800000}\selectfont \(\displaystyle 1.0\)}}
\end{pgfscope}%
\begin{pgfscope}%
\pgftext[left,bottom,x=0.250450in,y=0.632208in,rotate=90.000000]{{\rmfamily\fontsize{9.000000}{10.800000}\selectfont Transmitted energy ratio, \(\displaystyle \mathcal T\)}}
\end{pgfscope}%
\begin{pgfscope}%
\pgfsetrectcap%
\pgfsetroundjoin%
\pgfsetlinewidth{0.501875pt}%
\definecolor{currentstroke}{rgb}{0.000000,0.000000,0.000000}%
\pgfsetstrokecolor{currentstroke}%
\pgfsetdash{}{0pt}%
\pgfpathmoveto{\pgfqpoint{0.539608in}{2.574873in}}%
\pgfpathlineto{\pgfqpoint{2.533228in}{2.574873in}}%
\pgfusepath{stroke}%
\end{pgfscope}%
\begin{pgfscope}%
\pgfsetrectcap%
\pgfsetroundjoin%
\pgfsetlinewidth{0.501875pt}%
\definecolor{currentstroke}{rgb}{0.000000,0.000000,0.000000}%
\pgfsetstrokecolor{currentstroke}%
\pgfsetdash{}{0pt}%
\pgfpathmoveto{\pgfqpoint{2.533228in}{0.464787in}}%
\pgfpathlineto{\pgfqpoint{2.533228in}{2.574873in}}%
\pgfusepath{stroke}%
\end{pgfscope}%
\begin{pgfscope}%
\pgfsetrectcap%
\pgfsetroundjoin%
\pgfsetlinewidth{0.501875pt}%
\definecolor{currentstroke}{rgb}{0.000000,0.000000,0.000000}%
\pgfsetstrokecolor{currentstroke}%
\pgfsetdash{}{0pt}%
\pgfpathmoveto{\pgfqpoint{0.539608in}{0.464787in}}%
\pgfpathlineto{\pgfqpoint{2.533228in}{0.464787in}}%
\pgfusepath{stroke}%
\end{pgfscope}%
\begin{pgfscope}%
\pgfsetrectcap%
\pgfsetroundjoin%
\pgfsetlinewidth{0.501875pt}%
\definecolor{currentstroke}{rgb}{0.000000,0.000000,0.000000}%
\pgfsetstrokecolor{currentstroke}%
\pgfsetdash{}{0pt}%
\pgfpathmoveto{\pgfqpoint{0.539608in}{0.464787in}}%
\pgfpathlineto{\pgfqpoint{0.539608in}{2.574873in}}%
\pgfusepath{stroke}%
\end{pgfscope}%
\begin{pgfscope}%
\pgfpathrectangle{\pgfqpoint{0.539608in}{0.464787in}}{\pgfqpoint{1.993620in}{2.110085in}} %
\pgfusepath{clip}%
\pgfsetbuttcap%
\pgfsetroundjoin%
\definecolor{currentfill}{rgb}{1.000000,1.000000,1.000000}%
\pgfsetfillcolor{currentfill}%
\pgfsetlinewidth{0.501875pt}%
\definecolor{currentstroke}{rgb}{0.000000,0.000000,0.000000}%
\pgfsetstrokecolor{currentstroke}%
\pgfsetdash{}{0pt}%
\pgfsys@defobject{currentmarker}{\pgfqpoint{-0.020833in}{-0.020833in}}{\pgfqpoint{0.020833in}{0.020833in}}{%
\pgfpathmoveto{\pgfqpoint{0.000000in}{-0.020833in}}%
\pgfpathcurveto{\pgfqpoint{0.005525in}{-0.020833in}}{\pgfqpoint{0.010825in}{-0.018638in}}{\pgfqpoint{0.014731in}{-0.014731in}}%
\pgfpathcurveto{\pgfqpoint{0.018638in}{-0.010825in}}{\pgfqpoint{0.020833in}{-0.005525in}}{\pgfqpoint{0.020833in}{0.000000in}}%
\pgfpathcurveto{\pgfqpoint{0.020833in}{0.005525in}}{\pgfqpoint{0.018638in}{0.010825in}}{\pgfqpoint{0.014731in}{0.014731in}}%
\pgfpathcurveto{\pgfqpoint{0.010825in}{0.018638in}}{\pgfqpoint{0.005525in}{0.020833in}}{\pgfqpoint{0.000000in}{0.020833in}}%
\pgfpathcurveto{\pgfqpoint{-0.005525in}{0.020833in}}{\pgfqpoint{-0.010825in}{0.018638in}}{\pgfqpoint{-0.014731in}{0.014731in}}%
\pgfpathcurveto{\pgfqpoint{-0.018638in}{0.010825in}}{\pgfqpoint{-0.020833in}{0.005525in}}{\pgfqpoint{-0.020833in}{0.000000in}}%
\pgfpathcurveto{\pgfqpoint{-0.020833in}{-0.005525in}}{\pgfqpoint{-0.018638in}{-0.010825in}}{\pgfqpoint{-0.014731in}{-0.014731in}}%
\pgfpathcurveto{\pgfqpoint{-0.010825in}{-0.018638in}}{\pgfqpoint{-0.005525in}{-0.020833in}}{\pgfqpoint{0.000000in}{-0.020833in}}%
\pgfpathclose%
\pgfusepath{stroke,fill}%
}%
\begin{pgfscope}%
\pgfsys@transformshift{0.540605in}{2.574873in}%
\pgfsys@useobject{currentmarker}{}%
\end{pgfscope}%
\begin{pgfscope}%
\pgfsys@transformshift{0.640286in}{2.573066in}%
\pgfsys@useobject{currentmarker}{}%
\end{pgfscope}%
\begin{pgfscope}%
\pgfsys@transformshift{0.739967in}{2.560946in}%
\pgfsys@useobject{currentmarker}{}%
\end{pgfscope}%
\begin{pgfscope}%
\pgfsys@transformshift{0.839648in}{2.529364in}%
\pgfsys@useobject{currentmarker}{}%
\end{pgfscope}%
\begin{pgfscope}%
\pgfsys@transformshift{0.939329in}{2.471079in}%
\pgfsys@useobject{currentmarker}{}%
\end{pgfscope}%
\begin{pgfscope}%
\pgfsys@transformshift{1.039010in}{2.381463in}%
\pgfsys@useobject{currentmarker}{}%
\end{pgfscope}%
\begin{pgfscope}%
\pgfsys@transformshift{1.138691in}{2.258957in}%
\pgfsys@useobject{currentmarker}{}%
\end{pgfscope}%
\begin{pgfscope}%
\pgfsys@transformshift{1.238372in}{2.105225in}%
\pgfsys@useobject{currentmarker}{}%
\end{pgfscope}%
\begin{pgfscope}%
\pgfsys@transformshift{1.338053in}{1.924987in}%
\pgfsys@useobject{currentmarker}{}%
\end{pgfscope}%
\begin{pgfscope}%
\pgfsys@transformshift{1.437734in}{1.725559in}%
\pgfsys@useobject{currentmarker}{}%
\end{pgfscope}%
\begin{pgfscope}%
\pgfsys@transformshift{1.537415in}{1.516137in}%
\pgfsys@useobject{currentmarker}{}%
\end{pgfscope}%
\begin{pgfscope}%
\pgfsys@transformshift{1.637096in}{1.306895in}%
\pgfsys@useobject{currentmarker}{}%
\end{pgfscope}%
\begin{pgfscope}%
\pgfsys@transformshift{1.736777in}{1.107992in}%
\pgfsys@useobject{currentmarker}{}%
\end{pgfscope}%
\begin{pgfscope}%
\pgfsys@transformshift{1.836458in}{0.928571in}%
\pgfsys@useobject{currentmarker}{}%
\end{pgfscope}%
\begin{pgfscope}%
\pgfsys@transformshift{1.936139in}{0.775869in}%
\pgfsys@useobject{currentmarker}{}%
\end{pgfscope}%
\begin{pgfscope}%
\pgfsys@transformshift{2.035820in}{0.654505in}%
\pgfsys@useobject{currentmarker}{}%
\end{pgfscope}%
\begin{pgfscope}%
\pgfsys@transformshift{2.135501in}{0.566031in}%
\pgfsys@useobject{currentmarker}{}%
\end{pgfscope}%
\begin{pgfscope}%
\pgfsys@transformshift{2.235182in}{0.508775in}%
\pgfsys@useobject{currentmarker}{}%
\end{pgfscope}%
\begin{pgfscope}%
\pgfsys@transformshift{2.334863in}{0.478010in}%
\pgfsys@useobject{currentmarker}{}%
\end{pgfscope}%
\begin{pgfscope}%
\pgfsys@transformshift{2.434544in}{0.466414in}%
\pgfsys@useobject{currentmarker}{}%
\end{pgfscope}%
\begin{pgfscope}%
\pgfsys@transformshift{2.534225in}{0.464787in}%
\pgfsys@useobject{currentmarker}{}%
\end{pgfscope}%
\end{pgfscope}%
\begin{pgfscope}%
\pgfsetrectcap%
\pgfsetroundjoin%
\definecolor{currentfill}{rgb}{1.000000,1.000000,1.000000}%
\pgfsetfillcolor{currentfill}%
\pgfsetlinewidth{0.501875pt}%
\definecolor{currentstroke}{rgb}{0.000000,0.000000,0.000000}%
\pgfsetstrokecolor{currentstroke}%
\pgfsetdash{}{0pt}%
\pgfpathmoveto{\pgfqpoint{1.490812in}{1.782789in}}%
\pgfpathlineto{\pgfqpoint{2.477672in}{1.782789in}}%
\pgfpathlineto{\pgfqpoint{2.477672in}{2.519317in}}%
\pgfpathlineto{\pgfqpoint{1.490812in}{2.519317in}}%
\pgfpathlineto{\pgfqpoint{1.490812in}{1.782789in}}%
\pgfpathclose%
\pgfusepath{stroke,fill}%
\end{pgfscope}%
\begin{pgfscope}%
\pgfsetrectcap%
\pgfsetroundjoin%
\pgfsetlinewidth{1.003750pt}%
\definecolor{currentstroke}{rgb}{0.000000,0.000000,0.000000}%
\pgfsetstrokecolor{currentstroke}%
\pgfsetdash{}{0pt}%
\pgfpathmoveto{\pgfqpoint{1.535257in}{2.288751in}}%
\pgfpathlineto{\pgfqpoint{1.757479in}{2.288751in}}%
\pgfusepath{stroke}%
\end{pgfscope}%
\begin{pgfscope}%
\pgftext[left,bottom,x=1.790812in,y=2.365769in,rotate=0.000000]{{\rmfamily\fontsize{8.000000}{9.600000}\selectfont Analytical}}
\end{pgfscope}%
\begin{pgfscope}%
\pgftext[left,bottom,x=1.790812in,y=2.246281in,rotate=0.000000]{{\rmfamily\fontsize{8.000000}{9.600000}\selectfont  model,}}
\end{pgfscope}%
\begin{pgfscope}%
\pgftext[left,bottom,x=1.790812in,y=2.109270in,rotate=0.000000]{{\rmfamily\fontsize{8.000000}{9.600000}\selectfont  Eq. (12)}}
\end{pgfscope}%
\begin{pgfscope}%
\pgfsetbuttcap%
\pgfsetroundjoin%
\definecolor{currentfill}{rgb}{1.000000,1.000000,1.000000}%
\pgfsetfillcolor{currentfill}%
\pgfsetlinewidth{0.501875pt}%
\definecolor{currentstroke}{rgb}{0.000000,0.000000,0.000000}%
\pgfsetstrokecolor{currentstroke}%
\pgfsetdash{}{0pt}%
\pgfsys@defobject{currentmarker}{\pgfqpoint{-0.020833in}{-0.020833in}}{\pgfqpoint{0.020833in}{0.020833in}}{%
\pgfpathmoveto{\pgfqpoint{0.000000in}{-0.020833in}}%
\pgfpathcurveto{\pgfqpoint{0.005525in}{-0.020833in}}{\pgfqpoint{0.010825in}{-0.018638in}}{\pgfqpoint{0.014731in}{-0.014731in}}%
\pgfpathcurveto{\pgfqpoint{0.018638in}{-0.010825in}}{\pgfqpoint{0.020833in}{-0.005525in}}{\pgfqpoint{0.020833in}{0.000000in}}%
\pgfpathcurveto{\pgfqpoint{0.020833in}{0.005525in}}{\pgfqpoint{0.018638in}{0.010825in}}{\pgfqpoint{0.014731in}{0.014731in}}%
\pgfpathcurveto{\pgfqpoint{0.010825in}{0.018638in}}{\pgfqpoint{0.005525in}{0.020833in}}{\pgfqpoint{0.000000in}{0.020833in}}%
\pgfpathcurveto{\pgfqpoint{-0.005525in}{0.020833in}}{\pgfqpoint{-0.010825in}{0.018638in}}{\pgfqpoint{-0.014731in}{0.014731in}}%
\pgfpathcurveto{\pgfqpoint{-0.018638in}{0.010825in}}{\pgfqpoint{-0.020833in}{0.005525in}}{\pgfqpoint{-0.020833in}{0.000000in}}%
\pgfpathcurveto{\pgfqpoint{-0.020833in}{-0.005525in}}{\pgfqpoint{-0.018638in}{-0.010825in}}{\pgfqpoint{-0.014731in}{-0.014731in}}%
\pgfpathcurveto{\pgfqpoint{-0.010825in}{-0.018638in}}{\pgfqpoint{-0.005525in}{-0.020833in}}{\pgfqpoint{0.000000in}{-0.020833in}}%
\pgfpathclose%
\pgfusepath{stroke,fill}%
}%
\begin{pgfscope}%
\pgfsys@transformshift{1.646368in}{1.937153in}%
\pgfsys@useobject{currentmarker}{}%
\end{pgfscope}%
\end{pgfscope}%
\begin{pgfscope}%
\pgftext[left,bottom,x=1.790812in,y=1.971629in,rotate=0.000000]{{\rmfamily\fontsize{8.000000}{9.600000}\selectfont Geom. sim.}}
\end{pgfscope}%
\begin{pgfscope}%
\pgftext[left,bottom,x=1.790812in,y=1.829186in,rotate=0.000000]{{\rmfamily\fontsize{8.000000}{9.600000}\selectfont  \(\displaystyle a^+=a^-\)}}
\end{pgfscope}%
\end{pgfpicture}%
\makeatother%
\endgroup%

%% file: AnnihilationSignal_2_model.pgf
\begingroup%
\makeatletter%
\begin{pgfpicture}%
\pgfpathrectangle{\pgfpointorigin}{\pgfqpoint{2.750000in}{2.750000in}}%
\pgfusepath{use as bounding box}%
\begin{pgfscope}%
\pgfsetrectcap%
\pgfsetroundjoin%
\definecolor{currentfill}{rgb}{1.000000,1.000000,1.000000}%
\pgfsetfillcolor{currentfill}%
\pgfsetlinewidth{0.000000pt}%
\definecolor{currentstroke}{rgb}{1.000000,1.000000,1.000000}%
\pgfsetstrokecolor{currentstroke}%
\pgfsetdash{}{0pt}%
\pgfpathmoveto{\pgfqpoint{0.000000in}{0.000000in}}%
\pgfpathlineto{\pgfqpoint{2.750000in}{0.000000in}}%
\pgfpathlineto{\pgfqpoint{2.750000in}{2.750000in}}%
\pgfpathlineto{\pgfqpoint{0.000000in}{2.750000in}}%
\pgfpathclose%
\pgfusepath{fill}%
\end{pgfscope}%
\begin{pgfscope}%
\pgfsetrectcap%
\pgfsetroundjoin%
\definecolor{currentfill}{rgb}{1.000000,1.000000,1.000000}%
\pgfsetfillcolor{currentfill}%
\pgfsetlinewidth{0.000000pt}%
\definecolor{currentstroke}{rgb}{0.000000,0.000000,0.000000}%
\pgfsetstrokecolor{currentstroke}%
\pgfsetdash{}{0pt}%
\pgfpathmoveto{\pgfqpoint{0.539608in}{0.464787in}}%
\pgfpathlineto{\pgfqpoint{2.533228in}{0.464787in}}%
\pgfpathlineto{\pgfqpoint{2.533228in}{2.574873in}}%
\pgfpathlineto{\pgfqpoint{0.539608in}{2.574873in}}%
\pgfpathclose%
\pgfusepath{fill}%
\end{pgfscope}%
\begin{pgfscope}%
\pgfpathrectangle{\pgfqpoint{0.539608in}{0.464787in}}{\pgfqpoint{1.993620in}{2.110085in}} %
\pgfusepath{clip}%
\pgfsetrectcap%
\pgfsetroundjoin%
\pgfsetlinewidth{1.003750pt}%
\definecolor{currentstroke}{rgb}{0.000000,0.000000,0.000000}%
\pgfsetstrokecolor{currentstroke}%
\pgfsetdash{}{0pt}%
\pgfpathmoveto{\pgfqpoint{0.540605in}{0.968672in}}%
\pgfpathlineto{\pgfqpoint{0.640286in}{0.968672in}}%
\pgfpathlineto{\pgfqpoint{0.739967in}{0.968672in}}%
\pgfpathlineto{\pgfqpoint{0.839648in}{0.968672in}}%
\pgfpathlineto{\pgfqpoint{0.939329in}{0.968672in}}%
\pgfpathlineto{\pgfqpoint{1.039010in}{0.968672in}}%
\pgfpathlineto{\pgfqpoint{1.138691in}{0.968672in}}%
\pgfpathlineto{\pgfqpoint{1.238372in}{0.968672in}}%
\pgfpathlineto{\pgfqpoint{1.338053in}{0.968672in}}%
\pgfpathlineto{\pgfqpoint{1.437734in}{0.968672in}}%
\pgfpathlineto{\pgfqpoint{1.537415in}{0.968672in}}%
\pgfpathlineto{\pgfqpoint{1.637096in}{0.968672in}}%
\pgfpathlineto{\pgfqpoint{1.736777in}{0.968672in}}%
\pgfpathlineto{\pgfqpoint{1.836458in}{0.968672in}}%
\pgfpathlineto{\pgfqpoint{1.936139in}{0.968672in}}%
\pgfpathlineto{\pgfqpoint{2.035820in}{0.968672in}}%
\pgfpathlineto{\pgfqpoint{2.135501in}{0.968672in}}%
\pgfpathlineto{\pgfqpoint{2.235182in}{0.968672in}}%
\pgfpathlineto{\pgfqpoint{2.334863in}{0.968672in}}%
\pgfpathlineto{\pgfqpoint{2.434544in}{0.968672in}}%
\pgfpathlineto{\pgfqpoint{2.534225in}{0.968672in}}%
\pgfusepath{stroke}%
\end{pgfscope}%
\begin{pgfscope}%
\pgfpathrectangle{\pgfqpoint{0.539608in}{0.464787in}}{\pgfqpoint{1.993620in}{2.110085in}} %
\pgfusepath{clip}%
\pgfsetbuttcap%
\pgfsetroundjoin%
\pgfsetlinewidth{1.003750pt}%
\definecolor{currentstroke}{rgb}{0.000000,0.000000,0.000000}%
\pgfsetstrokecolor{currentstroke}%
\pgfsetdash{{6.000000pt}{6.000000pt}}{0.000000pt}%
\pgfpathmoveto{\pgfqpoint{0.540605in}{0.636415in}}%
\pgfpathlineto{\pgfqpoint{0.640286in}{0.636415in}}%
\pgfpathlineto{\pgfqpoint{0.739967in}{0.636415in}}%
\pgfpathlineto{\pgfqpoint{0.839648in}{0.636415in}}%
\pgfpathlineto{\pgfqpoint{0.939329in}{0.636415in}}%
\pgfpathlineto{\pgfqpoint{1.039010in}{0.636415in}}%
\pgfpathlineto{\pgfqpoint{1.138691in}{0.636415in}}%
\pgfpathlineto{\pgfqpoint{1.238372in}{0.636415in}}%
\pgfpathlineto{\pgfqpoint{1.338053in}{0.636415in}}%
\pgfpathlineto{\pgfqpoint{1.437734in}{0.636415in}}%
\pgfpathlineto{\pgfqpoint{1.537415in}{0.636415in}}%
\pgfpathlineto{\pgfqpoint{1.637096in}{0.636415in}}%
\pgfpathlineto{\pgfqpoint{1.736777in}{0.636415in}}%
\pgfpathlineto{\pgfqpoint{1.836458in}{0.636415in}}%
\pgfpathlineto{\pgfqpoint{1.936139in}{0.636415in}}%
\pgfpathlineto{\pgfqpoint{2.035820in}{0.636415in}}%
\pgfpathlineto{\pgfqpoint{2.135501in}{0.636415in}}%
\pgfpathlineto{\pgfqpoint{2.235182in}{0.636415in}}%
\pgfpathlineto{\pgfqpoint{2.334863in}{0.636415in}}%
\pgfpathlineto{\pgfqpoint{2.434544in}{0.636415in}}%
\pgfpathlineto{\pgfqpoint{2.534225in}{0.636415in}}%
\pgfusepath{stroke}%
\end{pgfscope}%
\begin{pgfscope}%
\pgfpathrectangle{\pgfqpoint{0.539608in}{0.464787in}}{\pgfqpoint{1.993620in}{2.110085in}} %
\pgfusepath{clip}%
\pgfsetbuttcap%
\pgfsetroundjoin%
\pgfsetlinewidth{1.003750pt}%
\definecolor{currentstroke}{rgb}{0.000000,0.000000,0.000000}%
\pgfsetstrokecolor{currentstroke}%
\pgfsetdash{{3.000000pt}{5.000000pt}{1.000000pt}{5.000000pt}}{0.000000pt}%
\pgfpathmoveto{\pgfqpoint{0.540605in}{1.402838in}}%
\pgfpathlineto{\pgfqpoint{0.640286in}{1.402838in}}%
\pgfpathlineto{\pgfqpoint{0.739967in}{1.402838in}}%
\pgfpathlineto{\pgfqpoint{0.839648in}{1.402838in}}%
\pgfpathlineto{\pgfqpoint{0.939329in}{1.402838in}}%
\pgfpathlineto{\pgfqpoint{1.039010in}{1.402838in}}%
\pgfpathlineto{\pgfqpoint{1.138691in}{1.402838in}}%
\pgfpathlineto{\pgfqpoint{1.238372in}{1.402838in}}%
\pgfpathlineto{\pgfqpoint{1.338053in}{1.402838in}}%
\pgfpathlineto{\pgfqpoint{1.437734in}{1.402838in}}%
\pgfpathlineto{\pgfqpoint{1.537415in}{1.402838in}}%
\pgfpathlineto{\pgfqpoint{1.637096in}{1.402838in}}%
\pgfpathlineto{\pgfqpoint{1.736777in}{1.402838in}}%
\pgfpathlineto{\pgfqpoint{1.836458in}{1.402838in}}%
\pgfpathlineto{\pgfqpoint{1.936139in}{1.402838in}}%
\pgfpathlineto{\pgfqpoint{2.035820in}{1.402838in}}%
\pgfpathlineto{\pgfqpoint{2.135501in}{1.402838in}}%
\pgfpathlineto{\pgfqpoint{2.235182in}{1.402838in}}%
\pgfpathlineto{\pgfqpoint{2.334863in}{1.402838in}}%
\pgfpathlineto{\pgfqpoint{2.434544in}{1.402838in}}%
\pgfpathlineto{\pgfqpoint{2.534225in}{1.402838in}}%
\pgfusepath{stroke}%
\end{pgfscope}%
\begin{pgfscope}%
\pgfpathrectangle{\pgfqpoint{0.539608in}{0.464787in}}{\pgfqpoint{1.993620in}{2.110085in}} %
\pgfusepath{clip}%
\pgfsetbuttcap%
\pgfsetroundjoin%
\pgfsetlinewidth{0.250937pt}%
\definecolor{currentstroke}{rgb}{0.000000,0.000000,0.000000}%
\pgfsetstrokecolor{currentstroke}%
\pgfsetdash{{1.000000pt}{3.000000pt}}{0.000000pt}%
\pgfpathmoveto{\pgfqpoint{0.539608in}{0.464787in}}%
\pgfpathlineto{\pgfqpoint{0.539608in}{2.574873in}}%
\pgfusepath{stroke}%
\end{pgfscope}%
\begin{pgfscope}%
\pgfsetbuttcap%
\pgfsetroundjoin%
\definecolor{currentfill}{rgb}{0.000000,0.000000,0.000000}%
\pgfsetfillcolor{currentfill}%
\pgfsetlinewidth{0.501875pt}%
\definecolor{currentstroke}{rgb}{0.000000,0.000000,0.000000}%
\pgfsetstrokecolor{currentstroke}%
\pgfsetdash{}{0pt}%
\pgfsys@defobject{currentmarker}{\pgfqpoint{0.000000in}{0.000000in}}{\pgfqpoint{0.000000in}{0.055556in}}{%
\pgfpathmoveto{\pgfqpoint{0.000000in}{0.000000in}}%
\pgfpathlineto{\pgfqpoint{0.000000in}{0.055556in}}%
\pgfusepath{stroke,fill}%
}%
\begin{pgfscope}%
\pgfsys@transformshift{0.539608in}{0.464787in}%
\pgfsys@useobject{currentmarker}{}%
\end{pgfscope}%
\end{pgfscope}%
\begin{pgfscope}%
\pgfsetbuttcap%
\pgfsetroundjoin%
\definecolor{currentfill}{rgb}{0.000000,0.000000,0.000000}%
\pgfsetfillcolor{currentfill}%
\pgfsetlinewidth{0.501875pt}%
\definecolor{currentstroke}{rgb}{0.000000,0.000000,0.000000}%
\pgfsetstrokecolor{currentstroke}%
\pgfsetdash{}{0pt}%
\pgfsys@defobject{currentmarker}{\pgfqpoint{0.000000in}{-0.055556in}}{\pgfqpoint{0.000000in}{0.000000in}}{%
\pgfpathmoveto{\pgfqpoint{0.000000in}{0.000000in}}%
\pgfpathlineto{\pgfqpoint{0.000000in}{-0.055556in}}%
\pgfusepath{stroke,fill}%
}%
\begin{pgfscope}%
\pgfsys@transformshift{0.539608in}{2.574873in}%
\pgfsys@useobject{currentmarker}{}%
\end{pgfscope}%
\end{pgfscope}%
\begin{pgfscope}%
\pgftext[left,bottom,x=0.457529in,y=0.328676in,rotate=0.000000]{{\rmfamily\fontsize{9.000000}{10.800000}\selectfont \(\displaystyle 0.0\)}}
\end{pgfscope}%
\begin{pgfscope}%
\pgfpathrectangle{\pgfqpoint{0.539608in}{0.464787in}}{\pgfqpoint{1.993620in}{2.110085in}} %
\pgfusepath{clip}%
\pgfsetbuttcap%
\pgfsetroundjoin%
\pgfsetlinewidth{0.250937pt}%
\definecolor{currentstroke}{rgb}{0.000000,0.000000,0.000000}%
\pgfsetstrokecolor{currentstroke}%
\pgfsetdash{{1.000000pt}{3.000000pt}}{0.000000pt}%
\pgfpathmoveto{\pgfqpoint{0.938332in}{0.464787in}}%
\pgfpathlineto{\pgfqpoint{0.938332in}{2.574873in}}%
\pgfusepath{stroke}%
\end{pgfscope}%
\begin{pgfscope}%
\pgfsetbuttcap%
\pgfsetroundjoin%
\definecolor{currentfill}{rgb}{0.000000,0.000000,0.000000}%
\pgfsetfillcolor{currentfill}%
\pgfsetlinewidth{0.501875pt}%
\definecolor{currentstroke}{rgb}{0.000000,0.000000,0.000000}%
\pgfsetstrokecolor{currentstroke}%
\pgfsetdash{}{0pt}%
\pgfsys@defobject{currentmarker}{\pgfqpoint{0.000000in}{0.000000in}}{\pgfqpoint{0.000000in}{0.055556in}}{%
\pgfpathmoveto{\pgfqpoint{0.000000in}{0.000000in}}%
\pgfpathlineto{\pgfqpoint{0.000000in}{0.055556in}}%
\pgfusepath{stroke,fill}%
}%
\begin{pgfscope}%
\pgfsys@transformshift{0.938332in}{0.464787in}%
\pgfsys@useobject{currentmarker}{}%
\end{pgfscope}%
\end{pgfscope}%
\begin{pgfscope}%
\pgfsetbuttcap%
\pgfsetroundjoin%
\definecolor{currentfill}{rgb}{0.000000,0.000000,0.000000}%
\pgfsetfillcolor{currentfill}%
\pgfsetlinewidth{0.501875pt}%
\definecolor{currentstroke}{rgb}{0.000000,0.000000,0.000000}%
\pgfsetstrokecolor{currentstroke}%
\pgfsetdash{}{0pt}%
\pgfsys@defobject{currentmarker}{\pgfqpoint{0.000000in}{-0.055556in}}{\pgfqpoint{0.000000in}{0.000000in}}{%
\pgfpathmoveto{\pgfqpoint{0.000000in}{0.000000in}}%
\pgfpathlineto{\pgfqpoint{0.000000in}{-0.055556in}}%
\pgfusepath{stroke,fill}%
}%
\begin{pgfscope}%
\pgfsys@transformshift{0.938332in}{2.574873in}%
\pgfsys@useobject{currentmarker}{}%
\end{pgfscope}%
\end{pgfscope}%
\begin{pgfscope}%
\pgftext[left,bottom,x=0.856253in,y=0.328676in,rotate=0.000000]{{\rmfamily\fontsize{9.000000}{10.800000}\selectfont \(\displaystyle 0.1\)}}
\end{pgfscope}%
\begin{pgfscope}%
\pgfpathrectangle{\pgfqpoint{0.539608in}{0.464787in}}{\pgfqpoint{1.993620in}{2.110085in}} %
\pgfusepath{clip}%
\pgfsetbuttcap%
\pgfsetroundjoin%
\pgfsetlinewidth{0.250937pt}%
\definecolor{currentstroke}{rgb}{0.000000,0.000000,0.000000}%
\pgfsetstrokecolor{currentstroke}%
\pgfsetdash{{1.000000pt}{3.000000pt}}{0.000000pt}%
\pgfpathmoveto{\pgfqpoint{1.337056in}{0.464787in}}%
\pgfpathlineto{\pgfqpoint{1.337056in}{2.574873in}}%
\pgfusepath{stroke}%
\end{pgfscope}%
\begin{pgfscope}%
\pgfsetbuttcap%
\pgfsetroundjoin%
\definecolor{currentfill}{rgb}{0.000000,0.000000,0.000000}%
\pgfsetfillcolor{currentfill}%
\pgfsetlinewidth{0.501875pt}%
\definecolor{currentstroke}{rgb}{0.000000,0.000000,0.000000}%
\pgfsetstrokecolor{currentstroke}%
\pgfsetdash{}{0pt}%
\pgfsys@defobject{currentmarker}{\pgfqpoint{0.000000in}{0.000000in}}{\pgfqpoint{0.000000in}{0.055556in}}{%
\pgfpathmoveto{\pgfqpoint{0.000000in}{0.000000in}}%
\pgfpathlineto{\pgfqpoint{0.000000in}{0.055556in}}%
\pgfusepath{stroke,fill}%
}%
\begin{pgfscope}%
\pgfsys@transformshift{1.337056in}{0.464787in}%
\pgfsys@useobject{currentmarker}{}%
\end{pgfscope}%
\end{pgfscope}%
\begin{pgfscope}%
\pgfsetbuttcap%
\pgfsetroundjoin%
\definecolor{currentfill}{rgb}{0.000000,0.000000,0.000000}%
\pgfsetfillcolor{currentfill}%
\pgfsetlinewidth{0.501875pt}%
\definecolor{currentstroke}{rgb}{0.000000,0.000000,0.000000}%
\pgfsetstrokecolor{currentstroke}%
\pgfsetdash{}{0pt}%
\pgfsys@defobject{currentmarker}{\pgfqpoint{0.000000in}{-0.055556in}}{\pgfqpoint{0.000000in}{0.000000in}}{%
\pgfpathmoveto{\pgfqpoint{0.000000in}{0.000000in}}%
\pgfpathlineto{\pgfqpoint{0.000000in}{-0.055556in}}%
\pgfusepath{stroke,fill}%
}%
\begin{pgfscope}%
\pgfsys@transformshift{1.337056in}{2.574873in}%
\pgfsys@useobject{currentmarker}{}%
\end{pgfscope}%
\end{pgfscope}%
\begin{pgfscope}%
\pgftext[left,bottom,x=1.254977in,y=0.328676in,rotate=0.000000]{{\rmfamily\fontsize{9.000000}{10.800000}\selectfont \(\displaystyle 0.2\)}}
\end{pgfscope}%
\begin{pgfscope}%
\pgfpathrectangle{\pgfqpoint{0.539608in}{0.464787in}}{\pgfqpoint{1.993620in}{2.110085in}} %
\pgfusepath{clip}%
\pgfsetbuttcap%
\pgfsetroundjoin%
\pgfsetlinewidth{0.250937pt}%
\definecolor{currentstroke}{rgb}{0.000000,0.000000,0.000000}%
\pgfsetstrokecolor{currentstroke}%
\pgfsetdash{{1.000000pt}{3.000000pt}}{0.000000pt}%
\pgfpathmoveto{\pgfqpoint{1.735780in}{0.464787in}}%
\pgfpathlineto{\pgfqpoint{1.735780in}{2.574873in}}%
\pgfusepath{stroke}%
\end{pgfscope}%
\begin{pgfscope}%
\pgfsetbuttcap%
\pgfsetroundjoin%
\definecolor{currentfill}{rgb}{0.000000,0.000000,0.000000}%
\pgfsetfillcolor{currentfill}%
\pgfsetlinewidth{0.501875pt}%
\definecolor{currentstroke}{rgb}{0.000000,0.000000,0.000000}%
\pgfsetstrokecolor{currentstroke}%
\pgfsetdash{}{0pt}%
\pgfsys@defobject{currentmarker}{\pgfqpoint{0.000000in}{0.000000in}}{\pgfqpoint{0.000000in}{0.055556in}}{%
\pgfpathmoveto{\pgfqpoint{0.000000in}{0.000000in}}%
\pgfpathlineto{\pgfqpoint{0.000000in}{0.055556in}}%
\pgfusepath{stroke,fill}%
}%
\begin{pgfscope}%
\pgfsys@transformshift{1.735780in}{0.464787in}%
\pgfsys@useobject{currentmarker}{}%
\end{pgfscope}%
\end{pgfscope}%
\begin{pgfscope}%
\pgfsetbuttcap%
\pgfsetroundjoin%
\definecolor{currentfill}{rgb}{0.000000,0.000000,0.000000}%
\pgfsetfillcolor{currentfill}%
\pgfsetlinewidth{0.501875pt}%
\definecolor{currentstroke}{rgb}{0.000000,0.000000,0.000000}%
\pgfsetstrokecolor{currentstroke}%
\pgfsetdash{}{0pt}%
\pgfsys@defobject{currentmarker}{\pgfqpoint{0.000000in}{-0.055556in}}{\pgfqpoint{0.000000in}{0.000000in}}{%
\pgfpathmoveto{\pgfqpoint{0.000000in}{0.000000in}}%
\pgfpathlineto{\pgfqpoint{0.000000in}{-0.055556in}}%
\pgfusepath{stroke,fill}%
}%
\begin{pgfscope}%
\pgfsys@transformshift{1.735780in}{2.574873in}%
\pgfsys@useobject{currentmarker}{}%
\end{pgfscope}%
\end{pgfscope}%
\begin{pgfscope}%
\pgftext[left,bottom,x=1.653701in,y=0.328676in,rotate=0.000000]{{\rmfamily\fontsize{9.000000}{10.800000}\selectfont \(\displaystyle 0.3\)}}
\end{pgfscope}%
\begin{pgfscope}%
\pgfpathrectangle{\pgfqpoint{0.539608in}{0.464787in}}{\pgfqpoint{1.993620in}{2.110085in}} %
\pgfusepath{clip}%
\pgfsetbuttcap%
\pgfsetroundjoin%
\pgfsetlinewidth{0.250937pt}%
\definecolor{currentstroke}{rgb}{0.000000,0.000000,0.000000}%
\pgfsetstrokecolor{currentstroke}%
\pgfsetdash{{1.000000pt}{3.000000pt}}{0.000000pt}%
\pgfpathmoveto{\pgfqpoint{2.134504in}{0.464787in}}%
\pgfpathlineto{\pgfqpoint{2.134504in}{2.574873in}}%
\pgfusepath{stroke}%
\end{pgfscope}%
\begin{pgfscope}%
\pgfsetbuttcap%
\pgfsetroundjoin%
\definecolor{currentfill}{rgb}{0.000000,0.000000,0.000000}%
\pgfsetfillcolor{currentfill}%
\pgfsetlinewidth{0.501875pt}%
\definecolor{currentstroke}{rgb}{0.000000,0.000000,0.000000}%
\pgfsetstrokecolor{currentstroke}%
\pgfsetdash{}{0pt}%
\pgfsys@defobject{currentmarker}{\pgfqpoint{0.000000in}{0.000000in}}{\pgfqpoint{0.000000in}{0.055556in}}{%
\pgfpathmoveto{\pgfqpoint{0.000000in}{0.000000in}}%
\pgfpathlineto{\pgfqpoint{0.000000in}{0.055556in}}%
\pgfusepath{stroke,fill}%
}%
\begin{pgfscope}%
\pgfsys@transformshift{2.134504in}{0.464787in}%
\pgfsys@useobject{currentmarker}{}%
\end{pgfscope}%
\end{pgfscope}%
\begin{pgfscope}%
\pgfsetbuttcap%
\pgfsetroundjoin%
\definecolor{currentfill}{rgb}{0.000000,0.000000,0.000000}%
\pgfsetfillcolor{currentfill}%
\pgfsetlinewidth{0.501875pt}%
\definecolor{currentstroke}{rgb}{0.000000,0.000000,0.000000}%
\pgfsetstrokecolor{currentstroke}%
\pgfsetdash{}{0pt}%
\pgfsys@defobject{currentmarker}{\pgfqpoint{0.000000in}{-0.055556in}}{\pgfqpoint{0.000000in}{0.000000in}}{%
\pgfpathmoveto{\pgfqpoint{0.000000in}{0.000000in}}%
\pgfpathlineto{\pgfqpoint{0.000000in}{-0.055556in}}%
\pgfusepath{stroke,fill}%
}%
\begin{pgfscope}%
\pgfsys@transformshift{2.134504in}{2.574873in}%
\pgfsys@useobject{currentmarker}{}%
\end{pgfscope}%
\end{pgfscope}%
\begin{pgfscope}%
\pgftext[left,bottom,x=2.052425in,y=0.328676in,rotate=0.000000]{{\rmfamily\fontsize{9.000000}{10.800000}\selectfont \(\displaystyle 0.4\)}}
\end{pgfscope}%
\begin{pgfscope}%
\pgfpathrectangle{\pgfqpoint{0.539608in}{0.464787in}}{\pgfqpoint{1.993620in}{2.110085in}} %
\pgfusepath{clip}%
\pgfsetbuttcap%
\pgfsetroundjoin%
\pgfsetlinewidth{0.250937pt}%
\definecolor{currentstroke}{rgb}{0.000000,0.000000,0.000000}%
\pgfsetstrokecolor{currentstroke}%
\pgfsetdash{{1.000000pt}{3.000000pt}}{0.000000pt}%
\pgfpathmoveto{\pgfqpoint{2.533228in}{0.464787in}}%
\pgfpathlineto{\pgfqpoint{2.533228in}{2.574873in}}%
\pgfusepath{stroke}%
\end{pgfscope}%
\begin{pgfscope}%
\pgfsetbuttcap%
\pgfsetroundjoin%
\definecolor{currentfill}{rgb}{0.000000,0.000000,0.000000}%
\pgfsetfillcolor{currentfill}%
\pgfsetlinewidth{0.501875pt}%
\definecolor{currentstroke}{rgb}{0.000000,0.000000,0.000000}%
\pgfsetstrokecolor{currentstroke}%
\pgfsetdash{}{0pt}%
\pgfsys@defobject{currentmarker}{\pgfqpoint{0.000000in}{0.000000in}}{\pgfqpoint{0.000000in}{0.055556in}}{%
\pgfpathmoveto{\pgfqpoint{0.000000in}{0.000000in}}%
\pgfpathlineto{\pgfqpoint{0.000000in}{0.055556in}}%
\pgfusepath{stroke,fill}%
}%
\begin{pgfscope}%
\pgfsys@transformshift{2.533228in}{0.464787in}%
\pgfsys@useobject{currentmarker}{}%
\end{pgfscope}%
\end{pgfscope}%
\begin{pgfscope}%
\pgfsetbuttcap%
\pgfsetroundjoin%
\definecolor{currentfill}{rgb}{0.000000,0.000000,0.000000}%
\pgfsetfillcolor{currentfill}%
\pgfsetlinewidth{0.501875pt}%
\definecolor{currentstroke}{rgb}{0.000000,0.000000,0.000000}%
\pgfsetstrokecolor{currentstroke}%
\pgfsetdash{}{0pt}%
\pgfsys@defobject{currentmarker}{\pgfqpoint{0.000000in}{-0.055556in}}{\pgfqpoint{0.000000in}{0.000000in}}{%
\pgfpathmoveto{\pgfqpoint{0.000000in}{0.000000in}}%
\pgfpathlineto{\pgfqpoint{0.000000in}{-0.055556in}}%
\pgfusepath{stroke,fill}%
}%
\begin{pgfscope}%
\pgfsys@transformshift{2.533228in}{2.574873in}%
\pgfsys@useobject{currentmarker}{}%
\end{pgfscope}%
\end{pgfscope}%
\begin{pgfscope}%
\pgftext[left,bottom,x=2.451149in,y=0.328676in,rotate=0.000000]{{\rmfamily\fontsize{9.000000}{10.800000}\selectfont \(\displaystyle 0.5\)}}
\end{pgfscope}%
\begin{pgfscope}%
\pgftext[left,bottom,x=0.468431in,y=0.133011in,rotate=0.000000]{{\rmfamily\fontsize{9.000000}{10.800000}\selectfont Normalized distance, \(\displaystyle L/\lambda=\Delta c t/\lambda\)}}
\end{pgfscope}%
\begin{pgfscope}%
\pgfpathrectangle{\pgfqpoint{0.539608in}{0.464787in}}{\pgfqpoint{1.993620in}{2.110085in}} %
\pgfusepath{clip}%
\pgfsetbuttcap%
\pgfsetroundjoin%
\pgfsetlinewidth{0.250937pt}%
\definecolor{currentstroke}{rgb}{0.000000,0.000000,0.000000}%
\pgfsetstrokecolor{currentstroke}%
\pgfsetdash{{1.000000pt}{3.000000pt}}{0.000000pt}%
\pgfpathmoveto{\pgfqpoint{0.539608in}{0.464787in}}%
\pgfpathlineto{\pgfqpoint{2.533228in}{0.464787in}}%
\pgfusepath{stroke}%
\end{pgfscope}%
\begin{pgfscope}%
\pgfsetbuttcap%
\pgfsetroundjoin%
\definecolor{currentfill}{rgb}{0.000000,0.000000,0.000000}%
\pgfsetfillcolor{currentfill}%
\pgfsetlinewidth{0.501875pt}%
\definecolor{currentstroke}{rgb}{0.000000,0.000000,0.000000}%
\pgfsetstrokecolor{currentstroke}%
\pgfsetdash{}{0pt}%
\pgfsys@defobject{currentmarker}{\pgfqpoint{0.000000in}{0.000000in}}{\pgfqpoint{0.055556in}{0.000000in}}{%
\pgfpathmoveto{\pgfqpoint{0.000000in}{0.000000in}}%
\pgfpathlineto{\pgfqpoint{0.055556in}{0.000000in}}%
\pgfusepath{stroke,fill}%
}%
\begin{pgfscope}%
\pgfsys@transformshift{0.539608in}{0.464787in}%
\pgfsys@useobject{currentmarker}{}%
\end{pgfscope}%
\end{pgfscope}%
\begin{pgfscope}%
\pgfsetbuttcap%
\pgfsetroundjoin%
\definecolor{currentfill}{rgb}{0.000000,0.000000,0.000000}%
\pgfsetfillcolor{currentfill}%
\pgfsetlinewidth{0.501875pt}%
\definecolor{currentstroke}{rgb}{0.000000,0.000000,0.000000}%
\pgfsetstrokecolor{currentstroke}%
\pgfsetdash{}{0pt}%
\pgfsys@defobject{currentmarker}{\pgfqpoint{-0.055556in}{0.000000in}}{\pgfqpoint{0.000000in}{0.000000in}}{%
\pgfpathmoveto{\pgfqpoint{0.000000in}{0.000000in}}%
\pgfpathlineto{\pgfqpoint{-0.055556in}{0.000000in}}%
\pgfusepath{stroke,fill}%
}%
\begin{pgfscope}%
\pgfsys@transformshift{2.533228in}{0.464787in}%
\pgfsys@useobject{currentmarker}{}%
\end{pgfscope}%
\end{pgfscope}%
\begin{pgfscope}%
\pgftext[left,bottom,x=0.319894in,y=0.424510in,rotate=0.000000]{{\rmfamily\fontsize{9.000000}{10.800000}\selectfont \(\displaystyle 0.0\)}}
\end{pgfscope}%
\begin{pgfscope}%
\pgfpathrectangle{\pgfqpoint{0.539608in}{0.464787in}}{\pgfqpoint{1.993620in}{2.110085in}} %
\pgfusepath{clip}%
\pgfsetbuttcap%
\pgfsetroundjoin%
\pgfsetlinewidth{0.250937pt}%
\definecolor{currentstroke}{rgb}{0.000000,0.000000,0.000000}%
\pgfsetstrokecolor{currentstroke}%
\pgfsetdash{{1.000000pt}{3.000000pt}}{0.000000pt}%
\pgfpathmoveto{\pgfqpoint{0.539608in}{0.886805in}}%
\pgfpathlineto{\pgfqpoint{2.533228in}{0.886805in}}%
\pgfusepath{stroke}%
\end{pgfscope}%
\begin{pgfscope}%
\pgfsetbuttcap%
\pgfsetroundjoin%
\definecolor{currentfill}{rgb}{0.000000,0.000000,0.000000}%
\pgfsetfillcolor{currentfill}%
\pgfsetlinewidth{0.501875pt}%
\definecolor{currentstroke}{rgb}{0.000000,0.000000,0.000000}%
\pgfsetstrokecolor{currentstroke}%
\pgfsetdash{}{0pt}%
\pgfsys@defobject{currentmarker}{\pgfqpoint{0.000000in}{0.000000in}}{\pgfqpoint{0.055556in}{0.000000in}}{%
\pgfpathmoveto{\pgfqpoint{0.000000in}{0.000000in}}%
\pgfpathlineto{\pgfqpoint{0.055556in}{0.000000in}}%
\pgfusepath{stroke,fill}%
}%
\begin{pgfscope}%
\pgfsys@transformshift{0.539608in}{0.886805in}%
\pgfsys@useobject{currentmarker}{}%
\end{pgfscope}%
\end{pgfscope}%
\begin{pgfscope}%
\pgfsetbuttcap%
\pgfsetroundjoin%
\definecolor{currentfill}{rgb}{0.000000,0.000000,0.000000}%
\pgfsetfillcolor{currentfill}%
\pgfsetlinewidth{0.501875pt}%
\definecolor{currentstroke}{rgb}{0.000000,0.000000,0.000000}%
\pgfsetstrokecolor{currentstroke}%
\pgfsetdash{}{0pt}%
\pgfsys@defobject{currentmarker}{\pgfqpoint{-0.055556in}{0.000000in}}{\pgfqpoint{0.000000in}{0.000000in}}{%
\pgfpathmoveto{\pgfqpoint{0.000000in}{0.000000in}}%
\pgfpathlineto{\pgfqpoint{-0.055556in}{0.000000in}}%
\pgfusepath{stroke,fill}%
}%
\begin{pgfscope}%
\pgfsys@transformshift{2.533228in}{0.886805in}%
\pgfsys@useobject{currentmarker}{}%
\end{pgfscope}%
\end{pgfscope}%
\begin{pgfscope}%
\pgftext[left,bottom,x=0.319894in,y=0.846527in,rotate=0.000000]{{\rmfamily\fontsize{9.000000}{10.800000}\selectfont \(\displaystyle 0.2\)}}
\end{pgfscope}%
\begin{pgfscope}%
\pgfpathrectangle{\pgfqpoint{0.539608in}{0.464787in}}{\pgfqpoint{1.993620in}{2.110085in}} %
\pgfusepath{clip}%
\pgfsetbuttcap%
\pgfsetroundjoin%
\pgfsetlinewidth{0.250937pt}%
\definecolor{currentstroke}{rgb}{0.000000,0.000000,0.000000}%
\pgfsetstrokecolor{currentstroke}%
\pgfsetdash{{1.000000pt}{3.000000pt}}{0.000000pt}%
\pgfpathmoveto{\pgfqpoint{0.539608in}{1.308822in}}%
\pgfpathlineto{\pgfqpoint{2.533228in}{1.308822in}}%
\pgfusepath{stroke}%
\end{pgfscope}%
\begin{pgfscope}%
\pgfsetbuttcap%
\pgfsetroundjoin%
\definecolor{currentfill}{rgb}{0.000000,0.000000,0.000000}%
\pgfsetfillcolor{currentfill}%
\pgfsetlinewidth{0.501875pt}%
\definecolor{currentstroke}{rgb}{0.000000,0.000000,0.000000}%
\pgfsetstrokecolor{currentstroke}%
\pgfsetdash{}{0pt}%
\pgfsys@defobject{currentmarker}{\pgfqpoint{0.000000in}{0.000000in}}{\pgfqpoint{0.055556in}{0.000000in}}{%
\pgfpathmoveto{\pgfqpoint{0.000000in}{0.000000in}}%
\pgfpathlineto{\pgfqpoint{0.055556in}{0.000000in}}%
\pgfusepath{stroke,fill}%
}%
\begin{pgfscope}%
\pgfsys@transformshift{0.539608in}{1.308822in}%
\pgfsys@useobject{currentmarker}{}%
\end{pgfscope}%
\end{pgfscope}%
\begin{pgfscope}%
\pgfsetbuttcap%
\pgfsetroundjoin%
\definecolor{currentfill}{rgb}{0.000000,0.000000,0.000000}%
\pgfsetfillcolor{currentfill}%
\pgfsetlinewidth{0.501875pt}%
\definecolor{currentstroke}{rgb}{0.000000,0.000000,0.000000}%
\pgfsetstrokecolor{currentstroke}%
\pgfsetdash{}{0pt}%
\pgfsys@defobject{currentmarker}{\pgfqpoint{-0.055556in}{0.000000in}}{\pgfqpoint{0.000000in}{0.000000in}}{%
\pgfpathmoveto{\pgfqpoint{0.000000in}{0.000000in}}%
\pgfpathlineto{\pgfqpoint{-0.055556in}{0.000000in}}%
\pgfusepath{stroke,fill}%
}%
\begin{pgfscope}%
\pgfsys@transformshift{2.533228in}{1.308822in}%
\pgfsys@useobject{currentmarker}{}%
\end{pgfscope}%
\end{pgfscope}%
\begin{pgfscope}%
\pgftext[left,bottom,x=0.319894in,y=1.268544in,rotate=0.000000]{{\rmfamily\fontsize{9.000000}{10.800000}\selectfont \(\displaystyle 0.4\)}}
\end{pgfscope}%
\begin{pgfscope}%
\pgfpathrectangle{\pgfqpoint{0.539608in}{0.464787in}}{\pgfqpoint{1.993620in}{2.110085in}} %
\pgfusepath{clip}%
\pgfsetbuttcap%
\pgfsetroundjoin%
\pgfsetlinewidth{0.250937pt}%
\definecolor{currentstroke}{rgb}{0.000000,0.000000,0.000000}%
\pgfsetstrokecolor{currentstroke}%
\pgfsetdash{{1.000000pt}{3.000000pt}}{0.000000pt}%
\pgfpathmoveto{\pgfqpoint{0.539608in}{1.730839in}}%
\pgfpathlineto{\pgfqpoint{2.533228in}{1.730839in}}%
\pgfusepath{stroke}%
\end{pgfscope}%
\begin{pgfscope}%
\pgfsetbuttcap%
\pgfsetroundjoin%
\definecolor{currentfill}{rgb}{0.000000,0.000000,0.000000}%
\pgfsetfillcolor{currentfill}%
\pgfsetlinewidth{0.501875pt}%
\definecolor{currentstroke}{rgb}{0.000000,0.000000,0.000000}%
\pgfsetstrokecolor{currentstroke}%
\pgfsetdash{}{0pt}%
\pgfsys@defobject{currentmarker}{\pgfqpoint{0.000000in}{0.000000in}}{\pgfqpoint{0.055556in}{0.000000in}}{%
\pgfpathmoveto{\pgfqpoint{0.000000in}{0.000000in}}%
\pgfpathlineto{\pgfqpoint{0.055556in}{0.000000in}}%
\pgfusepath{stroke,fill}%
}%
\begin{pgfscope}%
\pgfsys@transformshift{0.539608in}{1.730839in}%
\pgfsys@useobject{currentmarker}{}%
\end{pgfscope}%
\end{pgfscope}%
\begin{pgfscope}%
\pgfsetbuttcap%
\pgfsetroundjoin%
\definecolor{currentfill}{rgb}{0.000000,0.000000,0.000000}%
\pgfsetfillcolor{currentfill}%
\pgfsetlinewidth{0.501875pt}%
\definecolor{currentstroke}{rgb}{0.000000,0.000000,0.000000}%
\pgfsetstrokecolor{currentstroke}%
\pgfsetdash{}{0pt}%
\pgfsys@defobject{currentmarker}{\pgfqpoint{-0.055556in}{0.000000in}}{\pgfqpoint{0.000000in}{0.000000in}}{%
\pgfpathmoveto{\pgfqpoint{0.000000in}{0.000000in}}%
\pgfpathlineto{\pgfqpoint{-0.055556in}{0.000000in}}%
\pgfusepath{stroke,fill}%
}%
\begin{pgfscope}%
\pgfsys@transformshift{2.533228in}{1.730839in}%
\pgfsys@useobject{currentmarker}{}%
\end{pgfscope}%
\end{pgfscope}%
\begin{pgfscope}%
\pgftext[left,bottom,x=0.319894in,y=1.690561in,rotate=0.000000]{{\rmfamily\fontsize{9.000000}{10.800000}\selectfont \(\displaystyle 0.6\)}}
\end{pgfscope}%
\begin{pgfscope}%
\pgfpathrectangle{\pgfqpoint{0.539608in}{0.464787in}}{\pgfqpoint{1.993620in}{2.110085in}} %
\pgfusepath{clip}%
\pgfsetbuttcap%
\pgfsetroundjoin%
\pgfsetlinewidth{0.250937pt}%
\definecolor{currentstroke}{rgb}{0.000000,0.000000,0.000000}%
\pgfsetstrokecolor{currentstroke}%
\pgfsetdash{{1.000000pt}{3.000000pt}}{0.000000pt}%
\pgfpathmoveto{\pgfqpoint{0.539608in}{2.152856in}}%
\pgfpathlineto{\pgfqpoint{2.533228in}{2.152856in}}%
\pgfusepath{stroke}%
\end{pgfscope}%
\begin{pgfscope}%
\pgfsetbuttcap%
\pgfsetroundjoin%
\definecolor{currentfill}{rgb}{0.000000,0.000000,0.000000}%
\pgfsetfillcolor{currentfill}%
\pgfsetlinewidth{0.501875pt}%
\definecolor{currentstroke}{rgb}{0.000000,0.000000,0.000000}%
\pgfsetstrokecolor{currentstroke}%
\pgfsetdash{}{0pt}%
\pgfsys@defobject{currentmarker}{\pgfqpoint{0.000000in}{0.000000in}}{\pgfqpoint{0.055556in}{0.000000in}}{%
\pgfpathmoveto{\pgfqpoint{0.000000in}{0.000000in}}%
\pgfpathlineto{\pgfqpoint{0.055556in}{0.000000in}}%
\pgfusepath{stroke,fill}%
}%
\begin{pgfscope}%
\pgfsys@transformshift{0.539608in}{2.152856in}%
\pgfsys@useobject{currentmarker}{}%
\end{pgfscope}%
\end{pgfscope}%
\begin{pgfscope}%
\pgfsetbuttcap%
\pgfsetroundjoin%
\definecolor{currentfill}{rgb}{0.000000,0.000000,0.000000}%
\pgfsetfillcolor{currentfill}%
\pgfsetlinewidth{0.501875pt}%
\definecolor{currentstroke}{rgb}{0.000000,0.000000,0.000000}%
\pgfsetstrokecolor{currentstroke}%
\pgfsetdash{}{0pt}%
\pgfsys@defobject{currentmarker}{\pgfqpoint{-0.055556in}{0.000000in}}{\pgfqpoint{0.000000in}{0.000000in}}{%
\pgfpathmoveto{\pgfqpoint{0.000000in}{0.000000in}}%
\pgfpathlineto{\pgfqpoint{-0.055556in}{0.000000in}}%
\pgfusepath{stroke,fill}%
}%
\begin{pgfscope}%
\pgfsys@transformshift{2.533228in}{2.152856in}%
\pgfsys@useobject{currentmarker}{}%
\end{pgfscope}%
\end{pgfscope}%
\begin{pgfscope}%
\pgftext[left,bottom,x=0.319894in,y=2.112578in,rotate=0.000000]{{\rmfamily\fontsize{9.000000}{10.800000}\selectfont \(\displaystyle 0.8\)}}
\end{pgfscope}%
\begin{pgfscope}%
\pgfpathrectangle{\pgfqpoint{0.539608in}{0.464787in}}{\pgfqpoint{1.993620in}{2.110085in}} %
\pgfusepath{clip}%
\pgfsetbuttcap%
\pgfsetroundjoin%
\pgfsetlinewidth{0.250937pt}%
\definecolor{currentstroke}{rgb}{0.000000,0.000000,0.000000}%
\pgfsetstrokecolor{currentstroke}%
\pgfsetdash{{1.000000pt}{3.000000pt}}{0.000000pt}%
\pgfpathmoveto{\pgfqpoint{0.539608in}{2.574873in}}%
\pgfpathlineto{\pgfqpoint{2.533228in}{2.574873in}}%
\pgfusepath{stroke}%
\end{pgfscope}%
\begin{pgfscope}%
\pgfsetbuttcap%
\pgfsetroundjoin%
\definecolor{currentfill}{rgb}{0.000000,0.000000,0.000000}%
\pgfsetfillcolor{currentfill}%
\pgfsetlinewidth{0.501875pt}%
\definecolor{currentstroke}{rgb}{0.000000,0.000000,0.000000}%
\pgfsetstrokecolor{currentstroke}%
\pgfsetdash{}{0pt}%
\pgfsys@defobject{currentmarker}{\pgfqpoint{0.000000in}{0.000000in}}{\pgfqpoint{0.055556in}{0.000000in}}{%
\pgfpathmoveto{\pgfqpoint{0.000000in}{0.000000in}}%
\pgfpathlineto{\pgfqpoint{0.055556in}{0.000000in}}%
\pgfusepath{stroke,fill}%
}%
\begin{pgfscope}%
\pgfsys@transformshift{0.539608in}{2.574873in}%
\pgfsys@useobject{currentmarker}{}%
\end{pgfscope}%
\end{pgfscope}%
\begin{pgfscope}%
\pgfsetbuttcap%
\pgfsetroundjoin%
\definecolor{currentfill}{rgb}{0.000000,0.000000,0.000000}%
\pgfsetfillcolor{currentfill}%
\pgfsetlinewidth{0.501875pt}%
\definecolor{currentstroke}{rgb}{0.000000,0.000000,0.000000}%
\pgfsetstrokecolor{currentstroke}%
\pgfsetdash{}{0pt}%
\pgfsys@defobject{currentmarker}{\pgfqpoint{-0.055556in}{0.000000in}}{\pgfqpoint{0.000000in}{0.000000in}}{%
\pgfpathmoveto{\pgfqpoint{0.000000in}{0.000000in}}%
\pgfpathlineto{\pgfqpoint{-0.055556in}{0.000000in}}%
\pgfusepath{stroke,fill}%
}%
\begin{pgfscope}%
\pgfsys@transformshift{2.533228in}{2.574873in}%
\pgfsys@useobject{currentmarker}{}%
\end{pgfscope}%
\end{pgfscope}%
\begin{pgfscope}%
\pgftext[left,bottom,x=0.319894in,y=2.534595in,rotate=0.000000]{{\rmfamily\fontsize{9.000000}{10.800000}\selectfont \(\displaystyle 1.0\)}}
\end{pgfscope}%
\begin{pgfscope}%
\pgftext[left,bottom,x=0.250450in,y=0.632208in,rotate=90.000000]{{\rmfamily\fontsize{9.000000}{10.800000}\selectfont Transmitted energy ratio, \(\displaystyle \mathcal T\)}}
\end{pgfscope}%
\begin{pgfscope}%
\pgfsetrectcap%
\pgfsetroundjoin%
\pgfsetlinewidth{0.501875pt}%
\definecolor{currentstroke}{rgb}{0.000000,0.000000,0.000000}%
\pgfsetstrokecolor{currentstroke}%
\pgfsetdash{}{0pt}%
\pgfpathmoveto{\pgfqpoint{0.539608in}{2.574873in}}%
\pgfpathlineto{\pgfqpoint{2.533228in}{2.574873in}}%
\pgfusepath{stroke}%
\end{pgfscope}%
\begin{pgfscope}%
\pgfsetrectcap%
\pgfsetroundjoin%
\pgfsetlinewidth{0.501875pt}%
\definecolor{currentstroke}{rgb}{0.000000,0.000000,0.000000}%
\pgfsetstrokecolor{currentstroke}%
\pgfsetdash{}{0pt}%
\pgfpathmoveto{\pgfqpoint{2.533228in}{0.464787in}}%
\pgfpathlineto{\pgfqpoint{2.533228in}{2.574873in}}%
\pgfusepath{stroke}%
\end{pgfscope}%
\begin{pgfscope}%
\pgfsetrectcap%
\pgfsetroundjoin%
\pgfsetlinewidth{0.501875pt}%
\definecolor{currentstroke}{rgb}{0.000000,0.000000,0.000000}%
\pgfsetstrokecolor{currentstroke}%
\pgfsetdash{}{0pt}%
\pgfpathmoveto{\pgfqpoint{0.539608in}{0.464787in}}%
\pgfpathlineto{\pgfqpoint{2.533228in}{0.464787in}}%
\pgfusepath{stroke}%
\end{pgfscope}%
\begin{pgfscope}%
\pgfsetrectcap%
\pgfsetroundjoin%
\pgfsetlinewidth{0.501875pt}%
\definecolor{currentstroke}{rgb}{0.000000,0.000000,0.000000}%
\pgfsetstrokecolor{currentstroke}%
\pgfsetdash{}{0pt}%
\pgfpathmoveto{\pgfqpoint{0.539608in}{0.464787in}}%
\pgfpathlineto{\pgfqpoint{0.539608in}{2.574873in}}%
\pgfusepath{stroke}%
\end{pgfscope}%
\begin{pgfscope}%
\pgfpathrectangle{\pgfqpoint{0.539608in}{0.464787in}}{\pgfqpoint{1.993620in}{2.110085in}} %
\pgfusepath{clip}%
\pgfsetbuttcap%
\pgfsetroundjoin%
\definecolor{currentfill}{rgb}{1.000000,1.000000,1.000000}%
\pgfsetfillcolor{currentfill}%
\pgfsetlinewidth{0.501875pt}%
\definecolor{currentstroke}{rgb}{0.000000,0.000000,0.000000}%
\pgfsetstrokecolor{currentstroke}%
\pgfsetdash{}{0pt}%
\pgfsys@defobject{currentmarker}{\pgfqpoint{-0.020833in}{-0.020833in}}{\pgfqpoint{0.020833in}{0.020833in}}{%
\pgfpathmoveto{\pgfqpoint{0.000000in}{-0.020833in}}%
\pgfpathcurveto{\pgfqpoint{0.005525in}{-0.020833in}}{\pgfqpoint{0.010825in}{-0.018638in}}{\pgfqpoint{0.014731in}{-0.014731in}}%
\pgfpathcurveto{\pgfqpoint{0.018638in}{-0.010825in}}{\pgfqpoint{0.020833in}{-0.005525in}}{\pgfqpoint{0.020833in}{0.000000in}}%
\pgfpathcurveto{\pgfqpoint{0.020833in}{0.005525in}}{\pgfqpoint{0.018638in}{0.010825in}}{\pgfqpoint{0.014731in}{0.014731in}}%
\pgfpathcurveto{\pgfqpoint{0.010825in}{0.018638in}}{\pgfqpoint{0.005525in}{0.020833in}}{\pgfqpoint{0.000000in}{0.020833in}}%
\pgfpathcurveto{\pgfqpoint{-0.005525in}{0.020833in}}{\pgfqpoint{-0.010825in}{0.018638in}}{\pgfqpoint{-0.014731in}{0.014731in}}%
\pgfpathcurveto{\pgfqpoint{-0.018638in}{0.010825in}}{\pgfqpoint{-0.020833in}{0.005525in}}{\pgfqpoint{-0.020833in}{0.000000in}}%
\pgfpathcurveto{\pgfqpoint{-0.020833in}{-0.005525in}}{\pgfqpoint{-0.018638in}{-0.010825in}}{\pgfqpoint{-0.014731in}{-0.014731in}}%
\pgfpathcurveto{\pgfqpoint{-0.010825in}{-0.018638in}}{\pgfqpoint{-0.005525in}{-0.020833in}}{\pgfqpoint{0.000000in}{-0.020833in}}%
\pgfpathclose%
\pgfusepath{stroke,fill}%
}%
\begin{pgfscope}%
\pgfsys@transformshift{0.540605in}{2.574873in}%
\pgfsys@useobject{currentmarker}{}%
\end{pgfscope}%
\begin{pgfscope}%
\pgfsys@transformshift{0.640286in}{2.573358in}%
\pgfsys@useobject{currentmarker}{}%
\end{pgfscope}%
\begin{pgfscope}%
\pgfsys@transformshift{0.739967in}{2.563227in}%
\pgfsys@useobject{currentmarker}{}%
\end{pgfscope}%
\begin{pgfscope}%
\pgfsys@transformshift{0.839648in}{2.537071in}%
\pgfsys@useobject{currentmarker}{}%
\end{pgfscope}%
\begin{pgfscope}%
\pgfsys@transformshift{0.939329in}{2.487949in}%
\pgfsys@useobject{currentmarker}{}%
\end{pgfscope}%
\begin{pgfscope}%
\pgfsys@transformshift{1.039010in}{2.413281in}%
\pgfsys@useobject{currentmarker}{}%
\end{pgfscope}%
\begin{pgfscope}%
\pgfsys@transformshift{1.138691in}{2.310347in}%
\pgfsys@useobject{currentmarker}{}%
\end{pgfscope}%
\begin{pgfscope}%
\pgfsys@transformshift{1.238372in}{2.181822in}%
\pgfsys@useobject{currentmarker}{}%
\end{pgfscope}%
\begin{pgfscope}%
\pgfsys@transformshift{1.338053in}{2.030398in}%
\pgfsys@useobject{currentmarker}{}%
\end{pgfscope}%
\begin{pgfscope}%
\pgfsys@transformshift{1.437734in}{1.862412in}%
\pgfsys@useobject{currentmarker}{}%
\end{pgfscope}%
\begin{pgfscope}%
\pgfsys@transformshift{1.537415in}{1.686517in}%
\pgfsys@useobject{currentmarker}{}%
\end{pgfscope}%
\begin{pgfscope}%
\pgfsys@transformshift{1.637096in}{1.511480in}%
\pgfsys@useobject{currentmarker}{}%
\end{pgfscope}%
\begin{pgfscope}%
\pgfsys@transformshift{1.736777in}{1.346632in}%
\pgfsys@useobject{currentmarker}{}%
\end{pgfscope}%
\begin{pgfscope}%
\pgfsys@transformshift{1.836458in}{1.201047in}%
\pgfsys@useobject{currentmarker}{}%
\end{pgfscope}%
\begin{pgfscope}%
\pgfsys@transformshift{1.936139in}{1.083464in}%
\pgfsys@useobject{currentmarker}{}%
\end{pgfscope}%
\begin{pgfscope}%
\pgfsys@transformshift{2.035820in}{1.002215in}%
\pgfsys@useobject{currentmarker}{}%
\end{pgfscope}%
\begin{pgfscope}%
\pgfsys@transformshift{2.135501in}{0.968287in}%
\pgfsys@useobject{currentmarker}{}%
\end{pgfscope}%
\begin{pgfscope}%
\pgfsys@transformshift{2.235182in}{0.968131in}%
\pgfsys@useobject{currentmarker}{}%
\end{pgfscope}%
\begin{pgfscope}%
\pgfsys@transformshift{2.334863in}{0.968131in}%
\pgfsys@useobject{currentmarker}{}%
\end{pgfscope}%
\begin{pgfscope}%
\pgfsys@transformshift{2.434544in}{0.968131in}%
\pgfsys@useobject{currentmarker}{}%
\end{pgfscope}%
\begin{pgfscope}%
\pgfsys@transformshift{2.534225in}{0.968131in}%
\pgfsys@useobject{currentmarker}{}%
\end{pgfscope}%
\end{pgfscope}%
\begin{pgfscope}%
\pgfpathrectangle{\pgfqpoint{0.539608in}{0.464787in}}{\pgfqpoint{1.993620in}{2.110085in}} %
\pgfusepath{clip}%
\pgfsetbuttcap%
\pgfsetmiterjoin%
\definecolor{currentfill}{rgb}{1.000000,1.000000,1.000000}%
\pgfsetfillcolor{currentfill}%
\pgfsetlinewidth{0.501875pt}%
\definecolor{currentstroke}{rgb}{0.000000,0.000000,0.000000}%
\pgfsetstrokecolor{currentstroke}%
\pgfsetdash{}{0pt}%
\pgfsys@defobject{currentmarker}{\pgfqpoint{-0.020833in}{-0.020833in}}{\pgfqpoint{0.020833in}{0.020833in}}{%
\pgfpathmoveto{\pgfqpoint{-0.020833in}{-0.020833in}}%
\pgfpathlineto{\pgfqpoint{0.020833in}{-0.020833in}}%
\pgfpathlineto{\pgfqpoint{0.020833in}{0.020833in}}%
\pgfpathlineto{\pgfqpoint{-0.020833in}{0.020833in}}%
\pgfpathclose%
\pgfusepath{stroke,fill}%
}%
\begin{pgfscope}%
\pgfsys@transformshift{0.540605in}{2.574873in}%
\pgfsys@useobject{currentmarker}{}%
\end{pgfscope}%
\begin{pgfscope}%
\pgfsys@transformshift{0.640286in}{2.573187in}%
\pgfsys@useobject{currentmarker}{}%
\end{pgfscope}%
\begin{pgfscope}%
\pgfsys@transformshift{0.739967in}{2.561910in}%
\pgfsys@useobject{currentmarker}{}%
\end{pgfscope}%
\begin{pgfscope}%
\pgfsys@transformshift{0.839648in}{2.532082in}%
\pgfsys@useobject{currentmarker}{}%
\end{pgfscope}%
\begin{pgfscope}%
\pgfsys@transformshift{0.939329in}{2.477739in}%
\pgfsys@useobject{currentmarker}{}%
\end{pgfscope}%
\begin{pgfscope}%
\pgfsys@transformshift{1.039010in}{2.392857in}%
\pgfsys@useobject{currentmarker}{}%
\end{pgfscope}%
\begin{pgfscope}%
\pgfsys@transformshift{1.138691in}{2.277808in}%
\pgfsys@useobject{currentmarker}{}%
\end{pgfscope}%
\begin{pgfscope}%
\pgfsys@transformshift{1.238372in}{2.132925in}%
\pgfsys@useobject{currentmarker}{}%
\end{pgfscope}%
\begin{pgfscope}%
\pgfsys@transformshift{1.338053in}{1.962276in}%
\pgfsys@useobject{currentmarker}{}%
\end{pgfscope}%
\begin{pgfscope}%
\pgfsys@transformshift{1.437734in}{1.774302in}%
\pgfsys@useobject{currentmarker}{}%
\end{pgfscope}%
\begin{pgfscope}%
\pgfsys@transformshift{1.537415in}{1.575349in}%
\pgfsys@useobject{currentmarker}{}%
\end{pgfscope}%
\begin{pgfscope}%
\pgfsys@transformshift{1.637096in}{1.376832in}%
\pgfsys@useobject{currentmarker}{}%
\end{pgfscope}%
\begin{pgfscope}%
\pgfsys@transformshift{1.736777in}{1.187773in}%
\pgfsys@useobject{currentmarker}{}%
\end{pgfscope}%
\begin{pgfscope}%
\pgfsys@transformshift{1.836458in}{1.017236in}%
\pgfsys@useobject{currentmarker}{}%
\end{pgfscope}%
\begin{pgfscope}%
\pgfsys@transformshift{1.936139in}{0.871809in}%
\pgfsys@useobject{currentmarker}{}%
\end{pgfscope}%
\begin{pgfscope}%
\pgfsys@transformshift{2.035820in}{0.757684in}%
\pgfsys@useobject{currentmarker}{}%
\end{pgfscope}%
\begin{pgfscope}%
\pgfsys@transformshift{2.135501in}{0.678534in}%
\pgfsys@useobject{currentmarker}{}%
\end{pgfscope}%
\begin{pgfscope}%
\pgfsys@transformshift{2.235182in}{0.638924in}%
\pgfsys@useobject{currentmarker}{}%
\end{pgfscope}%
\begin{pgfscope}%
\pgfsys@transformshift{2.334863in}{0.636086in}%
\pgfsys@useobject{currentmarker}{}%
\end{pgfscope}%
\begin{pgfscope}%
\pgfsys@transformshift{2.434544in}{0.636086in}%
\pgfsys@useobject{currentmarker}{}%
\end{pgfscope}%
\begin{pgfscope}%
\pgfsys@transformshift{2.534225in}{0.636086in}%
\pgfsys@useobject{currentmarker}{}%
\end{pgfscope}%
\end{pgfscope}%
\begin{pgfscope}%
\pgfpathrectangle{\pgfqpoint{0.539608in}{0.464787in}}{\pgfqpoint{1.993620in}{2.110085in}} %
\pgfusepath{clip}%
\pgfsetbuttcap%
\pgfsetmiterjoin%
\definecolor{currentfill}{rgb}{1.000000,1.000000,1.000000}%
\pgfsetfillcolor{currentfill}%
\pgfsetlinewidth{0.501875pt}%
\definecolor{currentstroke}{rgb}{0.000000,0.000000,0.000000}%
\pgfsetstrokecolor{currentstroke}%
\pgfsetdash{}{0pt}%
\pgfsys@defobject{currentmarker}{\pgfqpoint{-0.020833in}{-0.020833in}}{\pgfqpoint{0.020833in}{0.020833in}}{%
\pgfpathmoveto{\pgfqpoint{-0.000000in}{-0.020833in}}%
\pgfpathlineto{\pgfqpoint{0.020833in}{0.020833in}}%
\pgfpathlineto{\pgfqpoint{-0.020833in}{0.020833in}}%
\pgfpathclose%
\pgfusepath{stroke,fill}%
}%
\begin{pgfscope}%
\pgfsys@transformshift{0.540605in}{2.574873in}%
\pgfsys@useobject{currentmarker}{}%
\end{pgfscope}%
\begin{pgfscope}%
\pgfsys@transformshift{0.640286in}{2.573657in}%
\pgfsys@useobject{currentmarker}{}%
\end{pgfscope}%
\begin{pgfscope}%
\pgfsys@transformshift{0.739967in}{2.565400in}%
\pgfsys@useobject{currentmarker}{}%
\end{pgfscope}%
\begin{pgfscope}%
\pgfsys@transformshift{0.839648in}{2.543796in}%
\pgfsys@useobject{currentmarker}{}%
\end{pgfscope}%
\begin{pgfscope}%
\pgfsys@transformshift{0.939329in}{2.504318in}%
\pgfsys@useobject{currentmarker}{}%
\end{pgfscope}%
\begin{pgfscope}%
\pgfsys@transformshift{1.039010in}{2.442758in}%
\pgfsys@useobject{currentmarker}{}%
\end{pgfscope}%
\begin{pgfscope}%
\pgfsys@transformshift{1.138691in}{2.359876in}%
\pgfsys@useobject{currentmarker}{}%
\end{pgfscope}%
\begin{pgfscope}%
\pgfsys@transformshift{1.238372in}{2.254841in}%
\pgfsys@useobject{currentmarker}{}%
\end{pgfscope}%
\begin{pgfscope}%
\pgfsys@transformshift{1.338053in}{2.132245in}%
\pgfsys@useobject{currentmarker}{}%
\end{pgfscope}%
\begin{pgfscope}%
\pgfsys@transformshift{1.437734in}{1.998689in}%
\pgfsys@useobject{currentmarker}{}%
\end{pgfscope}%
\begin{pgfscope}%
\pgfsys@transformshift{1.537415in}{1.859319in}%
\pgfsys@useobject{currentmarker}{}%
\end{pgfscope}%
\begin{pgfscope}%
\pgfsys@transformshift{1.637096in}{1.723688in}%
\pgfsys@useobject{currentmarker}{}%
\end{pgfscope}%
\begin{pgfscope}%
\pgfsys@transformshift{1.736777in}{1.599769in}%
\pgfsys@useobject{currentmarker}{}%
\end{pgfscope}%
\begin{pgfscope}%
\pgfsys@transformshift{1.836458in}{1.497605in}%
\pgfsys@useobject{currentmarker}{}%
\end{pgfscope}%
\begin{pgfscope}%
\pgfsys@transformshift{1.936139in}{1.427796in}%
\pgfsys@useobject{currentmarker}{}%
\end{pgfscope}%
\begin{pgfscope}%
\pgfsys@transformshift{2.035820in}{1.401958in}%
\pgfsys@useobject{currentmarker}{}%
\end{pgfscope}%
\begin{pgfscope}%
\pgfsys@transformshift{2.135501in}{1.401958in}%
\pgfsys@useobject{currentmarker}{}%
\end{pgfscope}%
\begin{pgfscope}%
\pgfsys@transformshift{2.235182in}{1.401958in}%
\pgfsys@useobject{currentmarker}{}%
\end{pgfscope}%
\begin{pgfscope}%
\pgfsys@transformshift{2.334863in}{1.401958in}%
\pgfsys@useobject{currentmarker}{}%
\end{pgfscope}%
\begin{pgfscope}%
\pgfsys@transformshift{2.434544in}{1.401958in}%
\pgfsys@useobject{currentmarker}{}%
\end{pgfscope}%
\begin{pgfscope}%
\pgfsys@transformshift{2.534225in}{1.401958in}%
\pgfsys@useobject{currentmarker}{}%
\end{pgfscope}%
\end{pgfscope}%
\begin{pgfscope}%
\pgftext[left,bottom,x=0.639289in,y=0.707849in,rotate=0.000000]{{\rmfamily\fontsize{9.000000}{10.800000}\selectfont \(\displaystyle a^-/a^+ = 1.2\)}}
\end{pgfscope}%
\begin{pgfscope}%
\pgftext[left,bottom,x=0.639289in,y=1.034912in,rotate=0.000000]{{\rmfamily\fontsize{9.000000}{10.800000}\selectfont \(\displaystyle a^-/a^+ = 1.5\)}}
\end{pgfscope}%
\begin{pgfscope}%
\pgftext[left,bottom,x=0.639289in,y=1.435828in,rotate=0.000000]{{\rmfamily\fontsize{9.000000}{10.800000}\selectfont \(\displaystyle a^-/a^+ = 2.0\)}}
\end{pgfscope}%
\begin{pgfscope}%
\pgfsetrectcap%
\pgfsetroundjoin%
\definecolor{currentfill}{rgb}{1.000000,1.000000,1.000000}%
\pgfsetfillcolor{currentfill}%
\pgfsetlinewidth{0.501875pt}%
\definecolor{currentstroke}{rgb}{0.000000,0.000000,0.000000}%
\pgfsetstrokecolor{currentstroke}%
\pgfsetdash{}{0pt}%
\pgfpathmoveto{\pgfqpoint{1.490681in}{1.937790in}}%
\pgfpathlineto{\pgfqpoint{2.477672in}{1.937790in}}%
\pgfpathlineto{\pgfqpoint{2.477672in}{2.519317in}}%
\pgfpathlineto{\pgfqpoint{1.490681in}{2.519317in}}%
\pgfpathlineto{\pgfqpoint{1.490681in}{1.937790in}}%
\pgfpathclose%
\pgfusepath{stroke,fill}%
\end{pgfscope}%
\begin{pgfscope}%
\pgfsetbuttcap%
\pgfsetroundjoin%
\definecolor{currentfill}{rgb}{1.000000,1.000000,1.000000}%
\pgfsetfillcolor{currentfill}%
\pgfsetlinewidth{0.501875pt}%
\definecolor{currentstroke}{rgb}{0.000000,0.000000,0.000000}%
\pgfsetstrokecolor{currentstroke}%
\pgfsetdash{}{0pt}%
\pgfsys@defobject{currentmarker}{\pgfqpoint{-0.020833in}{-0.020833in}}{\pgfqpoint{0.020833in}{0.020833in}}{%
\pgfpathmoveto{\pgfqpoint{0.000000in}{-0.020833in}}%
\pgfpathcurveto{\pgfqpoint{0.005525in}{-0.020833in}}{\pgfqpoint{0.010825in}{-0.018638in}}{\pgfqpoint{0.014731in}{-0.014731in}}%
\pgfpathcurveto{\pgfqpoint{0.018638in}{-0.010825in}}{\pgfqpoint{0.020833in}{-0.005525in}}{\pgfqpoint{0.020833in}{0.000000in}}%
\pgfpathcurveto{\pgfqpoint{0.020833in}{0.005525in}}{\pgfqpoint{0.018638in}{0.010825in}}{\pgfqpoint{0.014731in}{0.014731in}}%
\pgfpathcurveto{\pgfqpoint{0.010825in}{0.018638in}}{\pgfqpoint{0.005525in}{0.020833in}}{\pgfqpoint{0.000000in}{0.020833in}}%
\pgfpathcurveto{\pgfqpoint{-0.005525in}{0.020833in}}{\pgfqpoint{-0.010825in}{0.018638in}}{\pgfqpoint{-0.014731in}{0.014731in}}%
\pgfpathcurveto{\pgfqpoint{-0.018638in}{0.010825in}}{\pgfqpoint{-0.020833in}{0.005525in}}{\pgfqpoint{-0.020833in}{0.000000in}}%
\pgfpathcurveto{\pgfqpoint{-0.020833in}{-0.005525in}}{\pgfqpoint{-0.018638in}{-0.010825in}}{\pgfqpoint{-0.014731in}{-0.014731in}}%
\pgfpathcurveto{\pgfqpoint{-0.010825in}{-0.018638in}}{\pgfqpoint{-0.005525in}{-0.020833in}}{\pgfqpoint{0.000000in}{-0.020833in}}%
\pgfpathclose%
\pgfusepath{stroke,fill}%
}%
\begin{pgfscope}%
\pgfsys@transformshift{1.646237in}{2.414364in}%
\pgfsys@useobject{currentmarker}{}%
\end{pgfscope}%
\end{pgfscope}%
\begin{pgfscope}%
\pgftext[left,bottom,x=1.768459in,y=2.347697in,rotate=0.000000]{{\rmfamily\fontsize{8.000000}{9.600000}\selectfont \(\displaystyle a^-/a^+ = 1.5\)}}
\end{pgfscope}%
\begin{pgfscope}%
\pgfsetbuttcap%
\pgfsetmiterjoin%
\definecolor{currentfill}{rgb}{1.000000,1.000000,1.000000}%
\pgfsetfillcolor{currentfill}%
\pgfsetlinewidth{0.501875pt}%
\definecolor{currentstroke}{rgb}{0.000000,0.000000,0.000000}%
\pgfsetstrokecolor{currentstroke}%
\pgfsetdash{}{0pt}%
\pgfsys@defobject{currentmarker}{\pgfqpoint{-0.020833in}{-0.020833in}}{\pgfqpoint{0.020833in}{0.020833in}}{%
\pgfpathmoveto{\pgfqpoint{-0.020833in}{-0.020833in}}%
\pgfpathlineto{\pgfqpoint{0.020833in}{-0.020833in}}%
\pgfpathlineto{\pgfqpoint{0.020833in}{0.020833in}}%
\pgfpathlineto{\pgfqpoint{-0.020833in}{0.020833in}}%
\pgfpathclose%
\pgfusepath{stroke,fill}%
}%
\begin{pgfscope}%
\pgfsys@transformshift{1.646237in}{2.231633in}%
\pgfsys@useobject{currentmarker}{}%
\end{pgfscope}%
\end{pgfscope}%
\begin{pgfscope}%
\pgftext[left,bottom,x=1.768459in,y=2.164966in,rotate=0.000000]{{\rmfamily\fontsize{8.000000}{9.600000}\selectfont \(\displaystyle a^-/a^+ = 1.2\)}}
\end{pgfscope}%
\begin{pgfscope}%
\pgfsetbuttcap%
\pgfsetmiterjoin%
\definecolor{currentfill}{rgb}{1.000000,1.000000,1.000000}%
\pgfsetfillcolor{currentfill}%
\pgfsetlinewidth{0.501875pt}%
\definecolor{currentstroke}{rgb}{0.000000,0.000000,0.000000}%
\pgfsetstrokecolor{currentstroke}%
\pgfsetdash{}{0pt}%
\pgfsys@defobject{currentmarker}{\pgfqpoint{-0.020833in}{-0.020833in}}{\pgfqpoint{0.020833in}{0.020833in}}{%
\pgfpathmoveto{\pgfqpoint{-0.000000in}{-0.020833in}}%
\pgfpathlineto{\pgfqpoint{0.020833in}{0.020833in}}%
\pgfpathlineto{\pgfqpoint{-0.020833in}{0.020833in}}%
\pgfpathclose%
\pgfusepath{stroke,fill}%
}%
\begin{pgfscope}%
\pgfsys@transformshift{1.646237in}{2.048901in}%
\pgfsys@useobject{currentmarker}{}%
\end{pgfscope}%
\end{pgfscope}%
\begin{pgfscope}%
\pgftext[left,bottom,x=1.768459in,y=1.982235in,rotate=0.000000]{{\rmfamily\fontsize{8.000000}{9.600000}\selectfont \(\displaystyle a^-/a^+ = 2.0\)}}
\end{pgfscope}%
\end{pgfpicture}%
\makeatother%
\endgroup%